%% file: thesis.tex
\documentclass{bitsthesis}
\usepackage{epsfig}
\usepackage{times}
\usepackage{graphicx}
\usepackage{amssymb,amsmath} 
\usepackage{indentfirst}
\usepackage[small]{caption2}
\usepackage[colorlinks]{hyperref} 
\usepackage{soul,threeparttable,multirow,float,hhline}
\usepackage{booktabs}
\usepackage[babel,german=quotes]{csquotes} % Für gute Anführungszeichen

\hypersetup{
    colorlinks,%
    citecolor=blue,%
    filecolor=black,%
    linkcolor=blue,%
    urlcolor=magenta
}
\usepackage[sort]{natbib}
\bibpunct{[}{]}{,}{a}{}{,~}
\usepackage{textcomp}  % text version of symbols in TS1, e.g. \textrightarrow{}
\usepackage{amstext} % proper support of frequently used \text{} command
\newif\ifshowcitations\showcitationsfalse%
\newif\ifshowlinks\showlinksfalse%

\showcitationstrue
\ifshowlinks%
  \newcommand*{\inspireurl}[1]{\href{#1}{entry}}
\else
  \makeatletter
  \newcommand*{\inspireurl}[1]{\@bsphack\@esphack}
  \makeatother
\fi
\ifshowcitations%
  \newcommand*{\citations}[1]{\\* {\blue #1}}
\else
  \makeatletter
  \newcommand*{\citations}[1]{\@bsphack\@esphack}
  \makeatother
\fi

\usepackage{array}
\usepackage{wrapfig}
\graphicspath{{figures/}}
\usepackage{alltt}  % 

\usepackage{afterpage}

\newcommand\blankpage{%
    \null
    \thispagestyle{empty}%
    \addtocounter{page}{-1}%
    \newpage}
\def\be{\begin{eqnarray}}
\def\ee{\end{eqnarray}}

\def\ru1{\rule[-0.4truecm]{0mm}{1truecm}} %brod
\def\Im{\hbox{\rm Im\,}}

\def          % circa \ge
\simeq{{\ \lower2pt\hbox{$-$}\mkern-13mu \raise2pt \hbox{$\sim$}\ }} 
\def          % circa >/\ge
\simge{{\ \lower2pt\hbox{$\sim$}\mkern-13mu \raise2pt \hbox{$>$}\ }}

\global\arraycolsep=2pt %reduces the separation in eqnarrays
\input{epsf}

\def\eq#1\en{\begin{equation}#1\end{equation}}
\def\s[#1,#2]{[#1\stackrel{\star}{,}#2]}

\newlength{\myVSpace}% the height of the box
\makeatletter
\def\fmslash{\@ifnextchar[{\fmsl@sh}{\fmsl@sh[0mu]}}
\def\fmsl@sh[#1]#2{%
  \mathchoice
    {\@fmsl@sh\displaystyle{#1}{#2}}%
    {\@fmsl@sh\textstyle{#1}{#2}}%
    {\@fmsl@sh\scriptstyle{#1}{#2}}%
    {\@fmsl@sh\scriptscriptstyle{#1}{#2}}}
\def\@fmsl@sh#1#2#3{\m@th\ooalign{$\hfil#1\mkern#2/\hfil$\crcr$#1#3$}}
\makeatother
\usepackage{wasysym} 
\usepackage{mathrsfs} 
\usepackage{float}
\usepackage{amsfonts} 
\usepackage{amsbsy} 
\usepackage{amscd} 
\usepackage{pstricks} 
\usepackage{multirow}
\usepackage{slashed} 
\usepackage{tikz}
\usetikzlibrary{decorations.pathmorphing,decorations.markings}

\newcommand{\ms}{\scriptscriptstyle}
\newcommand{\beq}{\begin{equation}}
\newcommand{\eeq}{\end{equation}}
\newcommand{\bea}{\begin{eqnarray}}
\newcommand{\eea}{\end{eqnarray}}
\newcommand{\beas}{\begin{eqnarray*}}
\newcommand{\eeas}{\end{eqnarray*}}
\newcommand{\bi}{\begin{itemize}}
\newcommand{\ei}{\end{itemize}}
\usepackage{changepage}
\def\non {\nonumber\\}
\def\nonu {\nonumber}

\def\tev{\,{\ifmmode\mathrm {TeV}\else TeV\fi}}
\def\gev{\,{\ifmmode\mathrm {GeV}\else GeV\fi}}
\def\to{\rightarrow}

\usepackage{color}

\allowdisplaybreaks
\def\Re{{\cal R \mskip-4mu \lower.1ex \hbox{\it e}\,}}
\def\Im{{\cal I \mskip-5mu \lower.1ex \hbox{\it m}\,}}

\def\tev{\,{\ifmmode\mathrm {TeV}\else TeV\fi}}
\def\gev{\,{\ifmmode\mathrm {GeV}\else GeV\fi}}
\def\mev{\,{\ifmmode\mathrm {MeV}\else MeV\fi}}
\def\to{\rightarrow}

\renewcommand\descriptionlabel[1]{$\bullet$ \textbf{#1}}

\usepackage[english]{babel}
\usepackage{graphicx}
\usepackage{wrapfig}
\usepackage{lipsum}
 \usepackage{amssymb}
 \usepackage{slashed}
\begin{document}
% Reference Macros:  Enter parameters in order Vol, Page, Year
\def\issue(#1,#2,#3){ #1, #2 (#3)} % AIP format

\def\APP(#1,#2,#3){Acta Phys.\ Polon.\ \issue(#1,#2,#3)}
\def\ARNPS(#1,#2,#3){Ann.\ Rev.\ Nucl.\ Part.\ Sci.\ \issue(#1,#2,#3)}
\def\CPC(#1,#2,#3){Comp.\ Phys.\ Comm.\ \issue(#1,#2,#3)}
\def\CIP(#1,#2,#3){Comput.\ Phys.\ \issue(#1,#2,#3)}
\def\EPJC(#1,#2,#3){Eur.\ Phys.\ J.\ C\ \issue(#1,#2,#3)}
\def\EPJD(#1,#2,#3){Eur.\ Phys.\ J. Direct\ C\ \issue(#1,#2,#3)}
\def\IEEETNS(#1,#2,#3){IEEE Trans.\ Nucl.\ Sci.\ \issue(#1,#2,#3)}
\def\IJMP(#1,#2,#3){Int.\ J.\ Mod.\ Phys. \issue(#1,#2,#3)}
\def\JHEP(#1,#2,#3){JHEP \issue(#1,#2,#3)}
\def\JPG(#1,#2,#3){J.\ Phys.\ G \issue(#1,#2,#3)}
\def\MPL(#1,#2,#3){Mod.\ Phys.\ Lett.\ \issue(#1,#2,#3)}
\def\NP(#1,#2,#3){Nucl.\ Phys.\ \issue(#1,#2,#3)}
\def\NIM(#1,#2,#3){Nucl.\ Instrum.\ Meth.\ \issue(#1,#2,#3)}
\def\PL(#1,#2,#3){Phys.\ Lett.\ \issue(#1,#2,#3)}
\def\PRD(#1,#2,#3){Phys.\ Rev.\ D \issue(#1,#2,#3)}
\def\PRL(#1,#2,#3){Phys.\ Rev.\ Lett.\ \issue(#1,#2,#3)}
\def\PTP(#1,#2,#3){Progs.\ Theo.\ Phys. \ \issue(#1,#2,#3)}
\def\RMP(#1,#2,#3){Rev.\ Mod.\ Phys.\ \issue(#1,#2,#3)}
\def\SJNP(#1,#2,#3){Sov.\ J. Nucl.\ Phys.\ \issue(#1,#2,#3)}

%=====================================================================
% Include the prelude for Title page, abstract, table of contents, etc
% You need to modify it to contain your details then
% Include the technical part of the report\include{chapter1}
%=====================================================================

%================================================
%                  Chapters
%================================================
\include{prelude} %
\include{chapter1/introduction} % 
\include{chapter2/chapter2} %
\include{chapter3/chapter3} %
\include{chapter3/chapter3a}%
\include{chapter4/chapter4} % 
\include{chapter5/chapter5} % 
\include{chapter6/conclusion} %
\include{chapter6/future-work}%
%\include{chapter7/conclusion} %
%================================================
%                   Bibliography
%================================================
%\addcontentsline{toc}{chapter}{Bibliography}
%\renewcommand{\appendixname}{Bibliography}
%$\include{bibliography}%

\bibliographystyle{jmb}
\bibliography{thesis}{}
\end{document}

%% file: prelude.tex
% prelude.tex
%   - titlepage
%   - dedication (optional)
%   - approval sheet
%   - course certificate
%   - table of contents, list of tables and list of figures
%   - nomenclature
%   - abstract
%============================================================================

% This makes the page numbers Roman (i, ii, etc)
\clearpage\pagenumbering{roman}

% TITLE PAGE
%   - define \title{} \author{} 
\title{ Bayesian and Principal Component Analyses of Neutron Star Properties} %\\ \vspace{2mm} For Protons In Position Space}
%{Unveiling Neutron Star Properties: A Bayesian and Multivariate Analysis of Equations of State and Symmetry Energy Parameters}
%{Bayesian Analysis of Neutron Star Properties and Nuclear Matter Parameters: A Comprehensive Study of Equations of State and Symmetry Energy}
\author{\href{https://orcid.org/0000-0003-0103-5590}{ Naresh Kumar Patra \includegraphics[scale=0.06]{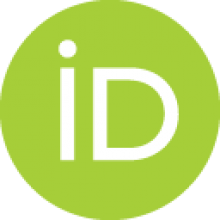}}}
\date{April - 2024}

%  - Roll number, required for title page, approval sheet, and
%    certificate of course work 
\rollnum{2019PHXF0033G} 

%   - The default degree is ``Doctor of Philosophy''
%     (unless the document style msthesis is specified
%      and then the default degree is ``Master of Science'')
%     Degree can be changed using the command \bitsbdegree{}
\bitsbdegree{ \textsc{DOCTOR OF PHILOSOPHY}}

%   - The default report type is preliminary report.
%      * for a PhD thesis, specify \thesis
\thesis
%      * for a M.Tech./M.Phil./M.Des./M.S. dissertation, specify \dissertation
%\dissertation
%      * for a DIIT/B.Tech./M.Sc.project report, specify \project
%\project
%      * for any other type, use  \reporttype{}
%\reporttype{ReportType}

%   - The default department is ``Unknown Department''
%     The department can be changed using the command \department{}
\department{DEPARTMENT OF PHYSICS}

%    - Set the guide's name
\setguide{Prof. Tarun Kumar Jha~}

%\setexguide{Prof. Bijay Kumar Agrawal~(Co-supervisor)}

%    - Set external guide (if you have one)
%\setexguide{Prof External Guide}

%   - once the above are defined, use \maketitle to generate the titlepage
\maketitle

%--------------------------------------------------------------------%
% DEDICATION
%   Dedications, if any, must be first page after title page.
\begin{dedication}
\begin{center}

{\Large{\bf  \textit{Thesis is dedicated to my Ph.D. advisors (Prof. Bijay Kumar Agrawal \\and Prof. Tarun Kumar Jha), my Father, Mother, Brothers, \\and my Friends}}}
% \\ and 
% \vspace{0.3in} \\
%\begin{figure}
%  \includegraphics[width=1.8in,height=2.2in]{Ramanujan.jpg}\\
%  \textit{Srinivasa Ramanujan (1887-1920)}
%\end{figure} 
\end{center}
\end{dedication}

%--------------------------------------------------------------------%
% APPROVAL SHEET
%   - for final thesis, you need Approval Sheet. So, uncomment the
%     \makeapproval command.
%     it should come after dedication, if dedication is
%     present. Otherwise it is the first page after title page.
%\makeapproval

%--------------------------------------------------------------------%
% CERTIFICATE OF COURSE WORK
%   - for final thesis, a course certificate is required.
%   - specify the  PhD joining date for the certificate.
%     Contact you department office or academic office if you do not
%     know it.

% \joiningdate{Aug 2012}
% 
% 
% 
% 
\begin{AnnexureIIIcertificate}
   %\begin{center}
   \vspace{0.8in}
   {\small 
   \begin{tabular}{ll}
  &\hspace{3.5in} \includegraphics[width=5.5cm,height=4.0cm,keepaspectratio]{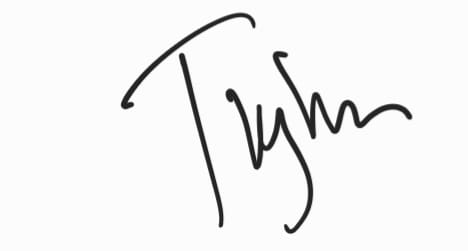} \vspace{-0.1in} \hspace{-1in}\\ 
 &\hspace{4.0in}Signature of the Supervisor  \\
 &\hspace{4.0in}Name:~ {\bf Prof.~Tarun~ Kumar~ Jha}  \\  
 &\hspace{4.0in}Designation: Associate professor   \\
 &\hspace{4.0in}Affiliation:~ BITS Pilani, Goa, India  \\
 &\hspace{4.0in}Date: April 24, 2024 \\
 \end{tabular}} % End of Course Certificate

\end{AnnexureIIIcertificate}

%--------------------------------------------------------------------%
% COPYRIGHT PAGE
%   - To include a copyright page use \copyrightpage
% \copyrightpage

%--------------------------------------------------------------------%
% ABSTRACT
% 
%------------------------------------------------------------------------------------------------%

   %\addcontentsline{toc}{chapter}{\Large Abstract}
   %\renewcommand{\appendixname}{\Large ABSTRACT}
  %\appendix
\newpage
%\begin{abstract}
\vspace{15mm}
\begin{center}
    \Large{\bf \textsc{Abstract}}
\end{center} 
\vspace{15mm}
A Bayesian method is used in this extensive work to generate a large set of minimally constrained equations of state (EOSs) for matters in neutron stars (NS). These EOSs are analyzed for their correlations with key NS properties, such as the tidal deformability, radius, and maximum mass, within the mass range of $1.2-2M_\odot$. The observed connections between the pressure of $\beta$-equilibrated matter and the properties of neutron stars at different densities offer significant insights into the behavior of NS matter in a nearly model-independent manner. The study also examines the influence of various factors on the correlation of symmetry energy parameters, such as slope and curvature parameters at saturation density ($\rho_0=0.16 ~\text{fm}^{-3}$) with the tidal deformability and radius of neutron stars. This study investigates the robustness of the observed correlations by considering the distributions and interdependence of symmetry energy parameters. Furthermore, the utilization of Principal Component Analysis (PCA) is employed to unveil the complicated relationship between various nuclear matter parameters and properties of neutron stars. This analysis highlights the importance of employing multivariate analysis techniques in order to comprehend the variety in tidal deformability and radius observed across distinct masses of NS. This comprehensive study aims to establish a connection between the parameters of nuclear matter and the properties of neutron stars, providing significant insights into the behavior of NS matter across different circumstances.

%\end{abstract}

%--------------------------------------------------------------------%
\begin{declaration} % Don't write anything here.The official formate already done in the thesis.cls file.

\end{declaration}

%  ----------------------------------------------------------------------%
 \begin{acknowledgments}

\vspace{-0.2in}
\textit{First of all, I would like to extend my heartfelt gratitude to my family, especially my parents and two older brothers, for their unwavering support throughout my Ph.D. journey. Their constant encouragement, sacrifices, and belief in my abilities have been the driving force behind my academic pursuit. Their love, guidance, and understanding have been my pillars of strength, and I am deeply thankful for their sacrifices and enduring faith in me. This thesis is a culmination of not just my efforts but also a testament to their enduring support and love.}

%about my co-supervisor
\textit{I owe immeasurable gratitude to Prof. Bijay Kumar Agrawal, who not only served as my project supervisor but also became a profound mentor and life advisor. It was under his guidance that I was first introduced to the captivating world of theoretical nuclear physics. I developed a strong affection for the subject because of Prof. Agrawal's insatiable inquiry and constant enthusiasm. His detailed input and astute suggestions were crucial in helping me polish my work and ensure that it was finished within the allotted time period. He helped me understand the complexities of my research topic as well as the importance of perseverance and dedication in academic work. However, his influence went far beyond academics. Through our interactions, I learned the invaluable lesson that the path of integrity and resilience is invariably the right one in life. Furthermore, I would like to extend my heartfelt acknowledgement to Mrs. Tanuja Agrawal, who offered invaluable assistance during challenging technical moments. Her support and guidance served as a beacon of inspiration in navigating the intricate puzzle that life often presents. No words can adequately express my gratitude for their enduring support and inspiration.}

\textit{I am immensely grateful to Dr. Tuhin Malik, not only for being an exceptional senior and collaborator but also for his mentorship, guidance, and unwavering belief in my research. Dr. Tuhin's insights, expertise, and dedication have been instrumental in shaping the success of our collaborative efforts. His willingness to share knowledge, brainstorm ideas, and provide valuable feedback has been invaluable to the progress of my Ph.D. I feel privileged to have had the opportunity to work alongside such a talented and supportive colleague. Dr. Tuhin, your contribution to my academic journey has been truly invaluable, and I thank you wholeheartedly for your mentorship and collaboration.}

%about collaborators
\textit{I take great pleasure in expressing my gratitude towards Prof. Constança Providência and Dr. Tuhin Malik for collaborating at different phases of this thesis work, who had also hosted me at the University of Coimbra, offering very warm hospitality during August 2023. 
I take this opportunity to thank all my other collaborators, Prof. Hiranmaya Mishra, Prof. Bharat Kumar Sharma, Prof. Arunava Mukherjee, Mr. Sk Md Adil Imam, Mr. Anagh Venneti, and Mr. Prafulla Saxsena.}

%about funding
\textit{I gratefully acknowledge the financial support provided by the Department of Science and Technology, Ministry of Science and Technology, India (DST-INSPIRE). The support was also extended for attending several schools, workshops, and conferences.}

%about my msc teachers
\textit{I wish to express my sincere gratitude to the professors who played a pivotal role in guiding my research journey during my master's studies. My favourite professor is Prof. R. Venkatachalapathy, whose competence, guidance, and dedication have shaped my academic career. Prof. Venkatachalapathy's insightful advice and continuous support shaped my research interests and talents. Dr. G. Viruthagiri and Dr. H. Saleem's skills and mentorship have greatly influenced my study. Their advice, support, and intellectual insights have improved my studies and expertise.}

%about my 12th teachers
\textit{I would like to extend my heartfelt acknowledgments to the teachers who played a significant role in shaping my early scientific journey during my twelfth class. Mrs. Abanti Sarangi, my favorite teacher, always inspired and motivated me. Her love for teaching and ability to simplify complex subjects left a lasting effect. She inspired me to love learning and the sciences throughout my academic career. I also appreciate my physics and chemistry teachers, Mr. Diptikant Das and Mr. Manasha Ranjan Puhan. Both teachers fostered my interest in the disciplines and gave me a solid science foundation. Their attention to teaching and patience helped me improve academically throughout those important years.}

%about best friends
\textit{Life has blessed me with a lot of good friends who often showed me the best ways to deal with a problem and become the best teachers. Manish, Ram, Harsh, Pupu, Ashmita, and Rajesh, without you, life would not be this beautiful. I have very fond memories of those birthday celebrations with them. I spent most of my time with Manish, Ram, and Harsh in the last five years of my life. The discussions with them and their suggestions gave me a clear view of my personal life during the Ph.D. journey. Satyanarayana, Rudra, and Satyabrata, it has been a pleasure having you in my life. Discussions with Satyanarayana (Bobulu) probably brought out the best of my way of life. Whatever doubts or difficulties I used to have during my life journey, Bobulu always had the best words to describe it.}

%about my supervisor
\textit{This thesis work is done in the Physics Department of BITS Pilani, K K Birla Goa Campus. Surely, it
would not have been possible without the assistance and support of many people present.
Foremost, I want to express my sincere gratitude to Prof. Tarun Kumar Jha from the Physics Department at BITS Pilani K K Birla Goa Campus, who oversaw my thesis. Prof. Jha has been a crucial part of my Ph.D. path, offering constant support and priceless advice each step of the way. My study has been motivated by his mentorship, and I am incredibly appreciative of his persistent guidance, ongoing support, and extraordinary patience. Prof. Jha's mastery of the subject has expanded my knowledge and motivated me to improve my research abilities. My Ph.D. experience has been forever changed by Prof. Tarun Kumar Jha's commitment to my academic and personal development, and I am incredibly grateful to have had the opportunity to work with him. I am really grateful for his support, which has helped shape my research and foster my intellectual growth.}

%about the department
\textit{I thank my Doctoral Advisory Committee members, Prof. Chandradew Sharma and Dr. Sunil Kumar Vattezhath, Department of Physics, for their insightful comments, useful suggestions, and encouragement that have helped me greatly to strengthen my research ideas from
various perspectives. I would also like to thank Prof. Radhika Vathsan (head of the department) and all the other faculty members of the department for their
continuous encouragement and support. I particularly thank my pre-Ph.D. coursework instructor Prof. Toby Joseph, Prof. Prasanta Kumar Das, Prof. Gaurav
Dar, Prof. Kinjal Banerjee, Prof. Prasad Anant Naik, Prof. Raghunath Anand Ratabole, and Prof. Arun Venkatesh Kulkarni, for their kind assistance during the courses.
I am also grateful to Prof. Senthamarai Kannan Ethirajulu (Department Research Committee (DRC) of the Physics Department) for his kindness and help during the five years of my department journey. 
I especially thank all my seniors of BITS Pilani, K K Birla Goa Campus, especially Dr. Tuhin Malik, Dr. Atanu, Dr. Debashree, Dr. Saumyen, Dr. Annu, Dr. Malavika, Dr. Payel, Dr. Sumit, Mr. Sourav, Mr. Kiran, and Mr. Abhay for their kind help and support of all sort. I would like to express my heartfelt thanks to all my Ph.D. friends of BITS Pilani, K K Birla Goa Campus. I
consider myself extremely fortunate to have a friend like Manish. I cannot thank him enough for his
help, unconditional support at tough times, and for everything.  It is said that work becomes more enjoyable with good friends at the workplace, and I have realized every bit of it because of them around! I also thank
Harsh, Ashmita, Mirnmoy, Gorav, Bhagya, Prashant, Manoj V, Manoj B, Sourish, Megha, Akshay, Vidyu, Dharmaraj, Sarga, Akhilesh, Sharad, Vrushali, Premchand, Karthik, Indra, Shreelaxmi, and Vaishnavi, for their support.}

%about institute
\textit{I would like to thank Prof. Ramgopal Rao (Vice Chancellor, BITS Pilani), Prof. Suman Kundu (Director, BITS Pilani, K K Birla Goa Campus), Prof. M. B. Srinivas (Dean, Academic Graduate Studies and Research Division, BITS Pilani Goa), and Prof. Bharat Deshpande (Associate Dean, Academic Graduate Studies and Research Division).
I thank our laboratory staff of the Department of Physics, Mr. Arun Naik, Mr. Rupesh Walke, Krishanu Pal, and Ms. Meenakshi Ambegar. I would also like to thank Mr. Pratap Behera of the Academic Graduate
Studies and Research Division Office and Ms. Niyata Barve of the Department of Accounts and Finance
for their kind help throughout the period of my Ph.D.}

\textit{I am also thankful to many research scholars and post-doctoral scholars all across the globe for
beneficial interactions and discussions. Of them, I particularly thank Dr. Harish Chandra Das (INFN, Italy), Dr. Prasad ( ICTS, Bangalore), Dr. Saraswati Pandey (BHU), Mr. Bikram Keshari Pradhan (IUCAA, Pune), Mrs. Tamanna Iqbal (University of Lucknow), Mrs. Gauri Tiwari (BHU), Mr. Jeet Amrit Pattnaik (SOA University), Mr. Pinku Routray(NIT Rourkela), Mr. Sagnik Charterjee(IISER Bhopal), Mr. Bramha Panigrahi (IIT Hyderabad), Mrs. Ipsita Nayak (IIT Hyderabad), Mr. Rajesh Kumar Roul (IIT Hyderabad), Mr. Satyasiban Dash (IIT Chennai), Mr. Smrutiranjan Sahu (GEU, Dehradun) and Dinesh Kumar Singha(CUH)  for also being nice friends.}

\textit{I am also thankful to my school and college friends, especially my best friends  Kiran, Hiranmaya, Shreetam, Pritam, Preeti, Himansu, Vipul, Priyabrata, Pritish, Kumar Deba, Chandan, Sabitri, Amlan, Suvendu, Nila, Anand Pravu, Tamil Arivu, and Jitendra for being my pillars of support during all hardships. Any amount of thanks or gratitude will not be enough for the unconditional love and support from my parents throughout the phase of my Ph.D. Above all, I am thankful to the almighty for giving me the strength to pursue my Ph.D. work through all odds and fulfill my dream.}

 \end{acknowledgments}
 
 \afterpage{\blankpage}
% %--------------------------------------------------------------------%
%\addcontentsline{toc}{chapter}{Contents}
 \renewcommand{\appendixname}{Appendix}
% \appendix
% % CONTENTS, TABLES, FIGURES
  \tableofcontents
  {\listoftables}
  {\listoffigures}
  
%=============================================================================
%                                Nomenclature
%=============================================================================
% NOMENCLATURE
\begin{nomenclature}
\vspace{-.3in}
\textbf{ \Large Physical constants}
 \begin{table}[h]
  \begin{center}
  \begin{tabular}{l l l} \\ \hline
  Quantity                         & Symbol            & Value \\ \hline
  speed of light in vacuum         & $c$               & $299,792,458\, m/s$               \\ 
  Planck constant                  & $h$               & $6.62607 \times 10^{-34}\, Js$    \\
  Planck constant, reduced         & $\hslash$         & $1.0545 \times 10^{-34}\, Js$     \\
  electron charge magnitude        & $e$               & $1.6021  \times 10^{-19}\, C$     \\
  conversion constant              & $\hslash c$       & $197.326\,MeV\,fm$                \\
  electron mass                    & $m_{e}$           & $0.510\,MeV/c^{2}$                \\
  proton mass                      & $m_{p}$           & $938.272\,MeV/c^{2}$              \\
  fine-structure constant          & $\alpha=e^{2}/4 \pi \epsilon_{0} \hslash c$ & $1/137.035$ \\
  Fermi coupling constant          & $G_{F}/(\hslash c)^{3}$  & $6.6740 \times 10^{-11}\,m^{3}Kg^{-1}s^{-2}$ \\  \hline
  \end{tabular}
  \end{center}
 \end{table}
\end{nomenclature}
 
%----------------------------------------------------------------
%ABBREVIATION
%----------------------------------------------------------------
\begin{abbreviation}
\vspace{-.3in}
 \begin{table}[h]
  %\begin{center}
\begin{tabular}{l l } 
ANM 	&	Asymmetric Nuclear matter	\\
BNS 	&	Binary Neutron Star	\\
BPS 	&	Baym-Pethick-Sutherland	\\
CMB 	&	Cosmic Microwave Background	\\
EFT 	&	Effective Field Theory	\\
EOS 	&	Equation of State	\\
GW      &   Gravitational-Wave\\
GEO 	&	Gravitational-Wave Observatory	\\
GTR 	&	General Theory of Relativity	\\
HIC     &   Heavy Ion Collisions \\
ISGMR 	&	Iso-scalar Giant Monopole Resonance	\\
IVGDR 	&	Iso-vector Giant Dipole Resonance	\\
LIGO 	&	Laser Interferometer Gravitational-Wave Observatory	\\
NICER 	&	Neutron Star Interior Composition Explorer	\\
CREX    & Coherent Reactions Experiment\\
PREX    & Parity Radius Experiment\\
NN 	    &	Nucleon-Nucleon	\\
NS 	    &	Neutron Star	\\
CCSN    & Core Collapse Supernova\\

\end{tabular} 
 %\end{center}
 \end{table}  
 \newpage
 \begin{table}[h]
 \begin{tabular}{l l}
PNS  & Proto-Neutron Star \\
SGR  & Soft Gamma Repeater \\
AXP  & Anomalous X-ray Pulsar \\
NMP  &   Nuclear Matter Parameter\\
PNM  & Pure Neutron Matter\\
BEM  & $\beta-$euilibrium Matter\\
SNM  &	Symmetric Nuclear Matter	\\
TOV  &	Tolman-Oppenheimer-Volkoff	\\
QCD  &	Quantum Chromodynamics	\\
pQCD & Perturbative Quantum Chromodynamics\\
QGP  &	Quark-Gluon Plasma	\\
BA   & Bayesian Analysis \\
PCA  & Principal Component Analysis\\
DNN  & Deep Neural Network \\
PDF  & probability density function
 \end{tabular}
 \end{table}
\end{abbreviation}
%----------------------------------------------------------------------
%KEYWORDS
%----------------------------------------------------------------------
%========================================================================
%                     List of Publications}
%========================================================================
\begin{publication}
\vspace{-18mm}
\begin{center}
\huge{\bf{List of Publications}}
\end{center}
\vspace{3mm}
\begin{enumerate}
  \item 
  {\bf "Establishing connection between neutron star properties and nuclear matter parameters through a comprehensive multivariate analysis"}\\
   {\textbf{N.K.Patra}, Prafulla Saxsena, B. K. Agrawal and T. K. Jha }\\
   {}Phys. Rev. D {\bf 108} {\bf 123015} (2023)\\
  \href{https://doi.org/10.1103/PhysRevC.108.123015}{DOI:10.1103/PhysRevD.108.123015}.
  \item
   {\bf "Systematic analysis of the impacts of symmetry energy parameters on neutron star properties"}\\
   {\textbf{N.K.Patra}, Anagh Venneti, Sk Md Adil Imam,  Arunava Mukherjee, and B.K. Agrawal }\\
 {}Phys. Rev. C {\bf 107} {\bf 055804} (2023)\\
  \href{https://doi.org/10.1103/PhysRevC.107.055804}{DOI:10.1103/PhysRevC.107.055804}.
  \item 
  {\bf "Effect of the $\sigma$-cut potential on the properties of neutron stars with or without a hyperonic core "}\\
  {\textbf{ N. K. Patra}, B. K. Sharma, A. Reghunath, A. K. H. Das, T. K. Jha}\\
 {}Phys. Rev. C {\bf 106} {\bf 055806} (2022)\\
 \href{https://doi.org/10.1103/PhysRevC.106.055806}{DOI:10.1103/PhysRevC.106.055806}.
  \item
  {\bf"Nearly model-independent constraints on dense matter equation of state in a Bayesian approach"} \\
 {\textbf{N.K.Patra},Sk Md Adil Imam, B.K. Agrawal, Arunava Mukherjee and Tuhin Malik  }\\
 %{}arXiv:2110.15776 [nucl-th]\\
 {}Phys. Rev. D {\bf 106} {\bf 043024} (2022)\\
  \href{https://doi.org/10.1103/PhysRevD.106.043024}{DOI:10.1103/PhysRevD.106.043024}.

  \item
  {\bf"Bayesian reconstruction of nuclear matter parameters from the equation of state of neutron star matter"} \\
 {Sk Md Adil Imam, \textbf{N.K.Patra}, C. Mondal, Tuhin Malik and B.K. Agrawal}\\
 %{}arXiv:2110.15776 [nucl-th]\\
{}Phys. Rev. C {\bf 105} {\bf015806} (2022)\\
\href{https://doi.org/10.1103/PhysRevC.105.015806}{DOI: 10.1103/PhysRevC.105.015806}.

 \item 
 {\bf"The Equation of State (EOS) for magnetized nuclear matter and tidal deformability in neutron star merger"}\\
{\textbf{N.K.Patra}, Tuhin Malik, Debashree Sen, T.K. Jha and Hiranmaya Mishra}\\
{}Astrophys. J.{\bf 900} {\bf49} (2020)\\
\href{https://doi.org/10.3847/1538-4357/aba8fc}{DOI: 10.3847/1538-4357/aba8fc}.
\end{enumerate}
\end{publication}
\newpage
\begin{publication}
\begin{center}
\LARGE{\bf{This thesis is based on following publications}}
\end{center}
\vspace*{3mm}
\begin{center}
\begin{enumerate}
 \item
  {\bf"Nearly model-independent constraints on dense matter equation of state in a Bayesian approach"} \\
 {\textbf{N.K.Patra},Sk Md Adil Imam, B.K. Agrawal, Arunava Mukherjee and Tuhin Malik  }\\
 %{}arXiv:2110.15776 [nucl-th]\\
 {}Phys. Rev. D {\bf 106} {\bf 043024} (2022)\\
  \href{https://doi.org/10.1103/PhysRevD.106.043024}{DOI:10.1103/PhysRevD.106.043024}.
  \item
   {\bf "Systematic analysis of the impacts of symmetry energy parameters on neutron star properties"}\\
   {\textbf{N.K.Patra}, Anagh Venneti, Sk Md Adil Imam,  Arunava Mukherjee, and B.K. Agrawal }\\
 {}Phys. Rev. C {\bf 107} {\bf 055804} (2023)\\
  \href{https://doi.org/10.1103/PhysRevC.107.055804}{DOI:10.1103/PhysRevC.107.055804}.
 \item 
  {\bf "Establishing connection between neutron star properties and nuclear matter parameters through a comprehensive multivariate analysis"}\\
   {\textbf{N.K.Patra}, Prafulla Saxsena, B. K. Agrawal and T. K. Jha }\\
   {}Phys. Rev. D {\bf 108} {\bf 123015} (2023)\\
  \href{https://doi.org/10.1103/PhysRevC.108.123015}{DOI:10.1103/PhysRevD.108.123015}.

\end{enumerate}

\end{center}

%\end{center}
% \vspace*{3mm}

\end{publication}
%======================================================================================
 \afterpage{\blankpage}

%-------------------------------------------------------------------------%
\cleardoublepage\pagenumbering{arabic} % Make the page numbers Arabic (1, 2, etc)
%--------------------------------------------------------------------%

%------------------------------------------------------------------------------------------------%

%--------------------------------------------------------------------%

% \cleardoublepage\pagenumbering{arabic} % Make the page numbers Arabic (1, 2, etc)

%% file: chapter1/introduction.tex
\chapter{Introduction}
\label{chap1}

\section{The Neutron Star Enigma: A Brief History}
\subsection{Astrophysical context}
The discovery of the atomic nucleus in 1911 by Ernest Rutherford was a major breakthrough in physics. Rutherford's discovery was a result of his careful observation of the scattering of alpha particles off of a thin gold foil formed by the scattering of alpha particles. A stunning discovery was made as a result of this experiment; previously, it had been believed that the atom was an unbreakable entity; however, it turned out that the atom was actually a complicated structure with a tiny and dense nucleus at its centre. It was discovered that the nucleus, which is considered to be the centre of the atom, contains positively charged protons in addition to electrically neutral neutrons. However, it wasn't until 1932 that James Chadwick conclusively identified and characterized the neutron, a particle having almost the same mass as a proton but no electric charge. Even though the nucleus plays a crucial role in defining the atom, it is incredibly small. It is measured in femtometers ($10^{-15}$ meters), whereas the size of a complete atom is on the order of angstroms ($10^{-10}$ meters), which is significantly bigger. The notation $^{A}_{Z} \mathrm{X}$, which is used to describe atoms, offers important information: The total of protons ($Z$) and neutrons ($N$) is represented by the mass number $A$, which is important information on the makeup of the atom composition ~\cite{Burcham:1995jg}.

Experimental findings that were generated from scattering experiments showed that the nuclear radius ($r$) may be approximated as $r\simeq r_0 \times A^{1/3}$. This was discovered as a result of the experiments. In this case, the constant denoted by $r_0$ measures around 1.15 fm. This approximation is based on the oversimplification that the nucleus behaves the same way as a traditional spherical system. Because of this, the volume of the nucleus may be calculated using the formula $V=\frac{4}{3} \pi r^3= \frac{4}{3} \pi (r_0 A^{1/3})^3 = \frac{4}{3} \pi (r_0^3 A) $. The average nucleon density within the nucleus, which is given by the symbol $\rho_0=A/V$, is remarkable in that it rarely changes, regardless of the nucleus's mass number $A$, and it tends to remain around $\rho_0\sim 0.16$ fm$^{-3}$. It represents a system with an equal number of neutrons and protons but with no Coulomb interactions, which is the saturation density of nuclear matter ~\cite{Das:1995sk}.

\subsection{History of Neutron Stars}
Landau made a revolutionary discovery in February of 1931 when he proposed the concept of compact stars ~\cite{Yakovlev:2012rd, Horvath2022}. He predicted the creation of stars made of nuclear matter that might be even smaller than white dwarfs. According to his theory, stars with masses greater than 1.5 times that of the Sun's mass would unavoidably have zones in which the laws of quantum statistics would not apply. But Landau concluded in the later section of his paper that the density of matter gets so great that atomic nuclei come into close proximity to one another and form a unique, massive nucleus ~\cite{Haensel:2007yy}. It is noteworthy that this research was released in February of 1932, just a few days following the neutron's discovery ~\cite{Chadwick:1932ma}.

In 1934, two years later, Baade and Zwicky began studying the massive energy discharges that come with supernova explosions ~\cite{Baade1934a, Baade1934b}. They interpreted these supernova explosions as the latter phases of regular stars exploding into objects made mostly of closely spaced neutrons, hence the term "neutron stars." It was later suggested that these stars would have extraordinarily high density and short radii. Furthermore, because neutrons may be packed more efficiently than atomic nuclei and electrons, it was suggested that the gravitational binding energy could be exceptionally great, surpassing normal nuclear packing fractions. As a result, it is possible to think of neutron stars as arrangements of the most stable kind of matter. On November 28, 1967, Jocelyn Bell made the astounding discovery that neutron stars are radio pulsars ~\cite{Hewish:1968bj, Pilkington:1968bk}.

\subsection{Formation and Structure of Neutron Stars}
\begin{figure}[!ht]
   \centering
      \includegraphics[width=0.6\textwidth]{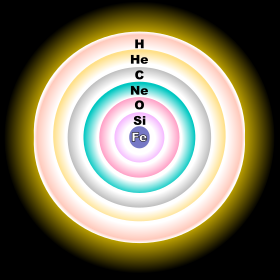}
  \caption{Schematic representation of onion-like nuclear fusion structures within a star. The image
is taken from ~\cite{Sutter:2020}}
  \label{intro:fig-Ns-onion}
\end{figure}

As we move from the realm of atomic nuclei to that of the cosmos, we are introduced to the mysterious and extraordinary objects that are known as neutron stars (NSs). These celestial phenomena, which are frequently spotted as pulsars, have captivated the imaginations of physicists and astronomers alike due to their peculiar qualities and the possibility that they will reveal insights into a variety of subfields of physics. 
During their evolution, stars with masses above $\sim$ 8 $M_\odot$ ($M_\odot$ being the mass of the Sun) undergo nuclear fusion processes, resulting in the fusion of core elements up to iron. Iron, being the most stable nuclide in the universe, imposes a limitation on nuclear fusion, preventing the continuation of fusion processes. As shown in Fig. \ref{intro:fig-Ns-onion}, massive stars have onion-like layered structures in their final stages. At low temperatures and densities, their cores are iron and neutron-rich iron-group nuclei ~\cite{Bethe:1979zd} surrounded by lighter elements, possibly inert hydrogen ~\cite{Woosley:2002zz}.
Electron degeneracy pressure supports the stratified core, which continues to grow by accretion as silicon shells are consumed. The core eventually exceeds the Chandrasekhar mass limit ($M_{Ch} \sim 1.44M_\odot$). After this, gravity overcomes electron degeneracy pressure ~\cite{Chandrasekhar:1931ih}, causing a Core Collapse Supernova (CCSN) explosion ~\cite{Janka:2006fh}. The aftermath of a CCSN event leaves a heated Proto-Neutron Star (PNS) with temperatures above $10^{10}$ K. If its mass reaches the maximum mass limit for neutron stars, the PNS may collapse further to produce a black hole, although this limit is unknown. After a few minutes, the heated PNS becomes a neutrino-transparent NS.

The masses of observed neutron stars typically range from around 1.4 to 2$M_\odot$, and their radii can range anywhere from $10-15$km. These values are a testament to the incredible density of neutron star material, where an entire solar mass can be packed into a sphere smaller than a typical city. Neutron stars have magnetic fields on their surfaces that can range from  $10^{12}$ Gauss to $10^{18}$ Gauss, depending on the star. Anomalous X-ray Pulsars (AXPs) and Soft Gamma Repeaters (SGRs) are two astronomical occurrences typically associated with magnetars, those with the most powerful magnetic fields \cite{Duncan:1992hi, Paczynski:1992zz, Patra:2020wjy}.
Consider the magnetars 1E 1048.1-5937 \cite{Gaensler:2005qk}and 1E 2259+586 \cite{Kuiper:2006is}, each with surface magnetic fields of about $10^{14}$ Gauss. Alternately, the neutron star 4U 0142+61 has an incredible surface field of $10^{16}$ Gauss, and the magnetar SGR 1806-20 has a surface field of around $10^{14}$ Gauss \cite{Kouveliotou:1998ze}, which is still an impressive value. These magnetars are fascinating illustrations of the harsh environments that can exist within neutron stars.

\begin{figure}[!ht]
   \centering
      \includegraphics[width=0.8\textwidth]{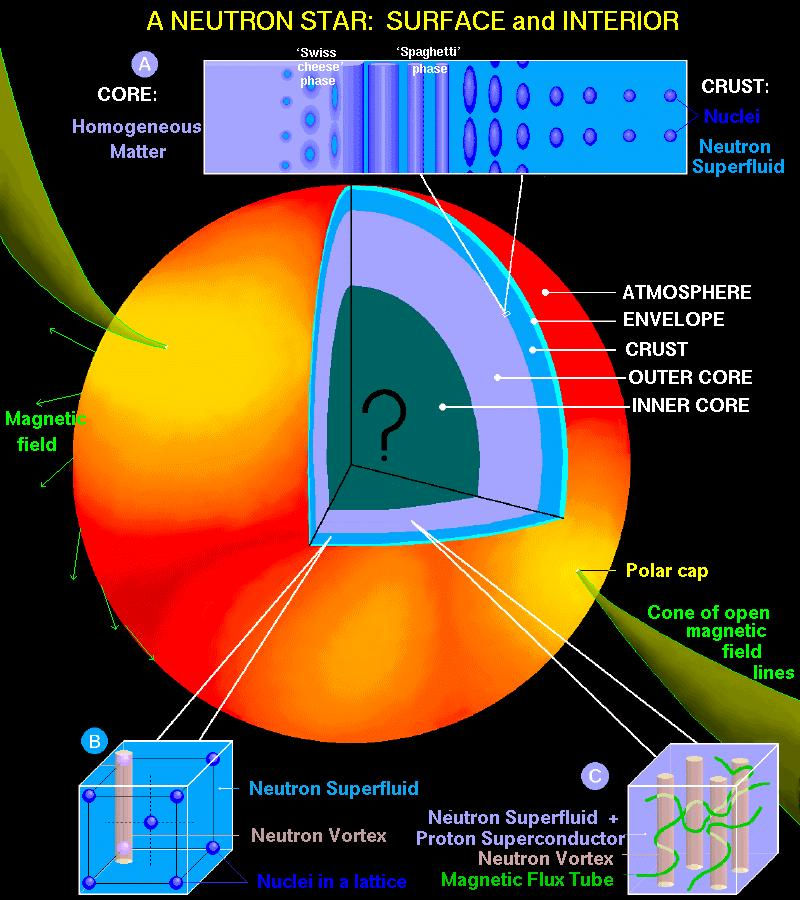}
  \caption{Diagram illustrating the interior of a neutron star. Image sourced from ~\cite{Page:2006ud}}
  \label{intro:fig-Ns}
\end{figure}

The internal structure of a cold neutron star is schematically depicted in Fig. \ref{intro:fig-Ns}. A neutron star's outermost layer is a few centimetres thin atmosphere followed by a few meters thick envelope. Thermal electromagnetic radiation originates in this envelope. This thermal emission from the surface layers of isolated neutron stars reveals surface temperatures. The observation of gravitationally red-shifted spectral lines also reveals neutron star mass-to-radius ratios. Unfortunately, most extremely young pulsars have a dominant non-thermal component that makes thermal radiation detection difficult. The surface temperature is typically too low to detect neutron stars that are older than roughly one million years ($\tau > 1$ Myr). In pulsars aged 10 to 100 thousand years ($\tau \sim 10-100$ kyr), thermal radiation from the surface of a neutron star can be dominant, especially at soft X-ray energies ~\cite{Haensel:2007yy}. A 1 km thick inhomogeneous crust lies under the outermost surface. The outer and inner crusts are separated at the neutron drip surface, a few hundred meters below the atmosphere. Atoms fully ionize and form a lattice in a relativistic electron gas inside the crust. If the chemical potential of neutrons is greater than their rest mass, a neutron gas could be present. As we approach the star's interior, stuff becomes more neutron-rich due to density. Non-spherical nuclei should be found at the lowermost layers of the inner crust. A viscous-free super-fluid state is created when the temperature of the inner crust falls below a critical limit ($T_c \sim 10^{10}$ K), forming Cooper pairs with free neutrons that have anti-aligned spins and zero orbital angular momentum.
 At half the saturation density ($\rho_0$), the equilibrium density of symmetric homogeneous nuclear matter, we reach the crust-core interface when nuclei disappear. The core has an outer core with a density range of 0.5$\rho_0$ to 2$\rho_0$ and an inner core with a baryon density ($\rho_B$) greater than 2$\rho_0$ ~\cite{Bandyopadhyay2022}. The outer core usually has neutrons, protons, electrons, and perhaps muons. The inner core's composition is unknown, with ideas including hyperons, boson condensates, and a phase transition to quark matter, all of which are conjectured to occur under extreme pressure and density conditions within neutron stars ~\cite{Lattimer:2004pg, Glendenning:1992vb}.

Astrophysicists and physicists are hoping that by diving into the features of these extraterrestrial objects, they will not only be able to further our grasp of the rules that govern the cosmos but will also be able to unveil the mysteries of matter under conditions that are inaccessible to testing on earth. Neutron stars are amazing cosmic laboratories that provide a platform for exploring the boundaries of physics and astrophysics. They also bridge the gap between the microscopic world of atomic nuclei and the immensity of the cosmos. Online catalogues and other astrophysical resources are available to researchers who are interested in doing in-depth examinations and looking up further examples.

\section{The equation of state of dense matter}
The internal composition of a neutron star depends on the hydrostatic equilibrium created by the gravitational force pulling matter inward and the outward pressure from neutron degeneracy. This insubstantial balance, based on Einstein's General Theory of Relativity (GR), helps us understand how these cosmic giants work on the inside.
To figure out what's going on inside a neutron star, we have to figure out how matter behaves in extreme situations. The theory of the equation of state (EOS) for infinite nuclear matter, which becomes important at the astounding high densities found in neutron stars, is central to this work. Here, it can't be ruled out that there is strange matter in the core. This core could be home to many different particles and resonances, such as $\Lambda^0$, $\Sigma^{-, 0, +}$, and $\Xi^{-, 0}$, or it could even go into the theoretical world of quarks \cite{Weber:1989uq, Hell:2014xva, Weissenborn:2011ut, Banik:2014qja, Patra:2022lds}.
In an atmosphere predominantly consisting of neutrons, the chemical potential of neutrons greatly exceeds that of protons, creating favourable conditions for phenomena such as $\beta-$ decay. This decay mechanism results in the transformation of some neutrons into protons and electrons, as demonstrated by reactions such as:
\bea 
n \rightarrow p + e^- + \bar{\nu}, \\
n + \nu \rightarrow p + e^- .
\eea 
In order to preserve charge neutrality within the context of a neutron star, the emergence of muons ($\mu$) may occur when the chemical potential of electrons attains a value equal to the rest mass of muons ($m_{\mu} = 106$ MeV). The charge neutrality requirement is written as follows for a given baryon density,
\bea 
\label{ch01}
\rho_p=\rho_e+\rho_{\mu}
\eea 
Here, $\rho_n$, $\rho_p$, and $\rho_\mu$ represent the number density of neutrons, protons, and muons, respectively. The condition of $\beta-$ equilibrium is simultaneously expressed as:
\bea 
\label{ch02}
\mu_n=\mu_p+\mu_e ~~~~~~~~~~~~~~{\rm and}~~~~~~~~~~\mu_e=\mu_\mu
\eea
Here, $\mu_n$, $\mu_p$, $\mu_e$, and $\mu_\mu$ represent the chemical potentials of neutrons, protons, electrons, and muons. In contrast to terrestrial experiments that enable the production and examination of high-density matter through heavy-ion collisions, the study of neutron stars presents a distinct challenge due to their pronounced asymmetry. To gain a comprehensive understanding of the nuclear equation of state, typically characterized as the relationship between energy (or pressure) and density, it is necessary to address the significant asymmetry inherent in neutron stars across the extensive range of densities they encompass.

The equation of state, which plays a vital role in comprehending the internal structure of neutron stars, can be divided into two distinct components. The first component relates to symmetric nuclear matter ($E(\rho,0)$), while the second component incorporates the density-dependent symmetry energy ($E_{\rm sym}(\rho)$). The process of decomposition results in an equation of state that may be expressed in the following form ~\cite{Ainsworth:1987hc, Wiringa:1988tp, Lattimer:2015nhk, Burgio:2021bzy}:
\bea
 E(\rho,\delta)\simeq E(\rho,0)+E_{\rm sym}(\rho)\delta^2,
 \label{eq:EOS}
\eea 
where,  $\delta  = \left(\frac{\rho_n -\rho_p}{\rho}\right)$ is isospin parameter that controls asymmetry of the medium, with $\rho_n$ and  $\rho_p$ being the neutron and proton densities, respectively. The condition of $\beta$-equilibrium and charge neutrality determine the value of $\delta$ at a given $\rho$.
 Once $\delta$ is known, the fraction of neutrons, protons, electrons, and muons can be
easily evaluated. $E(\rho,\delta)$ denotes the energy per nucleon at a given density $\rho$, which is the sum of the neutron and proton densities. 
By probing and constraining nuclear matter properties at saturation density, encompassing parameters such as the binding energy per nucleon $e_0=E(\rho,0)|_{\rho_0}$,  $J_0=E_{\rm sym}(\rho)|_{\rho_0}$ the symmetry energy coefficient, incompressibility coefficient $K_0=9 \rho_0^2 \left (\frac{\partial^2 E(\rho, 0)}{\partial
\rho^2}\right)_{\rho_0}$,  symmetry energy slope parameter $L_0=3 \rho_0 \left (\frac{\partial E_{\rm sym}(\rho)}{\partial \rho}\right)_{\rho_0}$, $K_{\rm sym,0}=9 \rho_0^2 \left (\frac{\partial^2 E_{\rm sym}(\rho)}{\partial
\rho^2}\right )_{\rho_0}$ the symmetry energy curvature parameter, skewness parameter $Q_0 [Q_{\rm sym,0}]=27 \rho_0^3 \left (\frac{\partial^3 E(\rho, 0)[E_{\rm sym}(\rho)]}{\partial \rho^3}\right)_{\rho_0}$ and others, we can gain insights into the density-dependent EOS for asymmetric matter. While calculating the EOS from first principles within Quantum Chromodynamics (QCD) remains a formidable challenge, especially at supra-saturation densities ($\rho >> \rho_0$), due to the nonperturbative nature of QCD, it represents a fundamental frontier at the intersection of nuclear physics, particle physics, and astrophysics, awaiting exploration.

\section{Data constraints from terrestrial and astrophysical sources}

The important constraints for the equation of state of nuclear matter and nucleon-nucleon interactions at low and high densities can be defined using data from terrestrial experiments such as those involving finite nuclei and Heavy Ion Collisions (HIC), Perturbative Quantum Chromodynamics (pQCD)  and astrophysical observations such as the properties of neutron stars. In the following sections, we summarise the empirical bounds that have been found so far.

\subsection{Finite nuclei}
The current nuclear matter parameters for symmetric nuclear matter (SNM) at the saturation density, denoted as $e_0$ and $K_0$, as well as the coefficients governing the density-dependent behaviour of symmetry energy, including $J_0$, $L_0$, and $K_{\rm sym,0}$, have been listed in Table \ref{intro:tab1}. Although these values pertain to infinite nuclear matter, they closely connect with finite nuclear matter observations.
The establishment of correlations between these parameters and the observables of finite nuclei at saturation density is one indirect way that researchers frequently use to gain precise insights into these parameters. These coefficients are closely related to SNM and symmetry energy, and associated properties include nuclear masses, isoscalar giant monopole resonances (ISGMR), and isovector giant dipole resonances (IVGDR) energies of finite nuclei.
The average values of $e_0$ and $\rho_0$, which are respectively $e_0=-15.88 \pm 0.24$ MeV and $\rho_0=0.163 \pm 0.005$ fm$^{-3}$, are obtained from a selection of the best nuclear equations of state that fit the selected nuclear data ~\cite{Dutra:2012mb, Dutra:2014qga}.
According to several studies ~\cite{Todd-Rutel:2005yzo, Avogadro:2013tza, Niksic:2008vp, Garg:2018uam, Agrawal:2005ix}, the nuclear incompressibility, denoted as $K_0$, is around $K_0 \approx 240 \pm 20$ MeV. This value is in good agreement with the centroids of the ISGMR that have been determined by experiment for several other nuclei, including $^{208}{\rm Pb}$, $^{90}{\rm Zr}$ and $^{144}{\rm Sm}$.
According to ~\cite{Jiang:2012zzb}, a thorough examination of the relationship between the experimental double differences in symmetry energies of finite nuclei and their mass numbers revealed the coefficient $J_0$ to be around $32.10 \pm 0.31$.

Microscopic calculations involving the neutron skin of heavy nuclei have been used recently to improve the determination of $L_0$ and have produced a value of $59 \pm 13$ MeV ~\cite{Agrawal:2012pq, Agrawal:2013hha}. In addition, investigations of the isovector giant dipole and quadrupole resonances in the $^{208}{\rm Pb}$ nucleus revealed that $L_0$ should be around $43 \pm 26$  and $37 \pm 18$, respectively ~\cite{Roca-Maza:2013mla, Roca-Maza:2012uor}. The slope parameter for symmetry energy has also been determined using different approaches on different nuclei. Neutron-skin thickness measurements on the $^{48}$Ca nucleus conducted by the Coherent Reactions Experiment (CREX) collaboration ~\cite{CREX:2022kgg} and on $^{208}$Pb by the Parity Radius Experiment (PREX-II) collaboration ~\cite{PREX:2021umo} have contributed to this endeavour.
Recently, the authors from ~\cite{Reed:2021nqk, Reed:2019ezm} carried out an extensive data analysis of the PREX-II experiment, yielding a result of $L_0=106 \pm 37$ MeV. Combining astronomical observations with PREX-II data, ~\cite{Essick:2021kjb} arrived at an estimate of $L_0=53^{+14}_{-15}$ MeV. Another investigation of the PREX-II data by ~\cite{Reinhard:2021utv} suggested a slightly lower value of $L_0=54 \pm 8$ MeV. Furthermore, the CREX experiment's data points to a range of possibilities for $L_0$, spanning from 0 to 51 MeV ~\cite{TAGAMI2022}.

When analyzing the density-dependent behaviour of symmetry energy at densities significantly higher than the saturation density ($\rho >> \rho_0$), the parameter $K_{\rm sym,0}$ is important. Despite this, it is still loosely limited because no direct probes can detect it. A wide range of values for $K_{\rm sym,0}$, ranging from $-700$ MeV to $400$ MeV, are suggested by several nuclear models ~\cite{Dutra:2012mb, Dutra:2014qga, Tsang:2020lmb, Ferreira:2021pni}. According to recent research ~\cite{Mondal:2017hnh}, there is a substantial link between $K_{\rm sym,0}$ and the pair $3 J_0 - L_0$, which limits $K_{\rm sym,0}$ to about $-111.8 \pm 71.3$ MeV.

\begin{table}
\label{intro:tab1}
\caption{The current nuclear parameters, including those related to symmetric nuclear matter such as $e_0$ and $K_0$, as well as the coefficients governing from the density-dependent of symmetry energy ($J_0$, $L_0$, and $K_{\rm sym,0}$), at saturation density $\rho_0$ are listed.}
\begin{center}
\setlength{\tabcolsep}{20pt} % General space between cols (6pt standard)
\renewcommand{\arraystretch}{1.1}
\begin{tabular}{ccc}
\hline \hline
NMP  & empirical value (MeV)  &   References  \\
\hline \hline
 $e_0$        &   $-15.88\pm 0.24$      &    ~\cite{Dutra:2012mb, Dutra:2014qga}          \\
 $K_0$        &   $240\pm20$ MeV        &    ~\cite{Garg:2018uam} \\
 $J_0$        &   $32.10\pm0.31$        &   ~\cite{Jiang:2012zzb} \\ %\hline 
 $L_0$                     &   $53 \pm 15$    &   ~\cite{Essick:2021kjb} \\ 
 $K_{\rm sym,0}$    &        $-111.8\pm71.8$                     &    ~\cite{Mondal:2017hnh}       \\ \hline \hline
\end{tabular}
\end{center}
\end{table}

\subsection{Heavy Ion Collisions and {\it ab-initio} calculations}
The current empirical constraints on the equation of state, excluding those associated with the saturation density $\rho_0$, are listed in Table \ref{intro:tab2}.
\begin{table}[!ht]
\caption{The current empirical constraints on the equation of state which include pressure ($P(\rho)$), energy per particle ($E(\rho)$), and symmetry energy ($E_{\rm sym}(\rho)$) corresponding to different nuclear matter configurations, such as symmetric nuclear matter (SNMX), pure neutron matter (PNMX), and the symmetry energy (SYMX). The limitations are presented with the corresponding density ranges from which they are generated. Here, X represents the number of multiple constraints on the same matter.}
  \label{intro:tab2}
  \begin{center}
\setlength{\tabcolsep}{6pt} % General space between cols (6pt standard)
 \renewcommand{\arraystretch}{1.1}
 \begin{tabular}{ l c c c c }
      \hline
      \hline
     
           & Quantity & Density region & Band/Range & References  \\ 
           &     & $(\text{fm}^{-3})$ & (MeV) & \\
      \hline 
      \hline
      SNM1 & $P(\rho)$ 	   &  $0.32~\text{to}~0.74$   &  HIC        &  ~\cite{Danielewicz:2002pu}  \\
      SNM2 & $P(\rho)$     &  $0.19~\text{to}~0.33$   &  Kaon exp.      &  ~\cite{Fuchs:2005zg}  \\
      \\
      PNM1 & $E(\rho)$  & 0.1            &   $10.9\pm0.5$     &  ~\cite{Brown:2013mga} \\
      PNM2 & $E(\rho)$   & $0.04~\text{to}~0.16$   &   N$^{3}$LO        &  ~\cite{Hebeler:2013nza} \\
      PNM3 & $P(\rho)$            & $0.04~\text{to}~0.16$   &   N$^{3}$LO        &  ~\cite{Hebeler:2013nza} \\
      PNM4 & $E(\rho)$   & $0.01~\text{to}~0.33$   &   N$^{3}$LO        &  ~\cite{Lattimer:2021emm} \\
      PNM5 & $P(\rho)$            & $0.01~\text{to}~0.33$   &   N$^{3}$LO        &  ~\cite{Lattimer:2021emm} \\
      \\
      
      SYM1 & $E_{\rm sym}(\rho)$       &   0.1              &   $24.1\pm0.8$    &  ~\cite{Trippa:2008gr}\\
      SYM2 & $E_{\rm sym}(\rho)$       & $0.01~\text{to}~0.19$       &   IAS,HIC         &  ~\cite{Danielewicz:2013upa, Tsang:2008fd} \\
      SYM3 & $E_{\rm sym}(\rho)$       &   $0.01~\text{to}~0.31$      &   ASY-EOS       &  ~\cite{Russotto:2016ucm}\\
      
      \hline
      \hline 
     \end{tabular}
     % \end{ruledtabular}
\end{center}     
\end{table}
The first two rows deal with the pressure of symmetric nuclear matter (SNM), which is determined by an analysis of directed and elliptic flow ~\cite{Danielewicz:2002pu} and kaon production in heavy ion collisions ~\cite{Fuchs:2005zg}.
The next five rows correspond to the energy density and pressure of pure neutron matter (PNM). The 'best-fit' Skyrme EDFs yield its energy density ($E(\rho)$) at a density of $\rho = 0.1$ fm$^{-3}$ ~\cite{Brown:2013mga}. From analytical calculations in terms of the effective degrees of freedom at low density, such as chiral effective theory, the uncertainty is negligible. The precise next-to-next-to-next-to-leading-order (N$^3$LO) calculation is usually fitted to the nucleon–deuteron scattering cross section or few-body observables, and even saturation properties of heavier nuclei ~\cite{Drischler:2021kxf}. The equation of state for pure neutron matter at low-density (0.04-0.16 fm$^{-3}$) was established using the next-to-next-to-next-to-leading order (N$^3$LO) framework within chiral effective field theory, as detailed in Ref. ~\cite{Hebeler:2013nza}. This EOS has been extended up to twice the saturation density $\rho_0$ in a subsequent reference, namely ~\cite{Lattimer:2021emm}.
The last three rows of the table are concerned with the symmetry energy $E_{\rm sym}(\rho)$ for different densities. 
The symmetry energy in $^{112}$Sn+$^{112}$Sn and $^{124}$Sn +$^{124}$Sn comes from the low-energy HIC simulations ~\cite{Tsang:2008fd}, Asy-EOS experiments at GSI ~\cite{Russotto:2016ucm}, and nuclear structure studies employing Isobaric Analogue States (IAS) ~\cite{Danielewicz:2013upa}. Additionally, the value of $E_{\rm sym}(\rho)$ at $\rho = 0.1$ fm$^{-3}$, obtained from a microscopic examination of IVGDR in $^{208}$Pb ~\cite{Trippa:2008gr}.

\subsection{Perturbative Quantum Chromodynamics}

The equation of state of neutron star matter, which describes the relationship between pressure and energy density in $\beta-$equilibrated matter interacting with quantum chromodynamics at temperature T=0, can be described in two ways. 
The Chiral effective field theory, in particular, offers the EOS with good precision in the area where matter is in the hadronic phase, between the well-studied NS crust and a density of roughly 1.1$\rho_0$ ~\cite{Tews:2012fj, Gandolfi:2009fj}.
Conversely, in the high-density limit, perturbative-QCD (pQCD) techniques, grounded in high-energy particle physics and founded on the concept of deconfined quark and gluon degrees of freedom ~\cite{Kurkela:2009gj, Gorda:2018gpy}, offer an equally accurate description, especially for quark matter EOS, at densities $\rho \geq 40\rho_0 \equiv \rho_{\rm pQCD}$. Also, as the conformal symmetry of QCD is restored at asymptotically high densities, perturbation theory can be used, and the speed of sound gets closer to the value found in conformal field theory and seen in ultra-relativistic fluids, which is $c^2_s=1/3$ from below.

\begin{figure}[!ht]
   \centering
      \includegraphics[width=0.8\textwidth]{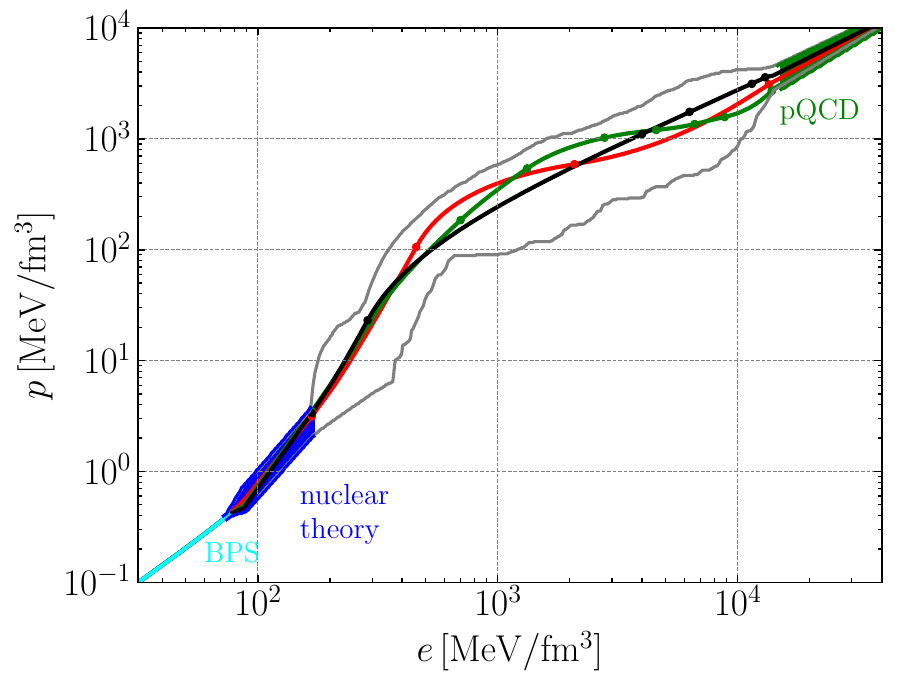}
  \caption{The pressure as a function of the energy density. The image is taken from ~\cite{Altiparmak:2022bke}.}
  \label{intro:fig-pqcd}
\end{figure}

Significantly, studies that used QCD principles found that at energy levels of about $E(\rho)\sim 750$ MeV.fm$^{-3}$, the EOS became much softer, and the speed of sound went down. As shown in Ref. ~\cite{Annala:2019puf}, these traits have been seen as signs of the beginning of a quark matter phase.
Figure \ref{intro:fig-pqcd} shows the relationship between pressure and energy density for the dense matter. Additionally, it includes the BPS EOS ~\cite{Baym:1971pw} (cyan solid line) employed at low densities. The uncertainties resulting from perturbative QCD ~\cite{Gorda:2018gpy, Fraga:2013qra} and nuclear theory ~\cite{Hebeler:2013nza} are represented by the green and blue bands, respectively. The grey region outlines the collective constraints-compliant EOSs, satisfying all available astrophysical criteria. In summary, this figure offers a comprehensive overview of the valid EOS for dense matter, spanning the entire density spectrum from low to high densities.

\subsection{Astrophysical Observations}
The upper boundaries for the maximum achievable mass and radius of neutron stars are theoretically determined by the properties of the nuclear equation of state throughout the complete spectrum of densities, including both low and high densities.
Similarly, precise measurements of neutron star mass and radius can put useful limitations on the equation of state of nuclear matter.  
However, due to the immense distances involved, the direct determination of the radius of neutron stars poses a formidable challenge. The observation of neutron stars with mass $\sim$ 2$M_\odot$ ~\cite{Antoniadis:2013pzd, NANOGrav:2017wvv} established a lower limit on the maximum mass that an EOS must predict.

The tidal deformability parameter of NS, which encodes the EOS information, has been inferred for the first time from a gravitational wave event GW170817 from a binary neutron star (BNS) merger with a total mass of $2.74_{-0.01}^{+0.04}M_\odot$ ~\cite{Abbott18a, Abbott2019}, observed by the advanced-LIGO~\cite{LIGOScientific:2014pky} and advanced-Virgo~\cite{VIRGO:2014yos} detectors. { The analysis of GW170817 predicts that the dimensionless tidal deformability for 1.4 $M_\odot$ NS, $\Lambda_{1.4}=190^{+390}_{-120}$ at the 90\% level.} Another subsequent event, GW190425, likely originating from the coalescence of BNSs, was observed~\cite{Abbott2020}. Future runs of LIGO-Virgo-KAGRA and detectors like the Einstein Telescope~\cite{Punturo:2010zz} and Cosmic Explorer~\cite{Reitze:2019iox} are expected to observe more BNS signals from coalescing neutron stars. Numerous theoretical studies of the NS properties have been prompted by the unprecedented constraints on the EOS that gravitational wave astronomy has promised to provide through the detailed analysis of gravitational wave parameter estimation ~~\cite{GW170817, Malik2018, De18, Fattoyev2018a, Forbes:2019xaz, Landry2019, Piekarewicz2019, Biswas2020, Abbott2020, Thi2021}.

Recently, two different groups of Neutron star Interior Composition Explorer (NICER) X-ray telescopes provided neutron star's mass and radius simultaneously for PSR J0030+0451 with $R=13.02^{+1.24}_{-1.06}$km for mass $1.44^{+0.15}_{-0.14} M_\odot$ ~\cite{Miller:2019cac} and $R=12.71^{+1.14}_{-1.19}$km for mass $1.34^{+0.15}_{-0.16} M_\odot$~\cite{Riley:2019yda}, which are the complementary constraints on the EoS. For heavier pulsar PSR J0740+6620, $R=13.7^{+2.6}_{-1.5}$km with mass $2.08 \pm 0.07 M_\odot$  ~\cite{Miller:2021qha} and $R=12.39^{+1.30}_{-0.98}$km with mass $2.072^{+0.067}_{-0.066} M_\odot$ ~\cite{Riley:2021pdl} were reported. Current observational lower bound on the maximum NS mass is $M_{\rm max}= 2.35\pm 0.17 M_\odot$ for the black-widow pulsar PSR J0952-0607 ~\cite{Romani:2022jhd} that exceeds any previous measurements, including $M_{\rm max} = 2.27^{+0.17}_{-0.15} M_\odot$ for PSR J2215-5135~\cite{Linares:2018ppq}. To support the NS with a mass higher than $2 M_\odot$, stiffer EOSs are required.

\section{Objectives}

The equations of state of $\beta$-equilibrated charge neutral matter and their connections to the properties of neutron stars have been studied for the last several decades~\cite{Oppenheimer:1939ne, Tolman:1939jz, Glendenning:1992dr}. The precise knowledge of the properties of NS and the data on Heavy ion collisions may constrain the behaviour of EOSs at supra-saturation densities ~\cite{Huth:2021bsp}. The behaviour of EOSs around $\rho_0$ may be important in determining the properties of such NSs. It has been shown that the radius of a neutron star with its mass in the range of $1-1.4 M_\odot$ is strongly correlated with the pressure for $\beta$-equilibrated matter at the densities $1-2\rho_0$ ~\cite{Lattimer:2000nx}. Similar analyses have been extended to the tidal deformability, which is also found to be strongly correlated with pressure at $2\rho_0$ ~\cite{Tsang:2019vxn, Tsang:2020lmb}. The EOS for $\beta$-equilibrated matter can be divided into two components: symmetric nuclear matter and density-dependent symmetry energy.
 It may be important to constrain them individually. Recently, it is shown in Ref. ~\cite{Imam:2021dbe} that the accurate knowledge of the equation of state of $\beta$-equilibrated matter may not be resolved appropriately into its two main components, symmetric nuclear matter, and density-dependent symmetry energy.

There have been several attempts to study the correlations of radius and tidal deformability of a neutron star with individual nuclear matter parameters which determine the density-dependence of symmetry energy~~\cite{Alam:2016cli, Carson:2018xri, Malik2018, Tsang:2019vxn, Guven:2020dok, Malik:2020vwo, Tsang:2020lmb, Malik_book, Reed:2021nqk, Pradhan:2022vdf, Pradhan:2022txg, Ghosh:2021bvw, Beznogov:2022rri}. The nuclear matter parameters, often drawn randomly from uncorrelated uniform or Gaussian distributions, are found to be weakly correlated with the NS properties. Several factors that affect the correlations of NS properties with the nuclear matter parameters are also summarized in Table III of Ref.~~\cite{Kunjipurayil:2022zah}. As of yet, it is unclear which nuclear matter parameters have the greatest impact on the neutron star's properties, such as tidal deformability and radius at canonical mass 1.4 $M_\odot$ ~\cite{Carson:2018xri, Pradhan:2022vdf, Pradhan:2022txg, Malik2018, Tsang:2019vxn, Malik:2020vwo, Tsang:2020lmb, Malik_book, Reed:2021nqk, Kunjipurayil:2022zah}. 
Therefore, the main objectives, in view of the recent problems, of this thesis are as follows,
\begin{itemize}
    \item To see the model-independent manner in the correlations of NS properties such as tidal deformability and radius with the pressure of $\beta$-equilibrated matter at densities higher than the saturation density ($\rho_0 = 0.16$ fm$^{-3}$). And to parametrize the pressure for $\beta$-equilibrated matter, around 2$\rho_0$, as a function of neutron star mass and the corresponding tidal deformability.
    
    \item To determine the principal factors influencing the correlations between NS properties and nuclear matter parameters.

    \item To find out the key nuclear matter parameters that have the greatest impact on the neutron star's properties, such as tidal deformability and radius at canonical mass 1.4 $M_\odot$.
\end{itemize}

\section{Organization of the Thesis}
The present thesis is organized in the following manner: After a brief introduction in Chapter \ref{chap1}, we proceed to examine Meta models, such as $\frac{n}{3}$ expansions and Taylor expansion, and Relativistic Mean Field (RMF) models, non-Relativistic models along with its utilization in the context of neutron stars in Chapter \ref{chap2}. In this chapter, we provide a thorough explanation of the formalism related to $\frac{n}{3}$ expansions, Taylor expansion, and RMF theory. The validity of these models is evaluated by applying them to a range of nuclear systems and comparing their numerical results with existing experimental data. The calculations on nuclear matter and neutron stars involve several parameter sets that have been generally recognized for their effectiveness in the present literature. These models thereafter form the basis for investigating various quantities that are discussed in the forthcoming chapters.

In Chapter-\ref{chap3}, we found that the tidal deformability and radius of NS with mass $1-2 M_\odot$  strongly correlated with the pressure of $\beta$-equilibrated matter at densities higher than the saturation density ($\rho_0 = 0.16$ ~fm$^{-3}$) in a nearly model-independent manner. We parameterized the pressure for $\beta$-equilibrated matter, around 2$\rho_0$, as a function of neutron star mass and the corresponding tidal deformability.

In Chapter-\ref{chap4}, we have systematically analyzed the factors affecting the correlations of slope and curvature parameters of symmetry energy at the saturation density ($\rho_0=0.16
\text{fm}^{-3}$) with the tidal deformability and stellar radius of non-spinning neutron stars in the mass range of $1.2 - 1.6 M_\odot$ using a large set of minimally constrained equations of state. These correlations are quite sensitive to the choice of the distributions of symmetry energy parameters and
their interdependence. 

In Chapter-\ref{chap5}, we have attempted to mitigate the challenge of connecting the neutron star properties with the nuclear matter parameters that describe equations of state. A Principal Component Analysis (PCA) is employed as a statistical tool to uncover the connection between multiple nuclear matter parameters and the tidal deformability as well as the radius of neutron stars within the mass range of $1.2-1.8M_\odot$. The contributions from iso-vector nuclear matter parameters to the tidal deformability and radius of NS decrease by $\sim$ 25\% with the increase in mass of NS 1.2$M_\odot$ to 1.8$M_\odot$ and accordingly those of iso-scalar nuclear matter parameters increase.

In Chapter-\ref{chap6},
 Finally, a summary and future perspective are given in the last chapter (Chapter-\ref{chap7}).

%% file: chapter2/chapter2.tex
\chapter{Formalism}
\label{chap2}

In nature, there are four known fundamental forces: electromagnetic, gravity, weak, and strong.
The strong force holds together the nucleons in an atom's nucleus by combining attractive and repulsive forces.
In the field of physics, it is frequently necessary to develop an equation to describe the behaviour of nuclear matter. This formula is known as the "equation of state." As "nuclear many-body dynamics," we refer to the study of the collective behaviour of a large number of particles within an atomic nucleus.
Now, physicists commonly use three basic types of models to generate these equations of state: (i) Microscopic Models, (ii) Phenomenological Models, and (iii) Meta-models, which offer additional avenues for approaching this problem and understanding nuclear matter.
The following is a quick summary of the theoretical frameworks used in this thesis to compute the EOS and neutron star parameters.

\section{Phenomenological Models}
Phenomenological models play a crucial role in developing the equation of state (EOS) for neutron stars, providing a theoretical framework to understand the connection between the internal structure of these dense objects and nuclear physics. These models are generally divided into (i) relativistic (\ref{rmf-model}) and (ii) non-relativistic (\ref{shf-model}) models. Relativistic models utilize general relativity to address the strong gravitational effects near neutron stars, especially in high-density situations. On the other hand, non-relativistic models take a classical approach, employing effective field theories and empirical fits for lower-density nuclear matter. Both types of models offer unique insights, contributing to a comprehensive understanding of neutron star physics by considering both relativistic and non-relativistic aspects.

\subsection{Relativistic Mean Field Models (RMF)}
\label{rmf-model}
The nucleon-nucleon interaction is governed by the exchange of particles known as mesons, especially when the nucleons are within a femtometer of each other. Within the framework of relativistic physics, this phenomenon serves as the basis for nuclear interaction. Notably, among the mesons, the $\sigma$, $\omega$, and $\rho$-mesons play crucial roles in this context, as previously discussed in Ref. ~\cite{Walecka:1974qa, Boguta:1977xi, Boguta:1981px, Serot:1997xg}.
The $\sigma$-mesons significantly influence the spin-orbit potential and are predominantly responsible for generating a robust central attractive force. In contrast, $\omega$-mesons are responsible for the short-distance manifestation of a repulsive force. It is essential to include $\rho$-mesons in the formalization to distinguish between protons and neutrons, as their primary distinction rests in their isospin projections. In this theoretical framework, nuclear interactions are mathematically described using $\mathcal{L}$-denoted Lagrangian density functions. The Euler-Lagrange equation regulating a field $\varphi$ is expressed as follows within the covariant formalism.
\begin{equation}
\partial_{\mu}\left(\frac{\partial \mathcal{L}}{\partial\left(\partial_{\mu}  \varphi\right)}\right)=\frac{\partial \mathcal{L}}{\partial  \varphi}
\end{equation}
and the stress-energy tensor $T^{\mu \nu}$ is,
\begin{equation}
T^{\mu \nu}=\frac{\partial \mathcal{L}}{\partial\left(\partial_{\mu}  \varphi\right)} \partial^{\nu}  \varphi-g^{\mu \nu} \mathcal{L}
\end{equation}
The metric tensor components in this case are denoted as $g^{\mu \nu}=\operatorname{Diag}[1,-1,-1,-1]$.  In the static case,
the energy $E$ and pressure $P$ of the system are then,
\begin{equation}
\begin{array}{c}{E=<T_{00}>} \\ {P=\frac{1}{3}<T_{i i}>}\end{array}
\end{equation}

In this work, we consider a more general, nonlinear finite-range Relativistic Mean-Field model, which is represented by the following Lagrangian density ~\cite{Dutra:2014qga},
\begin{eqnarray}
\mathcal{L}_{\rm NL} = \mathcal{L}_{\rm nm} + \mathcal{L}_\sigma +
\mathcal{L}_\omega
+ \mathcal{L}_\rho + \mathcal{L}_{\delta} + \mathcal{L}_{\sigma\omega\rho},
\label{dl}
\end{eqnarray}
where
\begin{align}
\mathcal{L}_{\rm nm} &= \overline{\psi}(i\gamma^\mu\partial_\mu - M)\psi 
+ g_\sigma\sigma\overline{\psi}\psi 
- g_\omega\overline{\psi}\gamma^\mu\omega_\mu\psi 
- \frac{g_\rho}{2}\overline{\psi}\gamma^\mu\vec{\rho}_\mu\vec{\tau}\psi
+ g_\delta\overline{\psi}\vec{\delta}\vec{\tau}\psi,
\\
\mathcal{L}_\sigma &= \frac{1}{2}(\partial^\mu \sigma \partial_\mu \sigma 
- m^2_\sigma\sigma^2) - \frac{A}{3}\sigma^3 - \frac{B}{4}\sigma^4,
\\
\mathcal{L}_\omega &= -\frac{1}{4}F^{\mu\nu}F_{\mu\nu} 
+ \frac{1}{2}m^2_\omega\omega_\mu\omega^\mu 
+ \frac{C}{4}(g_\omega^2\omega_\mu\omega^\mu)^2,
\\
\mathcal{L}_\rho &= -\frac{1}{4}\vec{B}^{\mu\nu}\vec{B}_{\mu\nu} 
+ \frac{1}{2}m^2_\rho\vec{\rho}_\mu\vec{\rho}^\mu,
\\
\mathcal{L}_\delta &= \frac{1}{2}(\partial^\mu\vec{\delta}\partial_\mu\vec{\delta} 
- m^2_\delta\vec{\delta}^2),
\end{align}
and
\begin{align}
\mathcal{L}_{\sigma\omega\rho} &= 
g_\sigma g_\omega^2\sigma\omega_\mu\omega^\mu
\left(\alpha_1+\frac{1}{2}{\alpha_1}'g_\sigma\sigma\right)
+ g_\sigma g_\rho^2\sigma\vec{\rho}_\mu\vec{\rho}^\mu
\left(\alpha_2+\frac{1}{2}{\alpha_2}'g_\sigma\sigma\right) 
\nonumber \\
&+ \frac{1}{2}{\alpha_3}'g_\omega^2 g_\rho^2\omega_\mu\omega^\mu
\vec{\rho}_\mu\vec{\rho}^\mu.
\label{lomegarho}
\end{align}
The term $\mathcal{L}_{\rm nm}$ in this Lagrangian density includes the kinetic part of nucleons and the terms corresponding to the nucleon-meson interactions between the mesons $\sigma$, $\delta$, $\omega$, and $\rho$. At the same time, the free and self-interaction components associated with each meson $j$ are contained in the term $\mathcal{L}_j$, where $j$ can take values from the set $\{ \sigma, \delta, \omega, \rho \}$. The Lagrangian density also includes the term $\mathcal{L}_{\sigma\omega\rho}$ to account for cross-interactions between the meson fields. Antisymmetric field tensors, $F_{\mu\nu}$ and $\vec{B}_{\mu\nu}$, must be included. $F_{\mu\nu}=\partial_\nu\omega_\mu-\partial_\mu\omega_\nu$ and $\vec{B}_{\mu\nu}=\partial_\nu\vec{\rho}_\mu-\partial_\mu\vec{\rho}_\nu - g_\rho (\vec{\rho}_\mu \times \vec{\rho}_\nu)$ are the definitions of these tensors, respectively. $M$ stands for the nucleon mass, and $m_j$ for the meson masses.

By employing the mean-field approximation, the meson fields are are given as,
\begin{eqnarray}
\sigma\rightarrow \left<\sigma\right>\equiv\sigma, \quad
\omega_\mu\rightarrow \left<\omega_\mu\right>\equiv\omega_0, \quad
\vec{\rho}_\mu\rightarrow \left<\vec{\rho}_\mu\right>\equiv \bar{\rho}_{0(3)}, \quad
\mbox{and}\quad 
\vec{\delta}\rightarrow\,\,<\vec{\delta}>\equiv\delta_{(3)},
\label{meanfield}
\end{eqnarray}
the following field equations are obtained in conjunction with the Euler-Lagrange equations:
,
\begin{align}
m^2_\sigma\sigma &= g_\sigma\rho_s - A\sigma^2 - B\sigma^3 
+g_\sigma g_\omega^2\omega_0^2(\alpha_1+{\alpha_1}'g_\sigma\sigma)
+g_\sigma g_\rho^2\bar{\rho}_{0(3)}^2(\alpha_2+{\alpha_2}'g_\sigma\sigma)\,\mbox{,}\quad 
\label{sigmaacm}\\
m_\omega^2\omega_0 &= g_\omega\rho - Cg_\omega(g_\omega \omega_0)^3 
- g_\sigma g_\omega^2\sigma\omega_0(2\alpha_1+{\alpha_1}'g_\sigma\sigma)
- {\alpha_3}'g_\omega^2 g_\rho^2\bar{\rho}_{0(3)}^2\omega_0, 
\label{omegaacm}\\
m_\rho^2\bar{\rho}_{0(3)} &= \frac{g_\rho}{2}\rho_3 
-g_\sigma g_\rho^2\sigma\bar{\rho}_{0(3)}(2\alpha_2+{\alpha_2}'g_\sigma\sigma)
-{\alpha_3}'g_\omega^2 g_\rho^2\bar{\rho}_{0(3)}\omega_0^2, 
\label{rhoacm} \\
m_\delta^2\delta_{(3)} &= g_\delta\rho_{s3}, 
\label{deltaacm}
\end{align}
In infinite nuclear matter, only the zero components of the four-vector fields remain nonzero due to translational invariance and rotational symmetry. Furthermore, our attention is limited to the third components of the isospin space vectors $\vec{\rho}_\mu$ and $\vec{\delta}$, considering rotational invariance about the third axis in the isospin space.

The scalar and vector densities are as follows,
\begin{align}
\rho_s &=\left<\overline{\psi}\psi\right>={\rho_s}_p+{\rho_s}_n,\quad
\rho_{s3}=\left<\overline{\psi}{\tau}_3\psi\right>={\rho_s}_p-{\rho_s}_n,
\label{rhos}\\
\rho &=\left<\overline{\psi}\gamma^0\psi\right> = \rho_p + \rho_n,\quad
\rho_3=\left<\overline{\psi}\gamma^0{\tau}_3\psi\right> = \rho_p - \rho_n=(2y-1)\rho,
\label{rho}
\end{align}
with 
\begin{eqnarray}
{\rho_s}_{p,n} &=& 
\frac{\gamma M^*_{p,n}}{2\pi^2}\int_0^{{k_F}_{p,n}}
\frac{k^2dk}{\sqrt{k^2+M^{*2}_{p,n}}} 
=\frac{\gamma 
(M^*_{p,n})^3q^2}{2\pi^2}\int_0^1\frac{\xi^2d\xi}{\sqrt{\xi^2+1/q^2}}
\nonumber \\
&=& \frac{\gamma (M^*_{p,n})^{3}}{4\pi^2}\left[q\sqrt{1+q^2}
- \mbox{ln}\left(q+\sqrt{1+q^2}\right)\right],\qquad
\label{rhospn}
\end{eqnarray}
and
\begin{eqnarray}
\rho_{p,n} = \frac{\gamma}{2\pi^2}\int_0^{{k_F}_{p,n}}k^2dk =
\frac{\gamma}{6\pi^2}{k_F^3}_{p,n},
\label{rhopn}
\end{eqnarray}
For $\xi = k/k_{F_{p,n}}$ and $q = k_{F_{p,n}}/M^*_{p,n}$, where protons and neutrons are denoted by the subscripts $p$ and $n$, respectively. When dealing with asymmetric matter, the proton fraction is defined as $y = \rho_p/\rho$, and the degeneracy factor is given the value $\gamma = 2$. In this case, the Fermi momentum is represented in units by ${k_{F_{p,n}}}$, where $\hbar$ and $c$ are both set to 1.

We can also define the
effective nucleon mass as 
\begin{eqnarray}
M_p^*=M-g_\sigma\sigma-g_\delta\delta_{(3)} \qquad\mbox{and}\qquad
M_n^*=M-g_\sigma\sigma+g_\delta\delta_{(3)}.
\label{effectivemasses}
\end{eqnarray}
The $\delta_{(3)}$ in symmetric nuclear matter is null since $\rho_{sp} = \rho_{sn}$, which results in $M_p^* = M_n^* = M^* = M - g_\sigma\sigma$.

The energy density and pressure of the asymmetric system can be obtained by using the energy-momentum tensor $T^{\mu\nu}$, which was estimated from the Lagrangian density in Eq.~(\ref{dl}).
\begin{eqnarray}
\mathcal{E}_{\rm NL} &=& \frac{1}{2}m^2_\sigma\sigma^2 
+ \frac{A}{3}\sigma^3 + \frac{B}{4}\sigma^4 - \frac{1}{2}m^2_\omega\omega_0^2 
- \frac{C}{4}(g_\omega^2\omega_0^2)^2 - \frac{1}{2}m^2_\rho\bar{\rho}_{0(3)}^2
+g_\omega\omega_0\rho+\frac{g_\rho}{2}\bar{\rho}_{0(3)}\rho_3
\nonumber \\
&+& \frac{1}{2}m^2_\delta\delta^2_{(3)} - g_\sigma g_\omega^2\sigma\omega_0^2
\left(\alpha_1+\frac{1}{2}{\alpha_1}'g_\sigma\sigma\right) 
- g_\sigma g_\rho^2\sigma\bar{\rho}_{0(3)}^2 
\left(\alpha_2+\frac{1}{2}{\alpha_2}' g_\sigma\sigma\right) \nonumber \\
&-& \frac{1}{2}{\alpha_3}'g_\omega^2 g_\rho^2\omega_0^2\bar{\rho}_{0(3)}^2
+ \mathcal{E}_{\mbox{\tiny kin}}^p + \mathcal{E}_{\mbox{\tiny kin}}^n,
\label{denerg}
\end{eqnarray}
where
\begin{eqnarray}
\mathcal{E}_{\mbox{\tiny kin}}^{p,n}&=&\frac{\gamma}{2\pi^2}\int_0^{{k_F}_{p,n}}k^2
(k^2+M^{*2}_{p,n})^{1/2}dk 
\label{decinnlw}= \frac{\gamma {k_F^4}_{p,n}}{2\pi^2}\int_0^1 
\xi^2(\xi^2+z^2)^{1/2}d\xi
\nonumber \\
&=&\frac{\gamma {k_F^4}_{p,n}}{2\pi^2}
\left[\left(1+\frac{z^2}{2}\right)\frac{\sqrt{1+z^2}}{4}
-\frac{z^4}{8}\mbox{ln}\left(\frac{1+\sqrt{1+z^2}}{z}\right)\right]
 \nonumber \\
&=& \frac{3}{4}{E_{F}}_{p,n}\rho_{p,n} + \frac{1}{4}M^{\ast}_{p,n}{\rho_{s}}_{p,n},
\end{eqnarray}
and
\begin{eqnarray}
P_{\rm NL} &=& - \frac{1}{2}m^2_\sigma\sigma^2 - \frac{A}{3}\sigma^3 -
\frac{B}{4}\sigma^4 + \frac{1}{2}m^2_\omega\omega_0^2 
+ \frac{C}{4}(g_\omega^2\omega_0^2)^2 + \frac{1}{2}m^2_\rho\bar{\rho}_{0(3)}^2
+ \frac{1}{2}{\alpha_3}'g_\omega^2 g_\rho^2\omega_0^2\bar{\rho}_{0(3)}^2
\nonumber \\
&-&\frac{1}{2}m^2_\delta\delta^2_{(3)} + g_\sigma g_\omega^2\sigma\omega_0^2
\left(\alpha_1+\frac{1}{2}{\alpha_1}'g_\sigma\sigma\right) 
+ g_\sigma g_\rho^2\sigma\bar{\rho}_{0(3)}^2 
\left(\alpha_2+\frac{1}{2}{\alpha_2}' g_\sigma\sigma\right) \nonumber \\
&+& P_{\mbox{\tiny kin}}^p + P_{\mbox{\tiny kin}}^n,\qquad
\label{pressure}
\end{eqnarray}
with
\begin{eqnarray}
P_{\mbox{\tiny kin}}^{p,n} &=& 
\frac{\gamma}{6\pi^2}\int_0^{{k_F}_{p,n}}\frac{k^4dk}{(k^2+M^{*2}_{p,n})^{1/2}} 
= \frac{\gamma {k_F^4}_{p,n}}{6\pi^2}\int_0^1 \frac{\xi^4d\xi}{(\xi^2+z^2)^{1/2}} 
\nonumber \\
&=&\frac{\gamma {k_F^4}_{p,n}}{6\pi^2}
\left[\left(1-\frac{3z^2}{2}\right)\frac{\sqrt{1+z^2}}{4}
+\frac{3z^4}{8}\mbox{ln}\left(\frac{1+\sqrt{1+z^2}}{z}\right)\right]
\nonumber \\
&=& \frac{1}{4}{E_{F}}_{p,n}\rho_{p,n} - \frac{1}{4}M^{\ast}_{p,n}{\rho_{s}}_{p,n},
\end{eqnarray}
and
\begin{equation}
{E_{F}}_{p,n} = \sqrt{{k_{F}}_{p,n}^{2}+(M^{\ast}_{p,n})^{2}}, ~~ z=M^{\ast}_{p,n}/{k_{F}}_{p,n}
\end{equation}

\subsection{Non-Relativistic Models}
\label{shf-model}
In nuclear physics, variational techniques combined with nucleon-nucleon potentials form the foundation of non-relativistic models. Regarding non-relativistic self-consistent techniques, the Skyrme-Hartree-Fock model (SHF) is a popular choice among them. In this context, the fundamental effective nucleon-nucleon contact is the Skyrme force, first described by Skyrme in 1959 ~\cite{Skyrme:1959zz}.
A zero-range, momentum-dependent method used to describe the general characteristics of finite nuclei is represented by this Skyrme force, which is expressed as an energy density functional. Its integration into Hartree-Fock calculations is made simpler by its zero-range character. Finite-range effects between nucleons are, however, explained by the momentum dependency present in this zero-range force. The density-dependent two-body interaction also accounts for the many-body interactions inside the Skyrme force formalism.
The  mathematical expression for the system's total energy ($E$) as a function of ${\cal H}(r)$ is:
\begin{equation}
 E=\int {\cal H}(r)d^3r 
 \end{equation}
 and the ${\cal H}(r)$ is expressed as  ~\cite{Vautherin:1971aw, Chabanat:1997qh},
 \begin{equation}
\label{c2_Hden}
{\cal H} = {\cal K} + {\cal H}_0  +{\cal H}_3+{\cal H}_{\rm eff} +{\cal H}_{\rm fin}
+ {\cal H}_{\rm so} +{\cal H}_{\rm sg}+{\cal H}_{\rm Coul}
\end{equation}
Here, we decompose the constituents of the overall energy using the kinetic energy term is denoted by ${\cal K}$, the zero-range term is denoted by ${\cal H}_0$, the density-dependent term is denoted by ${\cal H}_3$, the effective-mass term is denoted by ${\cal H}_{\rm eff}$; the finite-range term is denoted by ${\cal H}_{\rm fin}$, the spin-orbit term is contributed by ${\cal H}_{\rm so}$, the term resulting from tensor coupling with spin and gradient effects is denoted by ${\cal H}_{\rm Coul}$, as well as the energy density owing to Coulomb interaction. These elements are specifically described as follows within the Skyrme interaction framework:

\begin{equation}
{\cal H}_0 = \frac{1}{4}t_0\left [(2 + x_0)\rho^2 - (2x_0 + 1)(\rho_p^2 + \rho_n^2)\right ],
\end{equation}
\begin{equation}
{\cal H}_3 = \frac{1}{24}t_3\rho^\alpha\left [(2 + x_3)\rho^2 - (2x_3 + 1)(\rho_p^2 + \rho_n^2)\right ],
\end{equation}
\begin{equation}
{\cal H}_{\rm eff} = \frac{1}{8} \left [t_1 (2 + x_1) + t_2 (2 + x_2)\right ]
\tau\rho
+ \frac{1}{8}\left[ t_2(2x_2+1) -t_1(2x_1+1)\right] (\tau_p\rho_p + \tau_n\rho_n), 
\end{equation}
\begin{eqnarray}
{\cal H}_{\rm fin}&=&\frac{1}{32}\left [3t_1 (2 + x_1) - t_2 (2 + x_2)\right ](\nabla\rho)^2 \nonumber\\
&& - \frac{1}{32}\left[ 3t_1(2x_1+1) +  t_2(2x_2+1) \right] \left [ (\nabla\rho_p)^2 + (\nabla\rho_n)^2\right], \\
{\cal H}_{\rm so}&=&\frac{W_0}{2}\left [ {\bf J}\cdot\nabla\rho +
{\bf J_p}\cdot{\bf \nabla}\rho_p
+{\bf J_n}\cdot{\bf \nabla}\rho_n \right ],\\
{\cal H}_{\rm sg} &=&-\frac{1}{16}(t_1 x_1 + t_2 x_2){\bf J}^2 + 
\frac{1}{16}(t_1 - t_2) \left [{\bf J_p}^2 + {\bf J_n}^2 \right]. 
\label{c2_Hsg}
\end{eqnarray}

In this case, $ \rho = \rho_p + \rho_n$ is the total particle number density, $\tau$ is the total kinetic energy density, and ${\bf J}$ is the total spin density. The letters $p$ and $n$ stand for protons and neutrons, respectively. It is important to note that the number we used in our calculations was $\hbar^2/2m = 20.734$ MeV.fm$^2$.
It is important to note that ${\cal H}_{\rm fin}$, ${\cal H}_{\rm so}$, ${\cal H}_{\rm sg}$, and ${\cal H}_{\rm Coul}$ don't make any significant contributions in the case of uniform matter.
Also, it's important to remember that ${\cal H}(r)$ only depends on $\rho_p$ and $\rho_n$ and has nothing to do with the point $r$. So, this relationship can be written as ${\cal H}(r) \rightarrow {\cal E}(\rho_p, \rho_n)$.
We can write the energy per nucleon, $E(\rho_p, \rho_n)$, in terms of the total density $\rho$ and the asymmetry parameter $\delta=\frac{\rho_n - \rho_p}{\rho}$ like this:

\bea
E(\rho, \delta)&=& \frac{3}{5} \frac{\hbar^2}{2m}
 \left(\frac{3\pi^2 }{2}\right)^{2/3} \rho^{2/3} F_{5/3}+\frac{1}{8} t_0
\rho[2(x_0 +2)-(2x_0+1) F_2 ]\non
&+& \frac{1}{48}  t_3 \rho^{\sigma +1} [2(x_3 +2) -(2x_3 +1) F_2] \non
&+&  \frac{3}{40} \left(\frac{3\pi^2 }{2}\right)^{2/3}  \rho^{5/3} \left \{
[t_1 (x_1 +2) +t_2 (x_2+2 )]F_{5/3} + \frac{1}{2} [t_2 (2x_2 +1) - t_1
(2x_1+1 )]F_{8/3}\right \}\non
\label{c2_eos}\\
F_m(\delta) &=&  \frac{1}{2} [ (1+\delta)^m + (1 -\delta)^m ]
\eea
Once the Skyrme parameters are known, the equation of state for nuclear matter at a certain asymmetry $\delta$ can be calculated using the formula Eq. (\ref{c2_eos}). For convenience, we may also express the EOS in terms of various nuclear matter parameters computed at the saturation density $\rho_0$, provided that a reasonable approximation is used.

\section{Meta-Models}
\label{meta-model}
An additional modelling method for the nuclear equation of state consists of two different approaches: (a) using a Taylor expansion and (b) using a $ {\frac{n}{3}}$ expansion, both of which are based on the saturation density or a Fermi momentum expansion. These methods offer a unique chance to incorporate the strongest insights from nuclear physics into the nuclear EOS and reduce the number of free parameters. These extensions allow for a clean separation of the low-order derivatives, which are better defined by nuclear experiments, and the high-order derivatives, which are more constrained by neutron star data. This environment is often difficult to study through nuclear laboratory studies, and it is noteworthy that the higher-order parameters show increased sensitivity to the EOS at extremely high densities. The concept of a meta-model for the nucleonic equation of state is being introduced in this study, as it offers numerous notable advantages:
(i) It creates a unique mapping across a large number of current EOS models, each of which has a number of different input parameters.
(ii) This method is flexible and allows for smooth transitions between the current EOS models.
(iii) As a result, it might guide the choice of input parameters towards values that don't match any of the current EOS models.
(iv) The meta-model is a flexible framework that can easily store the large amount of nuclear physics information that comes from lab experiments as input parameters.
(v) It includes the results of complex { \it ab-initio} models in its parameter space, which makes it possible to get the EOS limits that these models put in place.
(vi) It makes it easier to figure out the error margins for both experiments and theories when used with a Bayesian framework, which turns them into confidence levels for astronomical observables.

The   energy per nucleon for neutron star matter $E(\rho, \delta)$ at a given  total
nucleon density $\rho$ and asymmetry $\delta$ can be decomposed into the energy per nucleon for the symmetric nuclear matter, $E(\rho,0)$
and the  density-dependent symmetry energy, $E_{\rm sym}(\rho)$ in the parabolic approximation as shown in Eq. (\ref{eq:EOS}),
In the following, we expand $E(\rho,0)$ and $E_{\rm sym}(\rho)$ appearing in Eq. (\ref{eq:EOS})  using (a) Taylor and (b) $\frac{n}{3}$ expansions.  The coefficients of expansion in the case of Taylor
correspond to the individual nuclear matter parameters. In the latter case, they are expressed as linear combinations of the nuclear matter parameters. 

\subsection{{Taylor expansion}}
\label{Taylor}
The $E(\rho,0)$ and $E_{\rm sym}(\rho)$ can be
expanded around the saturation density $\rho_0$ as  ~\cite{Chen:2005ti, Chen:2009wv, Newton:2014iha, Margueron:2017eqc, Margueron:2018eob},

\bea
E(\rho, 0)&=&	\sum_{n} \frac{a_n}{n!}\left(\frac{\rho-\rho_0}{3\rho_0}\right )^n, \label{eq:SNM_T} \\ 
E_{\rm sym}(\rho) &=&	\sum_{n} \frac{b_n}{n!}\left(\frac{\rho-\rho_0}{3\rho_0}\right )^n, \label{eq:sym_T}  
\eea
so that, 
\bea
E(\rho, \delta)&=&	\sum_{n} \frac{1}{n!}(a_n + b_n\delta^2)\left(\frac{\rho-\rho_0}{3\rho_0}\right )^n ,\label{eq:ebeta_T} 
\eea
where the coefficients $a_n$ and $b_n$ are the nuclear matter
parameters. We truncate the sum in Eqs. (\ref{eq:SNM_T}) and
(\ref{eq:sym_T}) at fourth order, i.e., $n = 0$ - $4$.  Therefore, the
coefficients $a_n$ and $b_n$ correspond,

\bea
a_n & \equiv & e_0, 0, K_0, Q_0, Z_0\label{eq:anm1}, \\
b_n  &\equiv & J_0, L_0, K_{\rm sym,0},Q_{\rm sym,0}, Z_{\rm sym,0}. \label{eq:bnm1}
\eea
In Eqs. (\ref{eq:anm1}) and (\ref{eq:bnm1}), $e_0 $ is the binding energy per nucleon, $K_0$ {the} incompressibility coefficient, $J_0$ {the} symmetry energy coefficient, its slope parameter $L_0$, $K_{\rm sym,0}$ {the} symmetry energy curvature parameter, $Q_0 [Q_{\rm sym,0}]$ and $Z_0 [Z_{\rm sym,0}]$ are related to third- and fourth-order density derivatives of  $E (\rho,0) $ [ $E_{\rm sym}(\rho)$], respectively. The subscript zero indicates that all the nuclear matter parameters are calculated at the saturation density.

It may be noticed from Eq. (\ref{eq:ebeta_T}) that the coefficients $a_n$ and $b_n$ may display some correlations among themselves, provided the asymmetry parameter depends weakly on the density.  Furthermore, Eq. (\ref{eq:ebeta_T}) may converge slowly at high densities, i.e., $\rho \gg 4\rho_0$. This situation is encountered for the heavier neutron stars. Neutron stars with a mass around $2M_\odot$, typically have central densities $\sim 4-6\rho_0$.

\subsection{${\frac{n}{3}}$ expansion}
\label{n/3}
An alternative expansion of $E(\rho,\delta)$ can be obtained by expanding $E(\rho,0)$ and
$E_{\rm sym}(\rho)$ as ~\cite{Cochet:2003ex, Agrawal:2006ea, Lattimer2015, Gil2017},

\bea
E(\rho,0) &=& \sum_{n=2}^6 (a'_{n-2}) \left(\frac{\rho }{\rho_0}\right )^{\frac{n}{3}},\label{eq:SNM_n3}\\
E_{\rm sym}(\rho) &=& \sum_{n=2}^6 (b'_{n-2}) \left(\frac{\rho }{\rho_0}\right )^{\frac{n}{3}},\label{eq:sym_n3}\\
E(\rho, \delta) &=& \sum_{n=2}^6 (a'_{n-2} +b'_{n-2} \delta^2 ) \left(\frac{\rho }{\rho_0}\right )^{\frac{n}{3}}.\label{eq:ebeta_n3}
\eea
 We refer {this} as {the} ${\frac{n}{3}}$ expansion. It is now evident from Eqs. (\ref{eq:SNM_n3}) and (\ref{eq:sym_n3}) that the coefficients of expansion are no longer the individual nuclear matter parameters, unlike in the case of Taylor's expansion.
The values of the nuclear matter parameters can be expressed in terms of the expansion coefficients $a'$ and $b'$ as, respectively,
\bea
\left (
\begin{matrix}
e_0\\
0\\
K_0 \\
Q_0\\
Z_0\\
\end{matrix}
\right ) &=& 
\left (
\begin{matrix}
1 &1 &1 & 1&1\\
2 &3 &4 & 5&6\\
-2  &  0 & 4 & 10&18\\
8  &  0 & -8 & -10 & 0\\
-56 & 0 & 40 & 40 & 0\\
\end{matrix}
\right )
\left (
\begin{matrix}
a'_0\\
a'_1\\
a'_2\\
a'_3\\
a'_4\\
\end{matrix}
\right ),\label{mat_a}\\
\left (
\begin{matrix}
J_0\\
L_0 \\
K_{\rm sym,0}\\
Q_{\rm sym,0}\\
Z_{\rm sym,0}\\
\end{matrix}
\right ) &=& 
\left (
\begin{matrix}
1 &1 &1 & 1&1\\
2  &  3 & 4 & 5&6\\
-2  &  0 & 4 & 10&18\\
8 & 0 & -8 & -10 & 0\\
-56 & 0 & 40 & 40 & 0\\ 

\end{matrix}
\right )
\left (
\begin{matrix}
b'_0\\
b'_1\\
b'_2\\
b'_3\\
b'_4\\
\end{matrix}
\right ).\label{mat_b}
\eea
The relations between the expansion coefficients and the nuclear matter parameters are governed by the nature of functional form for $E(\rho,0)$ and $E_{\rm sym}(\rho)$. The off-diagonal elements in the above matrices would vanish for the Taylor expansion of  $E(\rho,0)$ and $E_{\rm sym}(\rho)$ as given by Eqs. (\ref{eq:SNM_T}) and (\ref{eq:sym_T}), respectively. Therefore, each of the expansion coefficients is simply the individual nuclear matter parameter given by Eqs.  (\ref{eq:anm1}) and (\ref{eq:bnm1}). Inverting the matrices in Eqs.  (\ref{mat_a}) and (\ref{mat_b}) we have
\bea
a'_0&=& \frac{1}{24}(360 e_0 + 20K_0   + Z_0),\nonu\\
a'_1&=& \frac{1}{24}(-960 e_0 - 56K_0  - 4Q_0 - 4Z_0),\nonu\\
a'_2&=& \frac{1}{24}( 1080 e_0 + 60K_0 + 12Q_0 + 6Z_0),\nonu\\
a'_3&=& \frac{1}{24}(-576 e_0 - 32K_0 - 12Q_0 - 4Z_0),\nonu\\
a'_4&=& \frac{1}{24}( 120 e_0 + 8K_0 + 4Q_0  + Z_0),\label{eq:anm2}\\
b'_0&=& \frac{1}{24}(360J_0 -120L_0  + 20K_{\rm {sym,0}} + Z_{\rm {sym,0}}),\nonu\\
b'_1 &=& \frac{1}{24}(-960J_0 + 328L_0 - 56K_{\rm {sym,0}} - 4Q_{\rm {sym,0}}\nonumber  \\
&&    -4Z_{\rm {sym,0}}),\non
b'_2 &=& \frac{1}{24}(1080J_0 - 360L_0 + 60K_{\rm {sym,0}} + 12Q_{\rm {sym,0}} \nonumber \\
&&   + 6Z_{\rm {sym,0}}),  \non
b'_3 &=& \frac{1}{24}(-576J_0 + 192L_0 - 32K_{\rm {sym,0}} - 12Q_{\rm {sym,0}}  \nonumber \\
&&   -4Z_{\rm {sym,0}}),\non
b'_4 &=& \frac{1}{24}(120J_0 - 40L_0 + 8K_{\rm {sym,0}}  + 4Q_{\rm {sym,0}}  \nonumber \\
&&   + Z_{\rm {sym,0}}).\label{eq:bnm2}
\eea
Each of the coefficients $a'$ and $b'$ is a linear combination of nuclear matter parameters in such a way that the lower-order parameters may contribute dominantly at low densities. The effects of higher-order parameters become prominent with the increase in density.

\section{Speed-Of-Sound extension to higher densities} \label{H-eos}
We impose the causality condition on the speed of sound to construct the EOS beyond the density ($\rho_{\rm c_s}$), which is taken to be $1.5-2\rho_0$. The high-density part of the EOS ($\rho>\rho_{\rm c_s}$) joins smoothly to the one at the low density such that the velocity of the sound never exceeds the velocity of light and asymptotically approaches the conformal limit ($c_s^2$ = $\frac{1}{3}c^2$). The velocity of sound for $\rho>\rho_{\rm c_s}$ is given as ~\cite{Tews:2018kmu},   
\bea
\frac{c_s^2}{c^2} &=& \frac{1}{3} - c_1 exp\left[{-\frac{(\rho-c_2)^2}{n_b^2}}\right] + h_p exp\left[-\frac{(\rho-n_p)^2}{w_p^2}\right]\nonumber\\
&& \left[1 + erf(s_p \frac{\rho-n_p}{w_p})\right]. \label{eq-vs}
\eea 
where the peak height $h_p$ determines the maximum speed of sound, the position $n_p$ determines the density around which it happens, the width of the curve is controlled by $w_p$ and $n_b$, and the shape or skewness parameter $s_p$. For a given value of $n_b$, the parameters $c_1$ and $c_2$ are determined by the continuity of the speed of sound and its derivative at the density $\rho_{\rm c_s}$. The values of $n_b$, $h_p$, $w_p$, and $n_p$ are drawn from the uniform distribution with ranges in between 0.01-3.0 fm$^{-3}$,  0.0-0.9, 0.1-5.0 fm$^{-3}$, and ($\rho_{\rm c_s}$ + 0.08) - 5.0 fm$^{-3}$, respectively ~\cite{Tews:2018kmu}. We have taken $s_p$ equal to zero throughout our calculations as it does not affect the stiffness of EOS much. 

We construct the high-density equation of state starting from the density ($\rho_{\rm c_s}$), where the energy density ($\epsilon(\rho_{\rm c_s})$), the pressure (P($\rho_{\rm c_s}$)) and the derivative of energy density ($\epsilon^\prime(\rho_{\rm c_s})$) are known. The successive values of $\epsilon$ and P are obtained by assuming a step size  \(\Delta \rho=0.001\) fm$^{-3}$ as follows,
\bea
 \rho_{i+1} &=& \rho_i + \Delta \rho \label{eq-rhoE},\\
 \epsilon_{i+1} &=& \epsilon_i + \Delta \epsilon \nonumber\\
                &=& \epsilon_i + \Delta \rho \frac{\epsilon_i + P_i}{\rho_i}\label{eq-engE},\\
P_{i+1} &=&P_i + c_s^2(\rho_i) \Delta \epsilon  \label{eq-preE}.              
\eea
where, the index i = 0 refers to the density $\rho_{\rm c_s}$ . Note, in the Eq. (\ref{eq-engE}) $\Delta \epsilon$ has been evaluated using the thermodynamic relation $ P =  \rho \partial \epsilon/{\partial \rho}-\epsilon$ valid at zero temperature.

\section{Nuclear Matter Parameters}
\label{nmps}
The energy per nucleon at a given density $\rho$ and asymmetry $\delta$ can be decomposed into the energy per nucleon for the symmetric nuclear matter, $E(\rho,0)$
and the  density-dependent symmetry energy, $E_{\rm sym}(\rho)$, using a parabolic approximation as Eq. (\ref{eq:EOS}).
Using individual nuclear matter parameters of symmetric nuclear matter-energy  $E(\rho,0)$ and density-dependent symmetry energy $E_{\rm sym}(\rho)$ are expanded around $\rho_0$ as 
\bea
E(\rho, 0)&=&	e_0+\frac{1}{2}K_0\left (\frac{\rho-\rho_0}{3\rho_0}\right )^2
+\frac{1}{6} Q_0 \left(\frac{\rho-\rho_0}{3\rho_0}\right )^3 \nonumber \\ 
&+& \frac{1}{24} Z_0 \left(\frac{\rho-\rho_0}{3\rho_0}\right )^4, \label{eq:SNM_T1} \\ 
E_{\rm sym}(\rho) &=&	J_0 + L_0\left (\frac{\rho-\rho_0}{3\rho_0}\right )
+\frac{1}{2}K_{\rm sym,0}\left (\frac{\rho-\rho_0}{3\rho_0}\right )^2 \nonumber \\  
&+& \frac{1}{6}Q_{\rm sym,0}\left (\frac{\rho-\rho_0}{3\rho_0}\right )^3 + \frac{1}{24} Z_{\rm sym,0} \left(\frac{\rho-\rho_0}{3\rho_0}\right )^4. \label{eq:sym_T1}  
\eea 
In the above Eqs. (\ref{eq:SNM_T1}, \ref{eq:sym_T1}), $e_0$ is the binding energy per nucleon, $K_0$ is the incompressibility coefficient, $J_0$ is the symmetry energy coefficient, its slope parameter $L_0$, $K_{\rm sym,0}$ is the symmetry energy curvature parameter, $Q_0$ [$Q_{\rm sym,0}$] are the skewness parameter of  $E (\rho,0) $  $[E_{\rm sym}(\rho)]$ and $Z_0$ [$Z_{\rm sym,0}$] are the 4th derivative of $E (\rho,0) $  $[E_{\rm sym}(\rho)]$. These parameters are defined as follows,

\bea
e_0 &=&E(\rho, 0) _{\rho_0}\nonumber \\ 
K_0 &=& 9 \rho_0^2 \left (\frac{\partial^2 E(\rho, 0)}{\partial\rho^2}\right)_{\rho_0}\nonumber \\ 
Q_0 &=& 27 \rho_0^3 \left (\frac{\partial^3 E(\rho, 0)}{\partial
\rho^3}\right)_{\rho_0}\nonumber \\ 
Z_0 &=& 81 \rho_0^4 \left (\frac{\partial^4 E(\rho, 0)}{\partial
\rho^4}\right)_{\rho_0}\nonumber \\ 
J_0 &=&E_{\rm sym}(\rho) _{\rho_0}\nonumber \\ 
L_0&=&3 \rho_0 \left (\frac{\partial E_{\rm sym}(\rho)}{\partial \rho}\right
)_{\rho_0}\nonumber \\ 
K_{\rm sym,0}&=&9 \rho_0^2 \left (\frac{\partial^2 E_{\rm sym}(\rho)}{\partial
\rho^2}\right )_{\rho_0}\nonumber \\
Q_{\rm sym,0}&=&27 \rho_0^3 \left (\frac{\partial^3 E_{\rm sym}(\rho)}{\partial
\rho^3}\right )_{\rho_0}\nonumber \\
Z_{\rm sym,0}&=&81 \rho_0^4 \left (\frac{\partial^4 E_{\rm sym}(\rho)}{\partial
\rho^4}\right )_{\rho_0}\nonumber
\eea
\section{Structure and Dynamics of Neutron Star}
\label{ns-structure}
The stellar equation of state, which was carefully figured out in earlier sections ( \ref{rmf-model}, \ref{shf-model}, and \ref{meta-model}), is the basic idea on which models of neutron stars are built. It is a key part of the Tolman-Oppenheimer-Volkoff (TOV) equation, used to figure out physical observables important to neutron stars ~\cite{Tolman:1939jz, Oppenheimer:1939ne}. In the theory of general relativity, the TOV equation describes the hydrostatic stability of a spherically symmetric star that doesn't spin.
This section is all about estimating NS observables. First, we solve the hydrostatic equilibrium equations to find out how NS's mass and radius relate to each other. After that, we figure out the tidal deformability, which measures how much an NS is affected by tidal forces, like those that happen when two NSs merge at the end.

\subsection{Masses and Radii}
\label{ns-mr}
In general relativity, for non-rotating, spherically symmetric stars, the masses and radii of neutron stars are found as solutions to the hydrostatic equilibrium equations, which are written as follows ~\cite{Tolman:1939jz, Oppenheimer:1939ne}:
\begin{eqnarray}
\frac{dP}{dr}= - \frac{G \epsilon(r) m(r)}{c^2 r^2}\left[ 1+ \frac{P(r)}{\epsilon(r)}\right] 
\left[1+ \frac{4 \pi r^3 P(r)}{m(r) c^2}\right]\left[1- \frac{2 G m(r)}{c^2 r}\right],
\label{eqn:dpdr}
\end{eqnarray} 
\begin{eqnarray}
\frac{dm}{dr}= 4 \pi r^2 \epsilon(r),
\label{eqn:dmdr}
\end{eqnarray}
\begin{eqnarray}
 \frac{d\phi}{dr}=-\frac{1}{\epsilon(r)}\frac{dP}{dr}  \left(1 + \frac{P}{\epsilon(r)}\right)^{-1},
 \label{eqn:dphidr}
\end{eqnarray}
Here, G stands for the gravitational constant, P for pressure, $\epsilon=\rho c^2$ for the total energy density, and m(r) for the gravitational mass inside the sphere of radius r. Within the Schwarzschild metric, these parameters are defined by $ds^2 = c^2 dt^2 e^{2\phi} - e^{2\lambda} dr^2 - r^2 (d\theta^2 + \sin^2\theta d\phi^2)$.  The function $\phi(r)$ represents the gravitational potential, especially in the Newtonian limit. The variable $\lambda(r)$ is closely connected to the enclosed mass m(r), through a specific connection as,
\bea
e^{-\lambda}= \sqrt{1-\frac{2Gm}{rc^2}}.
\eea
The equation for hydrostatic equilibrium, Eq. \ref{eqn:dpdr}, is often called the TOV equation. The Eq. \ref{eqn:dmdr} is integrated over the range from r = 0 to the NS boundary at r = R, where R is the neutron star's radius. This gives us the neutron star's overall gravitational mass M. The Eq. (\ref{eqn:dphidr}) shows how the metric function $\phi(r)$ behaves in a relativistic setting.
In this case, we are mostly interested in solving Eqs. (\ref{eqn:dpdr}) and (\ref{eqn:dmdr}), which give us profiles for P(r), $\rho$(r), and m(r) for a given equation of state, written as P($\rho_B$), which has been talked about in detail in earlier sections (\ref{rmf-model}, \ref{shf-model}, and \ref{meta-model}).

We begin by picking a random number for the central mass density, $\rho_c$, in order to solve the hydrostatic equilibrium equations for a certain EOS. Then, we find the centre pressure, $P_c = P(\rho_{B,c})$, which is equal to r = 0, while keeping the boundary condition $m(r = 0) = 0$. Utilizing the Runge-Kutta method, we integrate up to r = R, where R is the radius that makes P(r = R) = 0. The pressure must stay below $5 \times 10^{-4}$ MeV/fm$^{3}$, which is the number we picked to ensure that the total mass results are all the same.
We use a prepared table to interpolate the mass density, $\rho$, at each step where the pressure changes. Finally, we get the curves P(r), $\rho$(r), and m(r) for a certain central mass density, $\rho_c$. This lets us figure out the NS mass, M, and its radius, R. According to ~\cite{Haensel:2007yy}, the canonical NS mass is about M = 1.4$M_\odot$, which gives a radius of R = 10 to 14 km. It is important to remember that the exact central density of a given NS is still unknown. It could be anywhere from about $4.6 \times 10^{14}$ to $4 \times 10^{15} g/cm^3$. So, finding the link between mass and radius is what we are most interested in. Using the numerical method we discussed earlier, we can determine this relationship by figuring out M and R for this range of central mass densities.

\subsection{Tidal deformability}
\label{ns-tidal}
The tidal deformability, a component of the gravitational signal received before the merger, is sensitive to the equation of state in a theoretical setting. As a result, gravitational wave observations may provide insightful new limitations on the EOS. Among these, the discovery of GW through the merging of two neutron stars, as demonstrated by the GW170817 event, is noteworthy since it has yielded important information about the tidal deformability of NS.
Consider a static quadrupolar external tidal field ${\cal E}_{tid}$ surrounded by a static, spherically symmetric NS. This arrangement causes the star to experience a quadrupole moment, which is expressed as follows:
\begin{equation}
Q_{tid}=-\lambda {\cal E}_{tid},
\end{equation}
Here, $\lambda$ corresponds to the tidal deformability and is related to the tidal Love number $k_2$ by,
\begin{equation}
\label{c2_eq2}
\lambda = \frac{2}{3}k_2 (\frac{Rc^2}{G})^5.
\end{equation}
The dimensionless tidal deformability is defined as,
\begin{equation}
\label{c2_eq3}
\Lambda = \frac{\lambda}{M^5}= \frac{2}{3}k_2 \beta^{-5},
\end{equation}
Here, we present the compactness parameter of the star, $\beta = GM/(Rc^2)$. The tidal Love number $k_2$ can be obtained by solving the following first-order differential equation, 
\bea
\label{eqn:dydr}
 \frac{dy}{dr}&=& -\frac{y^2}{r} - \frac{y-6}{r-2Gm/c^2}\nonumber\\
 &-&\frac{4\pi G}{c^2}r^2 \frac{(5-y)\rho+(9+y)P/c^2 + (P+\rho c^2)/c^2_s}{r-2Gm/c^2}\nonumber\\
 &+&\frac{1}{r}\left[ \frac{2G}{c^2}\frac{(m+4\pi pr^3/c^2)}{r-2Gm/c^2}\right]^2
\eea
The speed of sound is now introduced and is expressed as $c_s = \sqrt{\partial P/{\partial \epsilon}}$. The boundary constraint y(r = 0) = 2 applies to the function y = y(r). A simultaneous solution is obtained by solving equation (\ref{eqn:dydr}) in conjunction with the hydrostatic equilibrium equations, namely Eqs. (\ref{eqn:dpdr}) and (\ref{eqn:dmdr}).
The following is the expression for the tidal Love number:
\bea
\label{eqn:k2}
k_2&=&\frac{8}{5}\beta^5(1-2\beta)^2[2-y(R) + 2\beta(y(R)-1)]/a,
\eea
with
\bea
a &=& 6\beta[2 - y(R) + \beta(5y(R)-8)]\nonumber\\
&+& 4\beta^3 [13-11y(R) + \beta(3y(R)-2)+2\beta^2(1+y(R))]\nonumber\\
&+& 3(1-2\beta)^2[2-y(R) + 2\beta (y(R)-1)]ln(1-2\beta).
\eea

\section{Bayesian Estimation of Nuclear Matter Parameters}
\label{BA}
Bayesian inference is a common statistical approach that allows for estimating the joint posterior distribution of model parameters via using the updated prior beliefs in light of the likelihood function~\cite{Gelman2013, Buchner2014, Ashton2019}. Bayesian inference can refine our understanding of the nuclear equation of state in the field of neutron star physics, particularly from the nuclear physicist's point of view. This refinement is accomplished by adding data collected from various astrophysical studies, such as determining NS mass or assessing tidal deformability via gravitational wave measurements. In contrast, this model allows for applying nuclear physics limitations for predicting macroscopic observables associated with neutron stars.

This section focuses on the Bayesian method for determining the nuclear matter parameters. As detailed in Sec. \ref{BA-1}, we begin by reviewing the fundamental principles of Bayesian inference and then explain its application in the context of constraining the equation of state. Moreover, we shed light on the prior distribution of nuclear matter parameter values in Sec. \ref{BA-2}. The construction of the likelihood function, which is dependent on constraints derived from nuclear physics observations, neutron star observables, and constancy to physical constraints, is detailed in Sec. \ref{BA-3}.

\subsection{Principle of Bayesian Inference}
\label{BA-1}
The Bayesian inference is based on Bayes's theorem, which provides an expression for posterior probability, also known as conditional probability. It enables us to calculate the probability that a certain set B of values from a set of random variables A will occur.
\begin{equation}
   P(A |B ) =\frac{P(B|A) P(A)}{P(B)},\label{eq:BT-1} 
\end{equation}
Where $P(A)$ represents the prior probability associated with A, $P (B)$ represents the prior probability associated with B, and $P (B|A)$ indicates the conditional probability of B given A in this equation. The main idea of this method is to acknowledge that random variables are not limited to observational data but can also include unknown values from theoretical modeling, such as the EOS parameters in our case.
Using data for B, a parameter set represented by $\textbf{X}$ for A, and probability density functions (PDFs) for P, we can rewrite the Bayes theorem as follows:

\begin{equation}
p(\textbf{X} |data ) =\frac{{\mathcal L}(data|\textbf{X}) p(\textbf{X})}{p(data)},\label{eq:BT-2}
\end{equation}
The posterior distribution, represented as $P(\textbf{X}|data)$, is a true probability distribution. The denominator $p(data)$ acts as a normalization factor. This ensures that the distribution sum is equal to 1
The following is its mathematical expression:
\begin{equation}
    p(data) = \int {\mathcal L}(data|\textbf{X})p(\textbf{X})d\textbf{X}. \label{eq:BT-3}
\end{equation}
Because of this, Eq. (\ref{eq:BT-2}), which is a model-based form of Bayes' theorem, is often written as
\begin{equation}
    p(\textbf{X}|data) \propto {\mathcal L}(data|\textbf{X})p(\textbf{X}).
\end{equation}
The posterior distribution can be used to obtain the marginal one- and two-parameter posterior distributions. They each have the following definitions:
\bea
p(X_j|data) &=& \Biggl\{ \prod_{i\neq j}\int dX_i \Biggr\} p(\textbf{X}|data) \label{eq:BT-m1}\\
p(X_j, X_k|data) &=& \Biggl\{ \prod_{i\neq j,k}\int dX_i \Biggr\} p(\textbf{X}|data) \label{eq:BT-m2}
\eea
The prior distribution, shown by $p(\textbf{X})$, includes what we knew about the model parameters before we collected the "data," as well as any underlying biases that existed.  The likelihood function ${\mathcal L}(data|\textbf{X})$, on the other hand, shows how likely it is to see the data given the model values $\textbf{X}$. 
To put it simply, the probability distribution stores the link between the data and the model parameters. An unnormalized joint posterior probability density function is what we get when we multiply the prior by the likelihood function. The posterior PDF shows the conditional distribution of model parameters, marked as $\textbf{X}$, based on the data we can access. It is important to remember that the prior distribution significantly influences the posterior distribution. Therefore, it must be carefully considered. As such, selecting the appropriate prior is a crucial and intricate step in the process.

\subsection{Prior distribution of Nuclear Matter Parameters}
\label{BA-2}
We now discuss the prior distribution of nuclear matter parameters. The stellar EOSs are constructed based on the meta-modelling nuclear matter parameters discussed in Sec. \ref{meta-model}. We describe the parameter set as $\textbf{X}$, and it refers to the nuclear matter parameters such as $e_0, K_0, Q_0, Z_0, J_0, L_0, K_{\rm sym,0}, Q_{\rm sym,0}$ and $Z_{\rm sym,0}$. The prior distribution for $\textbf{X}$ is given by combining an uncorrelated ansatz with a uniform distribution for each individual parameter within the predefined interval, as shown in Table \ref{meth-tab1},

\bea
p(\textbf{X}) &=&  \prod_{i=1}^{2(N +1)+3} \large{U}(X_i^{min} , X_i^{max} ; X_i ) \label{eq:BT-p}
\eea
The parameter $X_i$, in this case, has a uniform distribution from $X_i^{min}$ to $X_i^{max}$. Adopting a flat prior means that all possible values of $\textbf{X}$ inside these intervals are equally probable.

\begin{table}[htp!]
\caption{\label{meth-tab1}The nuclear matter parameters in parameter set X, their respective derivative order N, and their minimum and maximum values (in MeV) are listed.} 
\centering
\begin{tabular}{cccccccc}
\hline \hline
  Parameters & N & Min. & Max. & Parameters & N & Min. & Max. \\[1.3ex]
  \hline 
  $e_0$ & 0 & -17 & -15 &$J_0$ & 0 & 28 & 36 \\[1.3ex]
       &   &       &     & $L_0$ & 1 & 0  & 100 \\[1.3ex]
  $K_0$ & 2 & 190 & 280 & $K_{\rm sym,0}$ & 2 & -400 & 200 \\[1.3ex]
  $Q_0$ & 3 & -1000 & 1000& $Q_{\rm sym,0}$ & 3 & -2000 & 2000 \\[1.3ex]
  $Z_0$ & 4 & -3000 & 3000& $Z_{\rm sym,0}$ & 4 & -5000 & 5000 \\[1.3ex]
 \hline\hline
\end{tabular}
\end{table}
The variation range for the model parameters reflects the degree of uncertainty associated with the EOS parameters. These empirical parameters can be classified into three distinct categories based on the extent to which they are experimentally constrained. The first group consists of the low-order iso-scalar parameters $e_0$ and $K_0$ and the iso-vector parameters $J_0$ of the density-dependent symmetry energy at saturation density. The relative uncertainties associated with the determinations of these parameters from nuclear experiments are below 15\%.
The second group consists of parameters that are not well-defined by current nuclear experiments. Parameters such as the iso-scalar skewness $Q_0$, the slope of the symmetry energy $L_0$, and the curvature parameter $K_{\rm sym,0}$ are characterized by significant uncertainties, although it is anticipated that greater precision will be achieved in the near future. For instance, the determination of $L_0$ is currently the subject of considerable interest and study ~\cite{Li:2014oda}.
The last group consists of parameters that are currently inaccessible to nuclear experiments. This includes the isovector skewness $Q_{\rm sym,0}$ and parameters of order N = 4, such as $Z_0$ and $Z_{\rm sym,0}$. Consequently, we explore a substantially very large range for these parameters. In this paper, we examine the effect of the pressure/energy density of pure neutron matter constraints ~\cite{Hebeler:2013nza, Lattimer:2021emm}  on the uncertainties associated with nuclear matter parameters. Using the same prior distribution for empirical parameters and likelihood function as in this thesis in Sec. \ref{BA-3}, we find that $L_0$ and $K_{\rm sym,0}$ exhibit a strong correlation and exert the greatest influence on neutron star properties, including radius and tidal deformability, especially at a canonical mass of 1.4 $M_\odot$.

\subsection{The Likelihood Function}
\label{BA-3}
The likelihood function shows how likely it is that the data will be seen with the model's parameters $\textbf{X}$. We used a Gaussian likelihood in our analysis, which is described as follows:
\bea
{\mathcal L}(data|\textbf{X})=&=&\prod_{j} 
\frac{1}{\sqrt{2\pi\sigma_{j}^2}}e^{-\frac{1}{2}\left(\frac{d_{j}-m_{j}({X)}}{\sigma_{j}}\right)^2}. 
\label{BA-eq:likelihood}  
\eea
Here, the index $ j$ runs over all the data, $d_j$ and $m_j$ are the data
and corresponding model values, respectively.  The $\sigma_j$ are the adopted
uncertainties. The selection of likelihood functions in Bayesian inference is crucial for the study of neutron stars because it has an immediate impact on the precision and dependability of parameter estimates. As a link between the theoretical model and the empirical measurements, the likelihood function basically measures the degree of agreement between model predictions and observed data. Choosing an appropriate likelihood function that accounts for the inherent uncertainties and features of the data is crucial in the context of neutron star physics. If the central limit theorem can be used, or if the data are regularly distributed, Gaussian likelihood functions are frequently used. However, other likelihood functions, including Poisson, Binomial, Exponential, Logistic, Multinomial, or Custom functions, might be more appropriate when the data distribution deviates from normality or shows notable outliers. In order to determine the robustness and informativeness of the Bayesian analysis in characterizing the features of neutron stars and the underlying nuclear equations of state, a thorough comprehension of the data and its statistical properties should be the basis for selecting the likelihood function.

\subsection{Physical Requirements and Constraints on Neutron Star Observables}
\label{BA-cons}

We use a low-density equation of state for pure neutron matter as constraints throughout our Bayesian analysis, which comes from a precise next-to-next-to-next-to-leading-order (N$^3$LO) calculation in chiral effective field theory ~\cite{Hebeler:2013nza}. As pseudodata, this EOS is used to build a basic likelihood function defined in Eq. (\ref{BA-eq:likelihood}). The $d$s and the $\sigma$s  in Eq. (\ref{BA-eq:likelihood}) are the pseudodata for the energy per neutron and the  corresponding uncertainties taken from
Ref. ~\cite{Hebeler:2013nza, Lattimer:2021emm}. Additionally, Drischler et al. ~\cite{Drischler:2021kxf} carried out a thorough validation to evaluate how well the meta-model, which is defined by a parameter set $\textbf{X}$, fits into the most recent chiral effective field theory calculations for both pure neutron matter (PNM) and symmetric nuclear matter (SNM).

In general, the following physical constraints apply to the functional's global density behaviour:
\begin{itemize}
    \item the causality condition, $c_s/c<1$;
    \item the thermodynamic stability of the EOS, $dp/d\rho > 0$,
    \item the maximum observed NS mass, $M_{max} (X) \geq M_{max}^{obs}$,
    \item the positiveness of the symmetry energy at all densities,
\end{itemize}
The $M_{max}(X)$ denotes the maximum mass supported by the EOS for a parameter set $\textbf{X}$, whereas $M_{max}^{obs}$ refers to the maximum neutron star mass observed. In the case of the heavier pulsar PSR J0740+6620, the maximum observed NS mass has been reported by two separate groups using Neutron star Interior Composition Explorer (NICER) X-ray telescopes. The masses reported are $2.08 \pm 0.07 M_\odot$ ~\cite{Miller:2021qha} and $2.072^{+0.067}_{-0.066} M_\odot$ ~\cite{Riley:2021pdl}, respectively. Since these measurements have large unpredictability issues, we should add this extra restriction to our analysis using an additional likelihood filter to the Eq. (\ref{BA-eq:likelihood}).

\subsection{Posterior Distribution of Nuclear Matter Parameters} \label{BA-posterior}
Next, we describe the collective posterior distribution over the nuclear matter parameters. Using two meta-models such as Taylor (in Sec. \ref{Taylor}) and $\frac{n}{3}$ (in Sec. \ref{n/3}) expansions, we compare the marginalized probability distributions while considering the isoscalar and isovector parameters. The Bayesian analysis is used to get the correlations between empirical parameters.

\begin{figure}[!ht]
   \centering
      \includegraphics[width=0.92\textwidth,height=5.5cm]{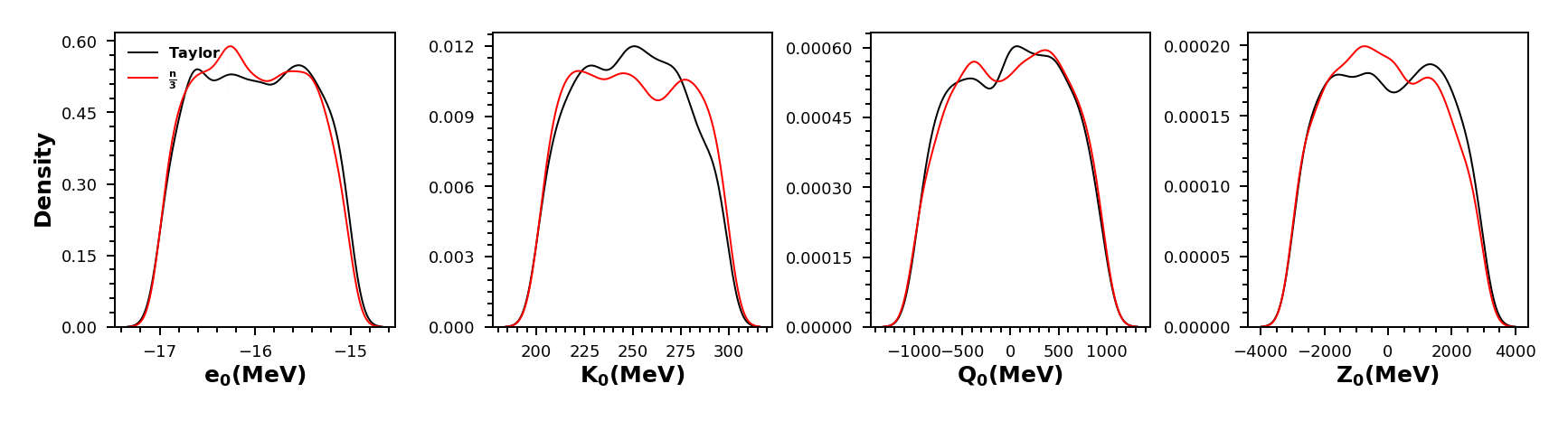}
      \includegraphics[width=0.96\textwidth,height=5.5cm]{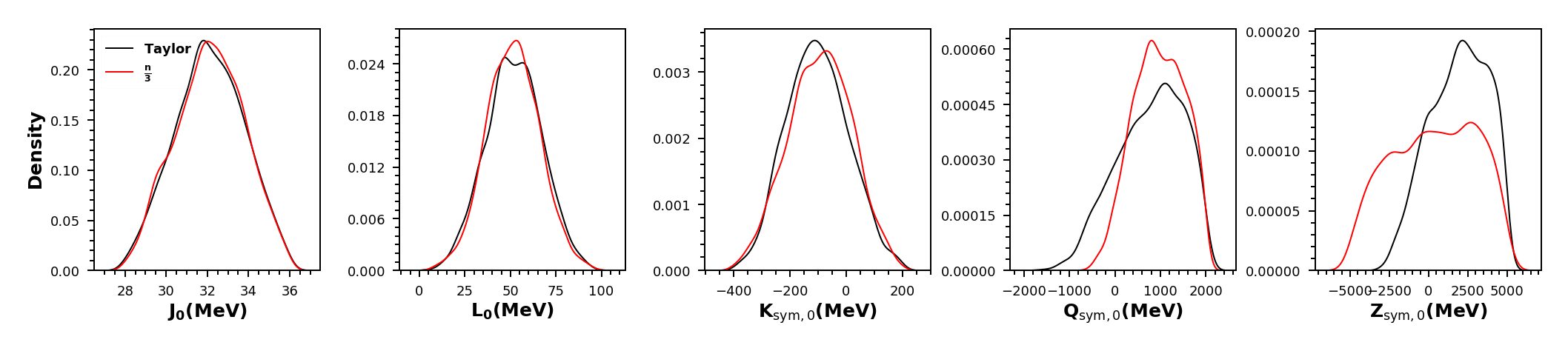}
  \caption{The marginalized posterior distributions of nuclear matter parameters. The black and red colour lines correspond to the Taylor and $\frac{n}{3}$ expansions.}
  \label{Meth:fig-dist-nmps}
\end{figure}

\begin{figure}[!ht]
   \centering
      \includegraphics[width=0.495\textwidth,height=8cm]{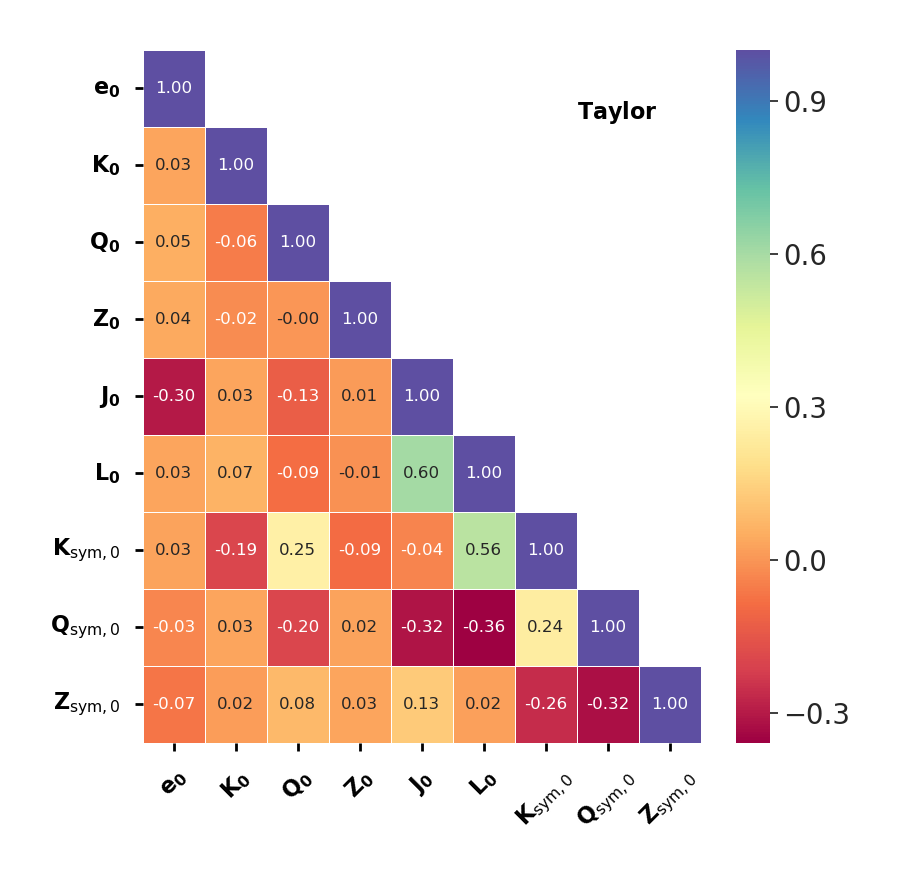}
      \includegraphics[width=0.495\textwidth,height=8cm]{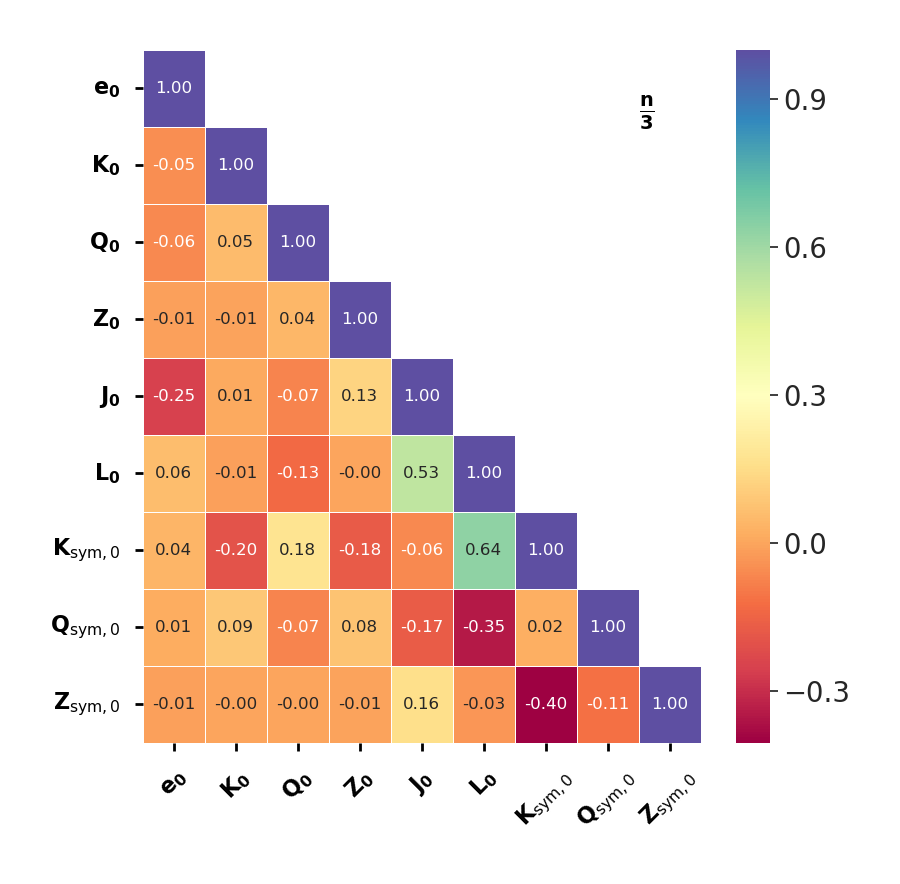}
  \caption{The correlation matrix for the nuclear matter parameters is shown. The left and right panels correspond to the Taylor and $\frac{n}{3}$ expansions.}
  \label{Meth:fig-matrix-nmps}
\end{figure}
Fig. \ref{Meth:fig-dist-nmps} depicts the marginalized posterior distributions of nuclear matter parameters, where both meta-models have been subjected to similar constraints. {For the constraints/data, we have used the low-density ($\rho=0.01-0.32$ fm$^{-3}$) EOS of pure neutron matter obtained from a N$^3$LO calculation in chiral effective field theory. In Likelihood, $d$s and $\sigma$s in Eq. (\ref{BA-eq:likelihood}) are the pseudodata for the energy per neutron and the corresponding uncertainties taken from figure-1 of Ref.~\cite{Lattimer:2021emm}.} It is clearly noticed in both the models the outputs of marginalized posterior distributions are almost the same except $Z_{\rm sym,0}$. The influence of the pure neutron matter constraint on the isovector parameters is particularly noteworthy, as can be observed in the figure. The Pearson's correlation coefficients among the nuclear matter parameters are shown in Fig. \ref{Meth:fig-matrix-nmps}. In both cases, the correlation coefficient values among the parameters are almost the same.  The correlations between the slope parameter $L_0$ and both $K_{\rm sym,0}$ and $J_0$ are notable, mostly due to the influence of the PNM constraint.

\section{Principal Component Analysis}
\label{PCA}
Principal Component Analysis (PCA) is an important statistical technique for analyzing multivariate data ~\cite{Wold1987, Aflalo2017, Al-Sayed:2015voa, shlens2014, Liu:2020ely, Acharya:2021lbe, Ali2023}. The PCA method analyzes the data of multiple dependent correlated variables to capture shared variation.
The objectives of PCA ~\cite{Abdi2010} are (i) Extract the most crucial information from the data, (ii) Reduce the dimensionality of the data by only maintaining the essential information, (iii) Examine the composition of the observations and variables, as encountered in literature.
In the present work, we considered  $K_0$, $Q_0$, $J_0$, $L_0$, $K_{\rm sym0}$ and  $Q_{\rm sym0}$ as our variables often referred to as features and the neutron stars properties such as tidal deformability and radius as our target variables. PCA  can verify our earlier hypothesis of selecting considered variables and prove the significance of these variables. 
In the process of PCA, the method computes Principal Components (PCs) that can be considered as new variables in other dimensions. PCs are the composition of linear combinations of original variables to capture the shared variation patterns. These PCs account for the amount of variation captured in data and return relative scores as eigenvalues in a sorted manner. So, the largest Eigenvalue associated with the first PC captures the largest variance, and so on. 

The methodology of PCA analysis is composed of the given steps:
(1) The covariance matrix is calculated from the given data. The covariance matrix measures the relationships between pairs of variables.
(2) Eigenvalue decomposition on the covariance matrix is carried out to obtain the eigenvectors and eigenvalues. 
(3) The eigenvectors represent the principal components, and the corresponding eigenvalues indicate the principal components' proportional variance captured.

To calculate the covariance matrix, we prepare our data $\bf{X}$ as a $I \times J$ matrix. We have $'I'$ samples that are represented by $'J'$ variables. We must standardize the data set by removing the mean and dividing by the standard deviation of each $\bf{X}$ column. This yields a standardized data matrix, such as $\bf{\dot{X}}$. The elements of the covariance matrix are determined as,
\bea
{{\bf{C}_{ij}}} &=& \frac{1}{n} \sum^n_{k=1} {\dot{\rm \bf{X}}_{ik}}{\dot{\rm \bf{X}}_{jk}}, 
\eea
where ${i,j}$ denotes the variables/features, and $\bf{\mathit{n}}$ runs over all the samples. 
The correlation matrix is used to determine the eigenvalues and eigenvectors. The order of eigenvalues provides the importance of the eigenvector. The most important principal component (PC1) is the eigenvector with the greatest corresponding eigenvalue. PC1 captures the highest variance among the data. The second component (PC2) must be orthogonal to the first component and capture the second-highest variance. In PC space, factor scores indicate observations (samples), which is the projection of data along with the PC components. The factor score matrix $\bf{F}$ is defined as,
\bea
\bf{F}= \bf{\dot{X}}\bf{V}. \label{eq-factor}
\eea
The matrix $\bf{V=\dot{X}A}$ is called a factor loading matrix, and matrix $\bf{A}$  contains eigenvectors. Matrix $\bf{F}$ gives the projections of observations on primary components, making it a projection matrix. 
Each PC's contribution to each original variable shows the captured variance. The importance of PC is decided based on the corresponding order of eigenvalue; hence, the contribution of the original variable also depends on the order of PC priority. 

\subsection{Computational details of PCA Analysis}\label{method-pca}

The technique of  Principal Component Analysis (PCA) is commonly used in data science for feature extraction and dimensionality reduction for a given target variable. Here are the steps to implement the PCA to identify key features
associated with a given target variable.

\begin{enumerate}
    \item Data Preprocessing: The initial step involves the preparation of the dataset.  The dataset should contain both the target variable and the features. 
    \item Standardization: The data should be standardized by adjusting it to have a mean of zero and a variance of unity. This stage is crucial for ensuring that all features are standardized to a similar scale, preventing any particular feature from exerting undue influence on the PCA process due to its greater magnitudes.
    
    \item The Covariance Matrix: Calculate the covariance matrix for the standardized dataset. The covariance matrix is a mathematical representation that captures both the variances and connections among different features. The computational complexity of the covariance matrix is $O(ND \times min(N, D))$, which results by multiplying two matrices of size $D \times N$ and $N \times D$, respectively. Here, $N$ is the number of samples, and $D$ is the dimensionality or simply the number of features.
    
    \item Eigenvalue decomposition: Determine the eigenvectors and eigenvalues of the covariance matrix. The eigenvectors represent the PCs, while the eigenvalues indicate the amount of variance explained by each PC. Arrange the eigenvalues in descending order to assign higher priority to the principal components that account for the greatest amount of variance. 
    
    \item Selection of PCs: Determine the number of main components that will be kept. The selection of this option depends upon our target to reduce dimensionality.  We can either select a certain number of top PCs or opt to keep a given proportion of the total variance (e.g., 95\%).
    \item Amplitude: An analysis is conducted to evaluate the square of the amplitude values of the feature on each PC. Features that have larger absolute loadings on a specific PC are regarded as having a greater contribution to that component.
    \item Percentage Contribution: The final step is to find the percentage contributions of all features to the target variable. The total contributions of a given feature are obtained by summing their contributions from each of the PCs weighted by the corresponding normalized eigenvalue. 
\end{enumerate}
The overall complexity of the PCA analysis is $O(ND\times min(N, D))$.
By following the above steps, One can identify the importance of PCs by reducing the dimensionality of the data and can extract the key parameters from the dataset for a specific target.

%% file: chapter3/chapter3.tex
\chapter{Bayesian Analysis of Dense Matter Equation of State}  
\label{chap3}
In this chapter, we obtained the EOSs for $\beta$-equilibrated matter (BEM)  using Taylor and $\frac{n}{3}$ expansions as discussed in the previous Chapter \ref{meta-model} in Eqs. (\ref{eq:ebeta_T}) and (\ref{eq:ebeta_n3}). The coefficients of the Taylor expansion are the individual nuclear matter parameters, whereas they correspond to linear combinations of nuclear matter parameters for the $\frac{n}{3}$ expansion.    
We have constructed marginalized posterior distributions for the nuclear matter parameters by applying a Bayesian approach to both the expansions considered. The nuclear matter parameters or the corresponding EOSs are consistent with a set of minimal constraints that includes basic properties of saturated nuclear matter and low-density ($\rho=0.08-0.16$~fm$^{-3}$) EOS for the pure neutron matter from  (N$^{3}$LO) calculation in chiral effective field theory \cite{Hebeler:2013nza}.
This large number of EOSs is employed to evaluate the properties of neutron stars, such as tidal deformability, radius, and maximum mass. The correlations of neutron star properties 
with the pressure of $\beta$-equilibrated matter at a given density are studied.  Most of these correlations are sensitive to the neutron star mass and EOS choice at a given density. Our results for the correlations of tidal deformability with pressure
for $\beta$-equilibrated matter are analogous to those obtained using a diverse set of nonrelativistic and relativistic mean-field models  (MFMs) that re-emphasize their model independence.
Such model-independent trends inspire us to parametrize the pressure for $\beta$-equillibrated matter around 2$\rho_0$   in terms of neutron star mass and the corresponding tidal deformability \cite{Patra:2022yqc}.

\section{Priors, Likelihood, and Filters}
We apply the Bayesian approach to obtain two large sets of EOSs corresponding to the Taylor and $\frac{n}{3}$ expansions.  The posterior
distributions for the NMPs are obtained by subjecting the EOSs to a  set of minimal constraints, which include some basic properties of nuclear matter evaluated at the saturation density $\rho_0$  and  EOS for the 
pure neutron matter at low density. The constraints on the nuclear matter parameters are incorporated through the priors and those
from the EOS for the pure neutron matter through the likelihood function. Not all the nuclear matter parameters are well-constrained.
Only a very few  low-order nuclear matter parameters  constrained within narrow bounds are the binding energy per nucleon
$\varepsilon_0=-16.0 \pm 0.3$ MeV , nuclear matter incompressibility coefficients $K_0=240 \pm 50$ MeV for the symmetric nuclear matter and symmetry energy coefficient $J_0=32.0 \pm 5 $ MeV. The values of $\varepsilon_0$ and $J_0$ are very well constrained by the binding energy of finite nuclei over a wide range of nuclear masses ~\cite{Chabanat98, Chabanat:1997qh, Malik:2019whk, Mondal:2015tfa, Mondal:2016roo, Sulaksono:2009rn}.
The value of $K_0$ is constrained from the experimental data on the
centroid energy of isoscalar giant monopole resonance in a few heavy
nuclei ~\cite{Garg:2018uam, Agrawal:2005ix}.  The values of $L_0$ have been
extracted from experimental data on variety of phenomena in the finite
nuclei as well as from neutron star observations. The model-independent
estimates of $L_0$ are expected to be derived from the measurement
of neutron-skin thickness in asymmetric nuclei. Recent measurement
of neutron-skin thickness in $^{208}$Pb nucleus yields $L_0 = 106\pm
37$ MeV~\cite{Reed:2021nqk}.  However, this value of $L_0$ has  only marginal overlap at the
lower side  with those determined using experimental data on iso-vector
giant dipole resonances in several nuclei~\cite{Roca-Maza:2015eza} and recent neutron
star observations~\cite{Essick:2021kjb}.  The remaining nuclear matter parameters,
$Q_0$, $Z_0$, $K_{\rm sym,0},
Q_{\rm sym,0}$ and $Z_{\rm sym,0}$  are constrained only poorly
~\cite{Tsang:2020lmb, Ferreira:2021pni, Dutra:2012mb, Dutra:2014qga, Mondal:2017hnh}.
The priors for the nuclear matter parameters employed in the present work
are listed in Table \ref{ch3-tab1}. The prior distributions of $\varepsilon_0,
K_0$ and  $J_0$ are assumed to be Gaussian with a smaller width,
whereas the other higher-order nuclear matter parameters correspond to   Gaussian distribution with a very
large width. We have also repeated our calculations with uniform priors for the higher order nuclear matter parameters, and the result for the median values are found to be practically unaltered, and uncertainties are modified marginally, up to 10\%(not shown). In what follows, we present only those results that are obtained with priors as listed in Table \ref{ch3-tab1}. 

\begin{table}[!ht]
\caption{\label{ch3-tab1} 
The prior distributions of the nuclear matter parameters. The nuclear matter parameters considered are the binding energy per nucleon ( $\varepsilon_0 $), incompressibility
coefficient ($K_0$), symmetry energy coefficient ($J_0$), its slope parameter ($L_0$),
symmetry energy curvature parameter ($K_{\rm sym,0}$) and  $Q_0 [Q_{\rm sym,0}]$ and $Z_0
[Z_{\rm sym,0}]$ are related to third and fourth-order density derivatives of  $E(\rho,0) $[ $E_{\rm sym}(\rho)$], respectively. All the nuclear matter parameters are evaluated at saturation density $\rho_0$ =  0.16 fm$^{-3}$.  The parameters of  Gaussian distribution (G) are the mean ($\mu$) and standard deviation ($\sigma$).} 
  \centering 
  \begin{tabular}{cccccccc}
    \hline \hline
NMPs &{Pr-Dist}&{$\mu$}&{$\sigma$} &NMPs &{Pr-Dist}&{$\mu$}&{$\sigma$}\\ [1.3ex]
(in MeV) &  &  & & (in MeV) &  &  & \\ [1.3ex]
%\cline{1-4}
\hline 
{$\varepsilon_0$} & G & -16 & 0.3 & {$J_0$} & G &32&5\\[1.3ex]
 &  &  &  & {$L_0$} &  G  &50   &50  \\[1.3ex] 
{$K_0$} & G &240 &  50 & {$K_{\rm sym,0}$} &G  &-100  &200 \\[1.3ex]
{$Q_0$} &   G  & -400 & 400 & {$Q_{\rm sym,0}$} & G& 550&400 \\[1.3ex]  
{$Z_0$} & G & 1500 & 1500& {$Z_{\rm sym,0}$} & G   &-2000& 2000\\
\hline\hline
\end{tabular}
\end{table}

We know that the direct application of the lattice QCD simulations
are challenging to hadronic physics at finite density due to sign
problems in Monte Carlo simulations. However, analytical calculations in terms of the effective degrees of freedom at low density ($\rho<\rho_0$), like chiral effective theory, are valid with negligible uncertainty. The
precise next-to-next-to-next-to-leading-order (N$^3$LO) calculation is
usually fitted to the nucleon–deuteron scattering cross section or
few-body observables, and even saturation properties of heavier nuclei
~\cite{Drischler:2021kxf}. The low-density EOS  for the pure neutron
matter obtained from a (N$^{3}$LO) calculation in chiral effective field
theory ~\cite{Hebeler:2013nza} is employed as pseudodata to obtain
a simple likelihood function as given by Eq. (\ref{BA-eq:likelihood}).
The $d$s and the $\sigma$s  in Eq. (\ref{BA-eq:likelihood}) are the pseudodata
for the energy per neutron and the corresponding uncertainties taken from
Ref. ~\cite{Hebeler:2013nza}. This has been employed in the past in many of the analyses as their pseudodata ~\cite{Ekstrom:2015rta, Lim:2018bkq, Lim:2019som, Malik:2022zol, Ghosh:2022lam}.
We have considered the values of energy per neutron 
over the density range $\rho = 0.08$ - $0.16$~fm$^{-3}$. At densities
lower than $0.08$~fm$^{-3}$, the neutron star matter is expected to be clusterized.

We  have filtered  the nuclear matter parameters
 by demanding  that (i)  pressure for the
$\beta$-equilibrated matter  should  increase monotonically with density
(thermodynamic stability), (ii) The speed of sound must not exceed the
speed of light (causality)  and (iii) maximum mass of neutron star must
exceeds $2M_\odot$ (observational constraint). The causality breaks down at higher density mostly for the  Taylor EOS.   In such cases, we use the stiffest EOS, $P(\epsilon)
= P_m + (\epsilon - \epsilon_m)$, where, $P_m$ and $\epsilon_m$ are the
pressure and corresponding energy density at which the causality breaks
~\cite{Glendenning:1992dr}.

\section{ Posterior distribution of nuclear matter parameters}\label{ch3-PD}

To undertake the correlation systematics as proposed, we need a large number of  EOSs with diverse behaviour and corresponding neutron star properties.    As discussed above, the posterior distributions for the nuclear matter parameters for the Taylor and $\frac{n}{3}$ expansions are obtained by subjecting the EOS  to a  set of minimal constraints. The joint posterior distribution of the NMPs for a given model depends on the product of the likelihood and the prior distribution of nuclear matter parameters (Eq. (\ref{eq:BT-2})).  The posterior distribution of each individual parameter is obtained by marginalizing the joint posterior distribution with the remaining model parameters. If the marginalized posterior distribution of a nuclear matter parameter is localized more than the corresponding prior distribution, then the nuclear matter parameter is said to be well constrained by the data used for model fitting. 
   
%\newpage
\begin{figure}
    \centering
    \includegraphics[width=\textwidth,height=0.80\textheight]{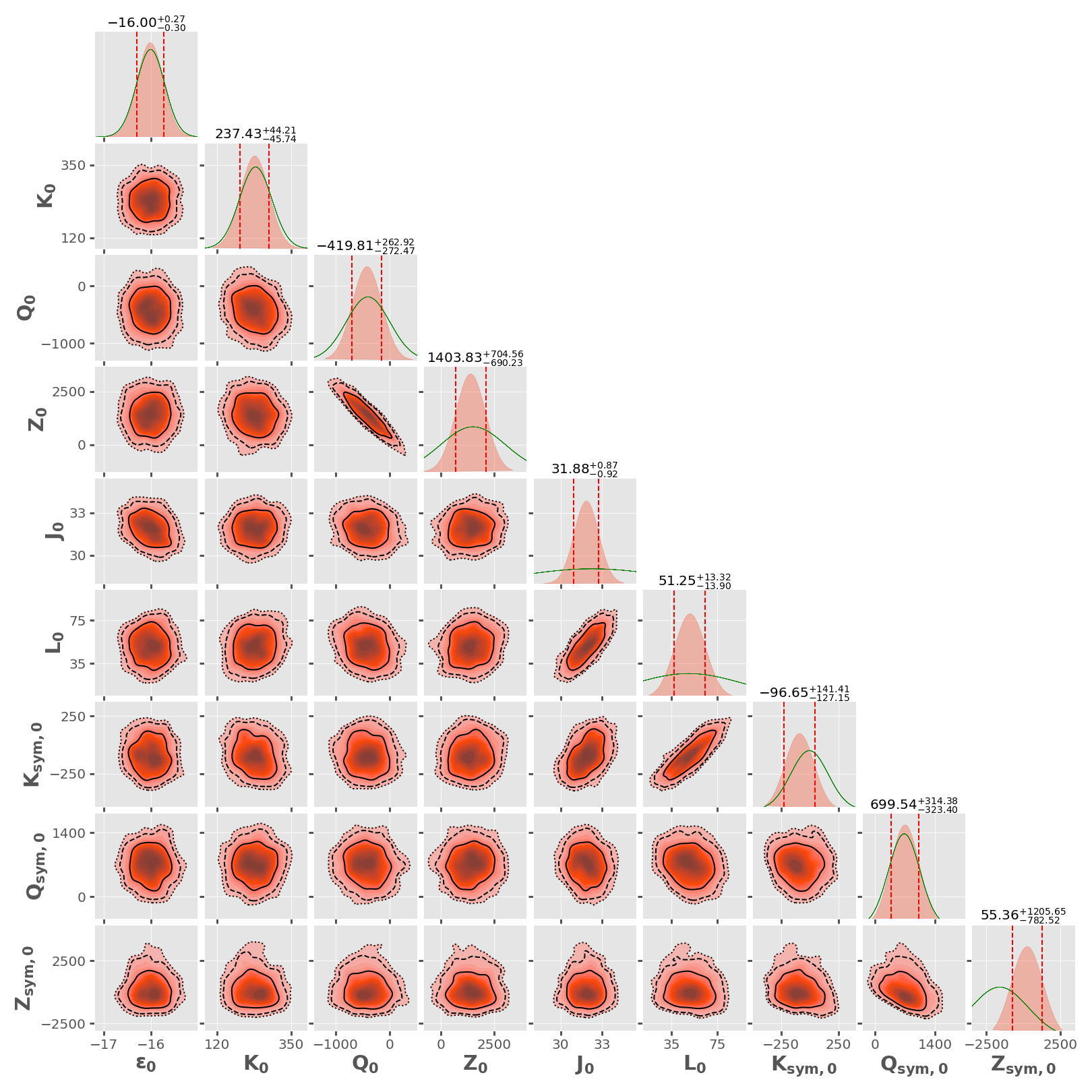}
       \caption{\label{ch3-fig1} Corner plots for the nuclear matter parameters (in MeV) obtained for Taylor expansions for the EOS of asymmetric nuclear matter. The one-dimensional marginalized posterior distributions (salmon) and the prior distributions (green lines) are displayed along the diagonal plots. The vertical lines represent the nuclear matter parameters' 68\% confidence interval.
 Along the off-diagonal plots, the confidence ellipses for two-dimensional posterior distributions are plotted with confidence intervals of 1$\sigma$, 2$\sigma$, and 3$\sigma$. The distributions of nuclear matter parameters are obtained by subjecting them to minimal constraints (see text for details). }
       
\end{figure}
%\newpage
% \vspace{0.5mm}
 \begin{figure}
    \centering
    \includegraphics[width=\textwidth, height=0.80\textheight]{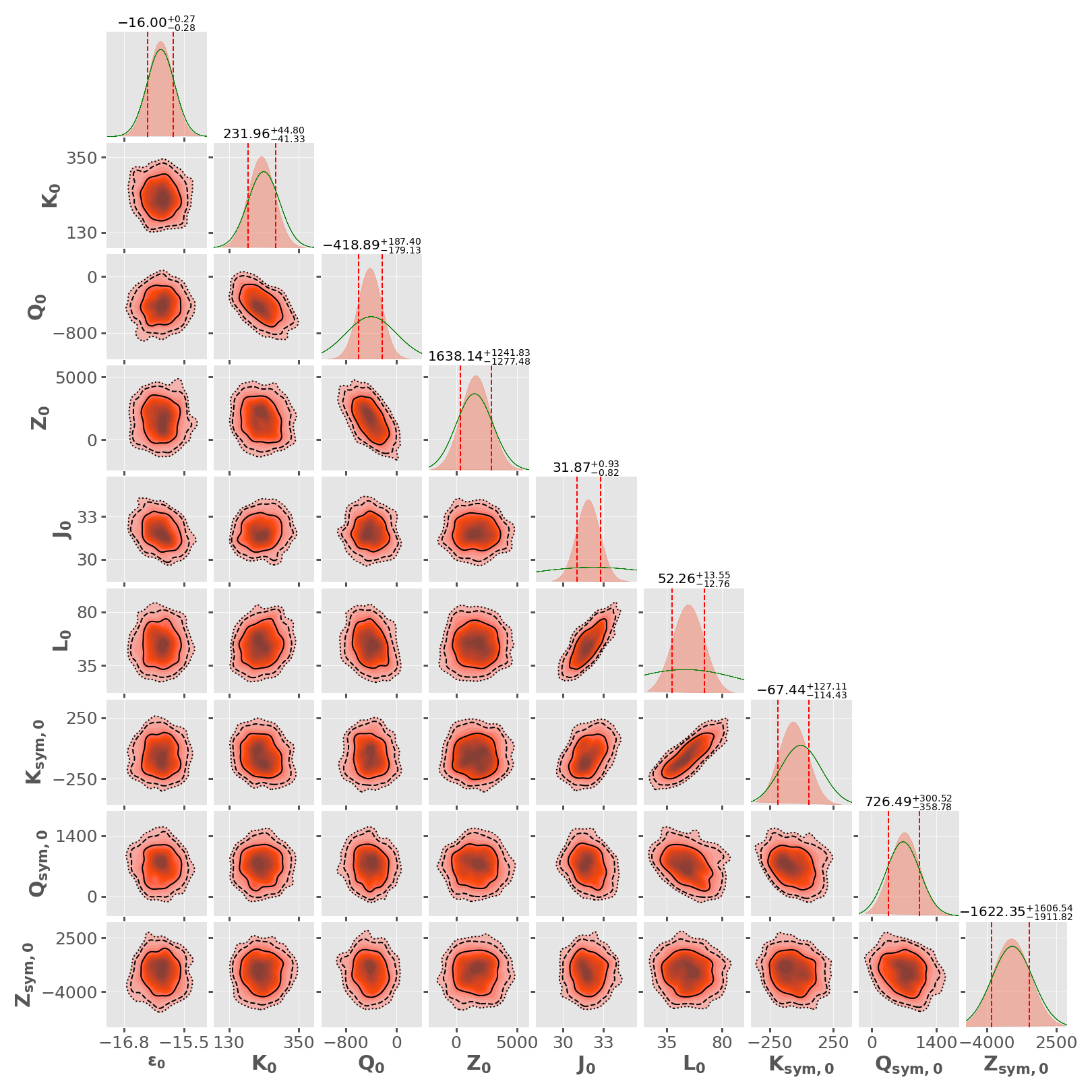} \caption{\label{ch3-fig2}
  The same as Fig. \ref{ch3-fig1}, but for $\frac{n}{3}$  expansions for the EOS of asymmetric nuclear matter. }
\end{figure}

The corner plots for the marginalized posterior distributions for the nuclear matter parameters in one and two dimensions obtained for Taylor and $\frac{n}{3}$ expansions are displayed in Figs.\ref{ch3-fig1} and \ref{ch3-fig2}, respectively.
The differences between the one-dimensional posterior distributions for the nuclear matter parameters and corresponding prior distributions reflect the role of low-density EOS for pure neutron matter in constraining the nuclear matter parameters. The EOS for the pure neutron matter mainly constraints the values of $J_0$, $L_0$ and $K_{\rm sym,0}$ and to some extent $Q_{\rm sym,0}$ and $Z_{\rm sym,0}$. The shapes and orientations of the confidence ellipses suggest that the correlations among most of the NMPs are weak. Most Strong correlations exist only between  $Q_0-Z_0$, $L_0-J_0$   and $L_0-K_{\rm sym,0}$  for both the expansions
with correlation coefficient r$\simeq$ 0.8. 
The $K_0 -Q_0$  correlation is  slightly better in case  of $\frac{n}{3}$ expansion (r$\sim$ -0.6) as compared to Taylor (r$\sim$ -0.18).
The median values of the nuclear matter parameters and the corresponding $68\% (90\%)$ confidence intervals obtained from the marginalized posterior distributions are listed in Table \ref{ch3-tab3}(see Appendix ). 
We also provide the values for the nuclear matter parameters obtained without the PNM constraints. The low-density pure neutron matter mainly constrains those nuclear matter parameters associated with the density dependence of symmetry energy. The median values of $L_0$ and $K_{\rm sym, 0}$, which determined the linear and quadratic density dependence of the symmetry energy, become smaller, suggesting softer symmetry energy "at high-density" with the inclusion of pure neutron matter constraints. Furthermore, the uncertainties on $L_0$ were reduced by more than 50\%. The median value of $Q_{\rm sym,0}$ remains more or less unaltered. 
From the recent measurement 
of the neutron-skin thickness 
for $^{208}$Pb nucleus (PREX-II)~\cite{PREX:2021umo,Reed:2021nqk},
 $\Delta R_{\rm skin}=0.283\pm0.071$ fm, 
 the value of
 $L_0$ has been determined to be $106\pm 37$ MeV ~\cite{Reed:2021nqk} . This value of $L_0$  agrees with the ones obtained in the
present work with PNM constraints only within 90\% confidence interval.

\section{Properties of Neutron Stars}

Once the EOS for the core and crust are known, the values of NS mass, radius, and tidal deformability corresponding to given central pressure can be obtained by solving Tolman-Oppenheimer-Volkoff equations ~\cite{Oppenheimer:1939ne, Tolman:1939jz}. The EOSs for the core region of a neutron star correspond to the $\beta$-equilibrated matter over the density range $0.5-8\rho_0$, are obtained from the posterior distributions of nuclear matter parameters for the Taylor and $\frac{n}{3}$ expansions. The core EOSs are matched to the crust EOSs to obtain the NS properties. The EOS for the outer crust is taken to be the one given by Baym, Pethick, and Sutherland  ~\cite{Baym:1971pw}. The inner crust that joins the inner edge of the outer crust and the outer edge of the core is assumed to be polytropic ~\cite{Carriere:2002bx}, $p(\varepsilon)= c_1 + c_2 \varepsilon^{\gamma}$. Here, the parameters $c_1$ and $c_2$ are determined in such a way that the EOS for the inner crust matches with the outer crust at one end ($\rho = 10^{-4}$ fm$^{-3}$) and with the core at the other end ($0.5\rho_0$ ). The polytropic index $\gamma$ is taken to be equal to 4/3. The radii of neutron stars with mass $\sim$ 1$M_\odot$ are more sensitive to the treatment of crust EOS  ~\cite{Fortin:2016hny}. It is demonstrated that the treatment of crust EOS employed in the present work may introduce the uncertainties of about 50-100 m in radii of NSs having mass 1.4$ M_\odot$. It is shown in Ref. ~\cite{Piekarewicz2019} that the choice of EOS for the inner crust does not significantly impact the values of tidal deformability, which depends on the Love number k$_{2}$ as well as the compactness parameter.

\begin{figure}[!ht]
\centering
\includegraphics[width=0.5\textwidth]{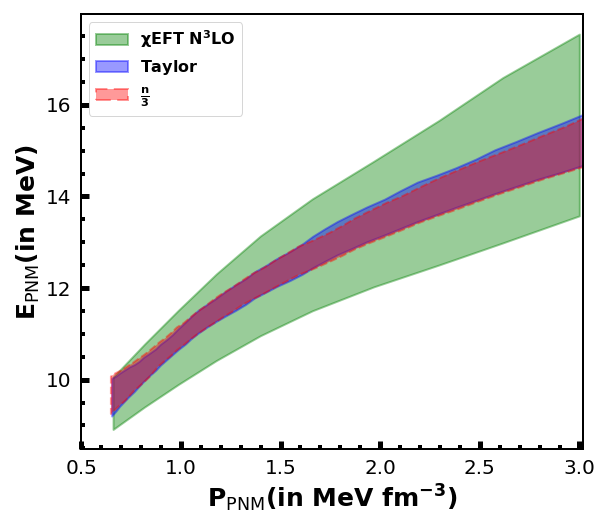}
\caption{ The equation of state for pure neutron matter is shown. The colored bands correspond to $\chi$EFT N$^3$LO (light green), 90\% confidence interval for the Taylor EOSs (light blue) and for $\frac{n}{3}$ EOSs (light magneta) obtained in our calculation (see text for details).}\label{ch3-fig3}
\end{figure}

{ The 90\% confidence interval for pure neutron matter Taylor EOSs (light blue) and  $\frac{n}{3}$ EOSs (light magenta) for the density range ( $0.5-1.0\rho_0$) is shown in Fig.~\ref{ch3-fig3}. For comparison, colored bands corresponding to  N$^3$LO (light green) from chiral effective field theory are displayed. Our EOSs lie almost in the middle of the N$^3$LO band. This indicates that our EOSs are well fitted with the pure neutron matter EoS from a precise next-to-next-to-leading-order calculation in chiral effective field theory.}

\begin{figure}[htp]
    \centering
    \includegraphics[width=0.48\textwidth]{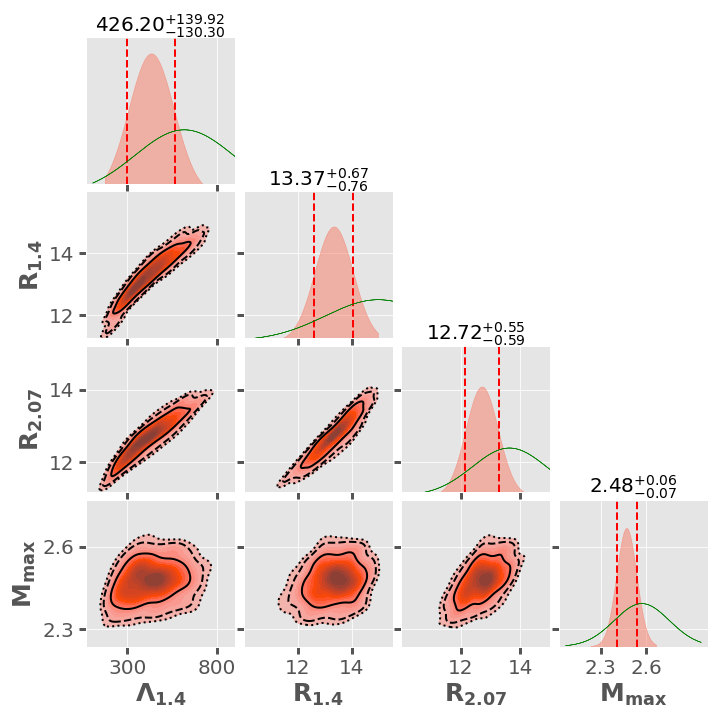}
    %\vspace{1.0mm}
    \includegraphics[width=0.48\textwidth]{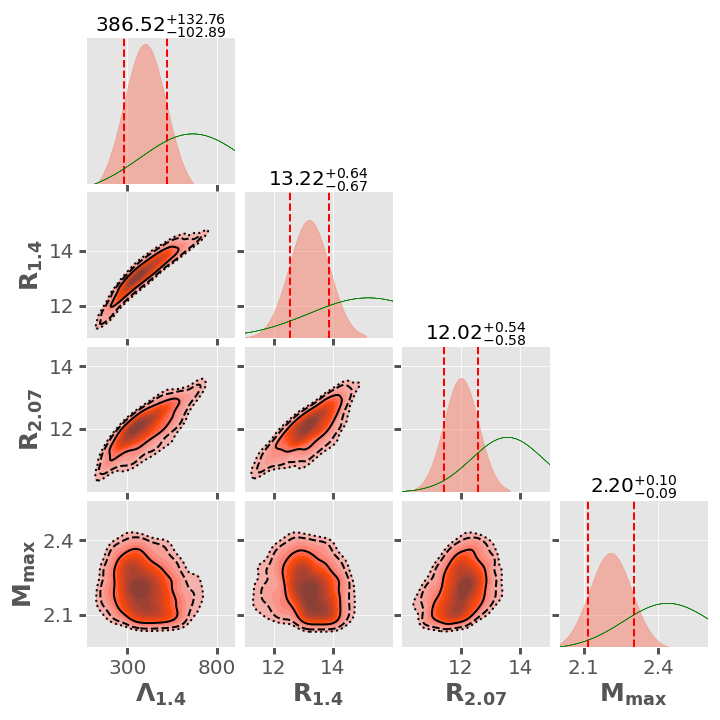}
    \caption{\label{ch3-fig4}Corner plots representing the marginalized posterior distributions (salmon)  of
the tidal deformability $\Lambda_{1.4}$, radii $R_{1.4}$ (km) and $R_{2.07}$ (km)  and the maximum mass $M_{\rm max}$ ($M_\odot$) for Taylor (left) and $\frac{n}{3}$ (right) expansions.  The green lines  represent   effective  priors obtained using the priors
for nuclear matter parameters (see also Table \ref{ch3-tab1}).}
\end{figure}

We have obtained the distributions of $\Lambda_{1.4}$, $R_{1.4}$, $R_{2.07}$ and $M_{\rm max}$ using the posterior distributions for the nuclear matter parameters corresponding to the Taylor and $\frac{n}{3}$ expansions. The corner plots for these NS properties are displayed in Fig.~\ref{ch3-fig4}. The effective priors for the NS properties, as shown by green lines, are
obtained using the priors for the nuclear matter parameters. The posterior
distributions of NS properties are narrower than the corresponding
effective priors, indicating the significance of the low-density EOS for
pure neutron matter. The posterior distributions of $\Lambda_{1.4}$ and $R_{1.4}$ for both the expansions
are quite close to each other. The differences begin to appear for the case of $R_{2.07}$, which becomes even larger for the maximum mass.  This is due to the fact that the Taylor EOSs are much more
stiffer than those for $\frac{n}{3}$. The dichotomy in the high-density
the behavior of the Taylor and $\frac{n}{3}$ expansions would help us to
understand the extent to which the correlations of the  EOSs with the
properties of NS, for masses in the range  $1$ - $2M_\odot$, are
model dependent. It is clear from off-diagonal plots that $\Lambda_{1.4}$ is strongly correlated with $R_{1.4}$, the correlation coefficient is  r$\sim$ 0.9. The $\Lambda_{1.4}$ and $R_{1.4}$ also display stronger correlations with $R_{2.07}$ (r $\sim$ 0.8) for the case of Taylor and somewhat moderate correlations (r $\sim$ 0.7) for the $\frac{n}{3}$ expansion.
 { As the EOSs are generated with the same models for each case and there is no phase transition appears, So we found a strong correlation between them. Otherwise, it will be weaker.}
 The maximum mass of a neutron star is almost uncorrelated with the other NS properties considered.

\begin{figure}[!ht]
    \centering
    \includegraphics[width=0.5\textwidth]{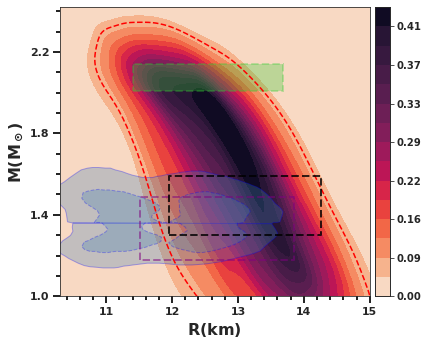}
    \caption{\label{ch3-fig5}
    The plot represents the joint probability distribution $P(M, R)$ as a function of mass and radius of neutron star obtained for $\frac{n}{3}$ expansion. The 90\% confidence interval is represented by the red dashed line. The 90\% (solid) and 50\% (dashed) confidence intervals of the LIGO-Virgo analysis for the BNS component from the GW170817 event are shown by the outer and inner gray shaded regions \cite{Abbott2018, Abbott18a, Abbott2021}. The constraints from the millisecond pulsar PSR J0030+0451 (purple \& black) NICER x-ray data \cite{Riley:2019yda, Miller:2019cac} and PSR J0740+6620 (green) \cite{Riley:2021pdl} are shown in the rectangular regions enclosed by dotted lines.
 }
\end{figure}

 We have  summarized in
 Table \ref{ch3-tab4} (see Appendix ) shows the median values of NS properties along with 68\% (90\%) confidence intervals.   Like in the case of nuclear matter parameters, the NS properties get significantly constrained by the EOS of pure neutron matter at low density. For instance, the median values of $\Lambda_{1.4}$ become smaller by about 15\%  and the associated uncertainties by about 40\%  with the pure neutron matter constraints. The median values of $R_{1.4}$ and the corresponding uncertainties also become noticeably smaller. 
The $R_{2.07}$ and $M_{\rm max}$ do not show significant changes with the inclusion of low-density pure neutron matter constraints. With the PNM constraints, the 90\% confidence interval of the neutron stars properties such as tidal deformability, radius, and mass overlap with the currently available bounds, $\Lambda_{1.4} \in [70,580]$ ~\cite{GW170817}, $R_{1.4} \in [11.41,13.61]$ km ~\cite{Miller:2021qha}, $R_{2.07} \in [11.8,13.1]$ km ~\cite{Riley:2021pdl} and $M_{\rm max} \geq 2.09 M_{\odot}$ ~\cite{Romani:2021xmb}.
The $M_{\rm max} = 2.48_{-0.07}^{+0.06} M_\odot$ obtained for the
Taylor EOSs are on the slightly higher side in comparison to the ones
derived by combining the GW170817 observations of merging of binary
neutron stars and quasiuniversal relation  ~\cite{Rezzolla_2018}. The observed electromagnetic emissions in the form of kilonova and the detection of a gamma-ray burst have been linked to the formation of a black hole and, thus, have been utilized to infer the maximum mass of a stable neutron star. However, such inference of the maximum mass is subjected to uncertainties originating from model dependence of postmerger dynamics. Recent observation of the GW190814 event, a neutron star black hole/binary neutron star merger, has triggered an assessment of the maximum mass of a stable neutron star \cite{GW190814}. While there are different opinions available in the literature, the nature of a compact object in the range of 2.5 - 2.67 $M_\odot$ being a neutron star or black hole seems to be an unsettled issue to date ~\cite{Tsokaros_2020, Lim:2020zvx, Drischler_2021, Li:2021crp, Rezzolla_2018, GW190814}. So the maximum mass ($M_{max}$) we got for the Taylor model supporting the static NS of mass greater than 2.5 M$_\odot$ may not be ruled out at present.

We obtain joint probability distribution  $P(M,R)$  for a given mass and radius
for both the  Taylor and $\frac{n}{3}$  expansions. They display qualitatively very similar trends.  In Fig. \ref{ch3-fig5}, 
we plot the
$P(M,R)$ obtained for the  $\frac{n}{3}$ expansion.
The red dashed line represents the 90\% confidence interval. The color gradient from orange to dark purple
represents the lowest to highest probability. The most probable values for $R_{1.4}$ and $R_{2.07}$ are approximately 13.3 and 12.3 km, respectively. The $P(M,R)$ is maximum for $M\sim 
1.4 - 2.0 
M_{\odot}$, $R \sim 12.4 - 13.4
$ km. The $90\%$ confidence interval has 
significant
 overlap with LIGO-Virgo 
 and NICER estimations.  It may be however pointed out that the main objective of the present work is to construct large sets of EOSs with diverse behavior to assess various correlation systematics as follows.

\section{Correlations of Neutron Star properties with EOS}
\label{ch3-subsec_correl}
\begin{figure}[!ht]
\centering
\includegraphics[width=0.7\textwidth]{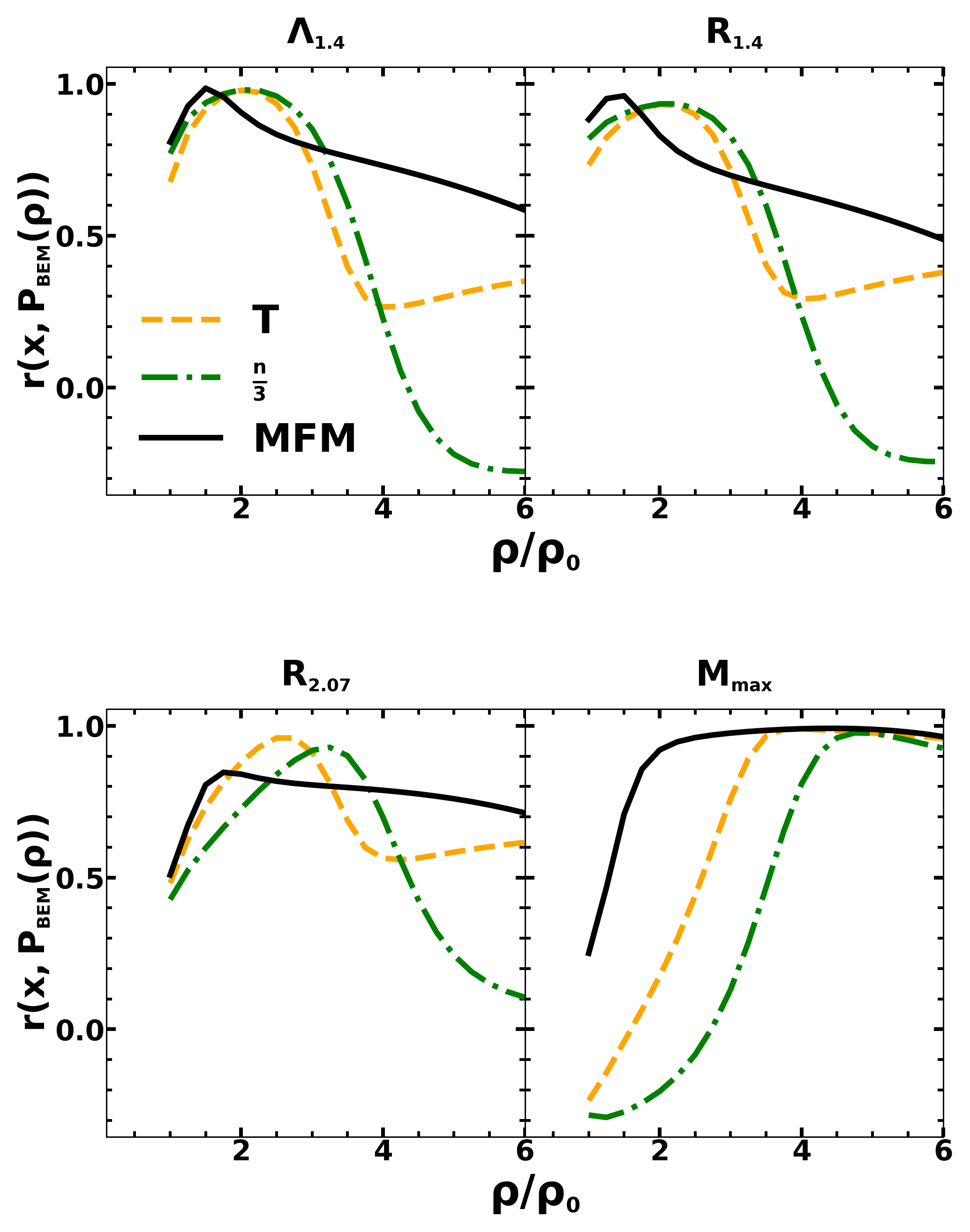}
\caption{\label{ch3-fig6}
The correlation coefficients r(x,$P_{\ms{\rm BEM}}(\rho)$), as approximated by both Taylor and $\frac{n}{3}$ expansion along with the mean-field theory calculations, is shown in this figure. Here, x represents either of the tidal deformability $\Lambda_{1.4}$, radii $R_{1.4}$, $R_{2.07}$, or maximum mass $M_{\rm max}$ of the neutron star, whereas, $P_{\ms{\rm BEM}}(\rho)$ represents the pressure for $\beta$-equilibrated matter  at a   density $\rho$. The calculations are performed with neutron star properties obtained using marginalized posterior distributions of nuclear matter parameters in Taylor and $\frac{n}{3}$ expansions. For the comparison, the results are also displayed for a diverse set of non-relativistic and relativistic microscopic mean-field models (MFMs).}
\end{figure}

We randomly select 1000 EOSs and corresponding NS properties from marginalized posterior distributions obtained for the Taylor as well as $\frac{n}{3}$ expansions. They are used to study the correlations of various NS properties with key quantities determining the behavior of the EOS. 
The correlations of $\Lambda_{1.4}$, $R_{1.4}$, $R_{2.07}$ and $M_{\rm max}$  with the pressure of $\beta$-equilibrated matter over a wide range of densities are evaluated. The values of correlation coefficients are plotted as a function of density in Fig.\ref{ch3-fig6}.
We also display the values of correlation coefficients for NS properties with the pressure of $\beta$-equilibrated matter calculated using unified EOSs for a diverse set of 41 non-relativistic and relativistic microscopic mean-field models (MFMs) \cite{Fortin:2016hny}. The various NS properties considered show strong correlations with $P_{\ms{\rm BEM}}(\rho)$ around a particular density.
The density at which the correlation is at its maximum increases with the NS mass. The values of $\Lambda_{1.4}$  and $R_{1.4}$ are strongly correlated with $P_{\ms{\rm BEM}}(\rho)$ at density $\sim$ $1.5-2.5\rho_0$. The $R_{2.07}$ is strongly correlated with $P_{\ms{\rm BEM}}(\rho)$ around 3$\rho_0$. The $M_{\rm max}$ is strongly correlated with $P_{\ms{\rm BEM}}(\rho)$  around 4.5$\rho_0$.
Our results for the Taylor and $\frac{n}{3}$ expansions for the region of maximum correlations are in line with those obtained using a diverse set of mean-field models, except for the case of $R_{2.07}$. Thus, it seems possible that the EOS  over a range of densities beyond $\rho_0$ can be constrained in a nearly model-independent manner with the help of various NS observables.

\begin{figure}[!ht]
\centering
\includegraphics[width=0.7\textwidth]{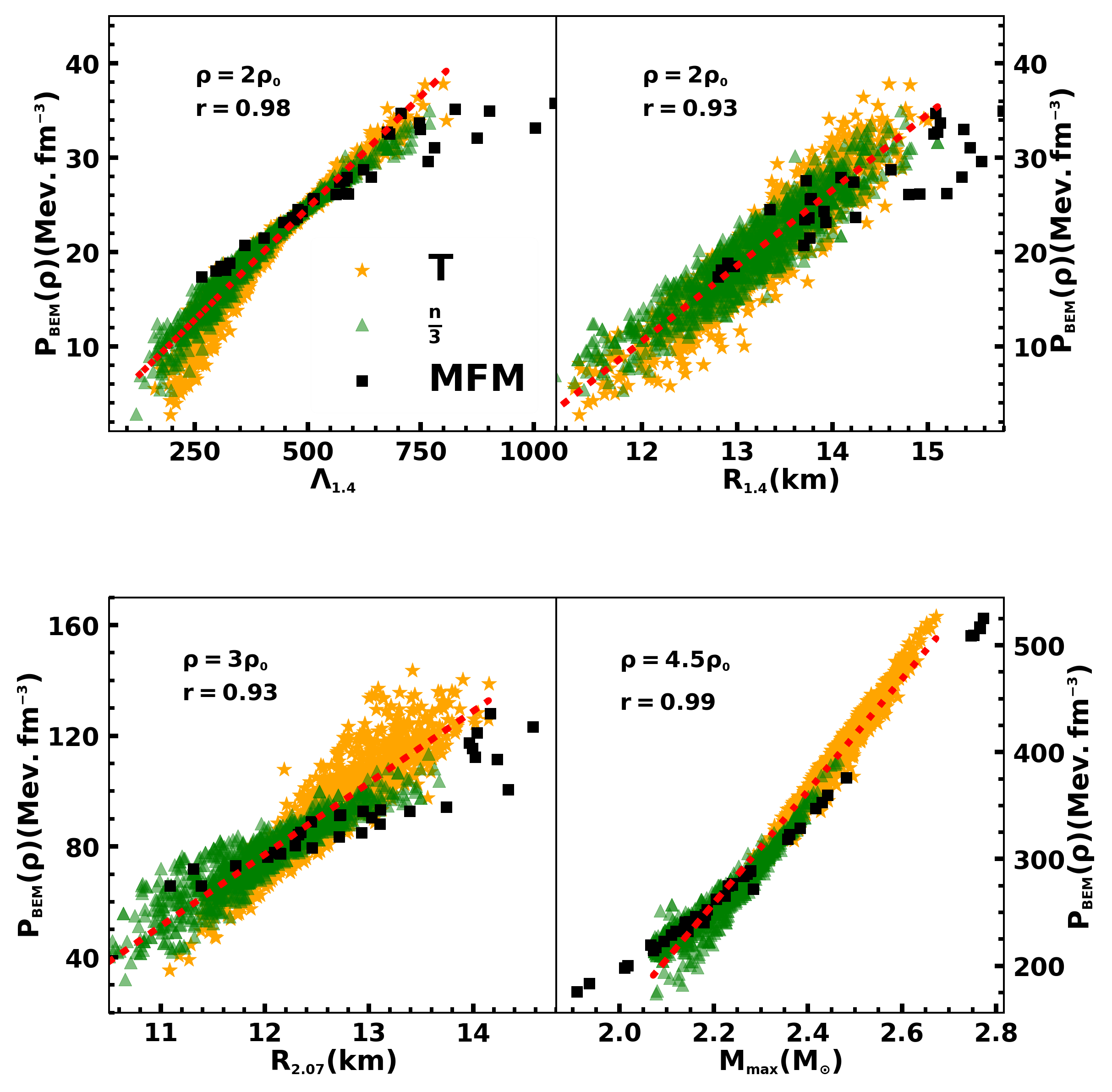}
\caption{\label{ch3-fig7}The variations of pressure for $\beta$-equilibrated matter [$P_{\ms{\rm BEM}}(\rho)$] at selected densities versus tidal deformability $\Lambda_{1.4}$, radii $R_{1.4}$ and $R_{2.07}$ and maximum mass $M_{\rm max}$ of neutron star. The red dashed lines are obtained by linear regression [see Eq.~(\ref{ch3-linear_regr}) in Sec.\ref{ch3-subsec_correl}].}
\end{figure}

\begin{figure}[!ht]
    \centering
    \includegraphics[width=0.5\textwidth]{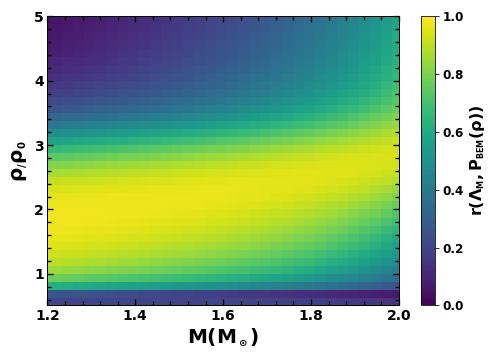}
    \caption{\label{ch3-fig8} Dependence of correlation coefficients between tidal deformability ($\Lambda_M$)  and the pressure of $\beta$-equilibrated matter ($P_{\ms{\rm BEM}}(\rho)$) on neutron star mass (M) and density ($\rho$) is depicted in this plot. Here $\rho_0$=0.16 fm$^{-3}$ is used only for scaling purposes. }
\end{figure}

In Table \ref{ch3-tab5}(see the Appendix ), we list the values of correlation coefficients obtained between the NS properties and the EOS at some selected densities. The correlation coefficients are obtained using  100 and 1000 EOSs, corresponding to Taylor and $\frac{n}{3}$ expansions,
randomly selected from the posterior distributions. We also present the results which are obtained by combining 1000 EOSs corresponding to each of the expansions. The values of correlation coefficients for the combined set of EOSs are close to those obtained separately. The values of the correlation coefficients are close to those obtained for mean-field models, which are listed in the second column.
We plot in Fig. \ref{ch3-fig7}, the variations of $P_{\ms{\rm BEM}}(\rho)$, at selected densities, with $\Lambda_{1.4}$, $R_{1.4}$, $R_{2.07}$ and $M_{\rm max}$ for which the correlations are stronger. We compare our results with those obtained from a diverse set of mean-field models. The correlation lines obtained by combining results of the Taylor and $\frac{n}{3}$ expansions are also plotted to estimate the values of $P_{\ms{\rm BEM}}(\rho)$ at selected densities with the help of NS properties. The equations for the correlation lines are obtained using linear regression as,
\bea
\label{ch3-linear_regr}
\frac{ P_{\ms{\rm BEM}}(2\rho_0)}{\rm MeV \rm fm^{-3}} &=& (0.96 \pm 0.10)   + (0.0473 \pm 0.0002)\Lambda_{1.4},\non
 \frac{ P_{\ms{\rm BEM}}(2\rho_0)}{\rm MeV \rm fm^{-3}} &=&  (-85.63 \pm 0.89) +  (8.01 \pm 0.06)\frac{ R_{1.4}}{\rm
km},\non
 \frac{ P_{\ms{\rm BEM}}(3\rho_0)}{\rm MeV \rm fm^{-3}} &=&
(-233.16 \pm 2.85) + (25.86 \pm 0.23) \frac{R_{2.07}}{ \rm km},\non
 \frac{ P_{\ms{\rm BEM}}(4.5\rho_0)}{\rm MeV \rm  fm^{-3}} &=&
 (-895.85 \pm 4.00) +  (524.75 \pm 1.70) \frac{M_{ \rm max}}{M_{\odot}}.
\eea

\begin{table}[!ht]
    \centering
    \caption{\label{ch3-tab2} The median values and associated 68\%(90\%) uncertainties for the parameters, appearing in Eq. (\ref{ch3-eq_para}), obtained from their marginalized posterior distributions. The values of parameters $b_0$, $b_1$, and $b_2$ as listed are scaled up by a factor of 10.}
    \setlength{\tabcolsep}{1pt}
    \renewcommand{\arraystretch}{0.1}
    \begin{adjustwidth}{-1.7cm}{}
    \nointerlineskip\leavevmode
    \begin{tabular}{ccccccc}
    \hline\hline
    Pressure & $a_0$ & $a_1$ & $a_2$ & $b_0$ & $b_1$ & $b_2$\\[1.5ex]
    ({in MeV fm$^{-3}$}) &  &   &  &  & & \\
    \hline
     $P_{\ms{\rm BEM}}(1.5\rho_0)$  & $0.544^{+0.031(0.050)}_{-0.029(0.060)}$ & $1.869^{+0.158(0.260)}_{-0.161(0.309)}$ & $7.451^{+0.237(0.390)}_{-0.234(0.450)}$ & $0.176^{+0.001(0.001)}_{-0.001(0.002)}$ & $0.740^{+0.004(0.007)}_{-0.004(0.008)}$ &  $1.152^{+0.013(0.021)}_{-0.012(0.025)}$ \\[1.6ex]
     $P_{\ms{\rm BEM}}(2\rho_0)$ &  $0.146^{+0.030(0.050)}_{-0.030(0.059)}$  & $-0.598^{+0.163(0.269)}_{-0.159(0.320)}$ & $27.909^{+0.233(0.397)}_{-0.240(0.469)}$ & $0.493^{+0.001(0.001)}_{-0.001(0.002)}$  & $2.234^{+0.004(0.007)}_{-0.004(0.008)}$ & $3.728^{+0.012(0.021)}_{-0.012(0.025)}$ \\[1.6ex]
     $P_{\ms{\rm BEM}}(2.5\rho_0)$ & $7.345^{+0.030(0.050)}_{-0.030(0.060)}$  & $-15.102^{+0.161(0.272)}_{-0.167(0.321)}$ & $68.411^{+0.239(0.396)}_{-0.238(0.475)}$ & $0.906^{+0.001(0.001)}_{-0.001(0.002)}$ &  $4.518^{+0.004(0.007)}_{-0.004(0.008)}$ & $8.115^{+0.012(0.021)}_{-0.012(0.025)}$ \\[1.6ex]
     \hline\hline
    
    \end{tabular}
    \end{adjustwidth}
\end{table}

\begin{figure}[!ht]
    \centering
    %\hspace{-0.5cm}
    \includegraphics[width=0.7\textwidth]{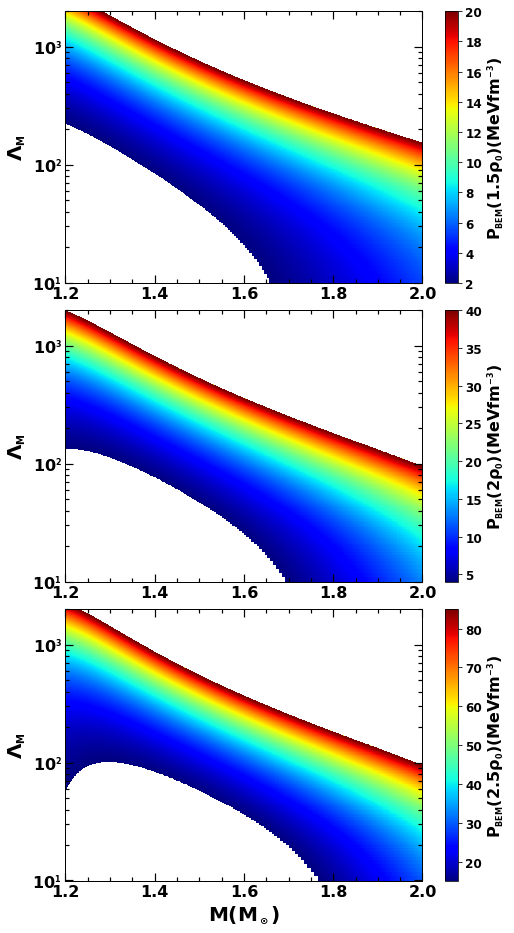}
    \caption{\label{ch3-fig9} The median values of pressure for $\beta$-equilibrated matter is shown here as a function of neutron star mass and its tidal deformability at densities 1.5$\rho_0$ (top), 2.0$\rho_0$ (middle) and 2.5$\rho_0$ (bottom).}
    
\end{figure}

We extend our analysis for the correlations of the pressure for the $\beta$-equilibrated matter with tidal deformability over a wide range of neutron stars' mass. In Fig. \ref{ch3-fig8}, we display color-coded graph for the correlations of tidal deformability of neutron star for the mass $1.2-2.0 M_\odot$ with the pressure for $\beta$-equilibrated matter at densities $0.5-5\rho_0$ (r($\Lambda_M, P_{\ms{\rm BEM}}(\rho)$)). One can easily obtain the value of the correlation coefficient as a function of density at a given NS mass. The $P_{\ms{\rm BEM}}(\rho)$ at $\rho \sim 1.5-2.5\rho_0$ are strongly correlated ($r \sim 0.8-1$ ) with tidal deformability for NS masses in the range $1.2-2.0 M_\odot$. Hence,
$P_{\ms{\rm BEM}} (\rho)$ can be parametrized at a given $\rho$ as,
\bea
\frac{ P_{\ms{\rm BEM}}(\rho)}{\rm MeV \rm fm^{-3}}
 &=&a(M) +b(M) \Lambda_M, \label{ch3-eq_para}
\eea
with  mass-dependent coefficients $a(M)$ and $b(M)$ expanded as
\bea
a(M) &=&  (a_0 + a_1(M-M_0) + a_2(M-M_0)^2), \\
b(M) &=&  ( b_0 + b_1(M-M_0) + b_2(M-M_0)^2), \label{ch3-eq_para2}
\eea
respectively, where $M_0$ is taken to be 1.4$M_\odot$ and the values of $a_i$ and $b_i$ are estimated using a Bayesian approach with the help of  $P_{\ms{\rm BEM}} (\rho)$ and tidal deformability obtained for Taylor and $\frac{n}{3}$ expansions. For a given $\rho$, the Eq. (\ref{ch3-eq_para}) is fitted using the tidal deformability corresponding to NS mass $1.2-2.0 M_\odot$. The priors for $a_i$ and $b_i$ are taken to be uniform in the range of -100 to 100. The calculations are performed for $\rho$= 1.5, 2.0, and 2.5 $\rho_0$. All the $a_i$s are strongly correlated with corresponding $b_i$s.
The median values of parameters $a_i$ and $b_i$ and associated uncertainties are summarized in Table \ref{ch3-tab2}. It may be noticed that the values of $a_0$ and $b_0$ for the case of
$P_{\ms{\rm BEM}}  (2\rho_0)$ are not the same as those in Eq. (\ref{ch3-linear_regr}).  This may be partly due to the  strong correlations between 
 $a_0$ and $b_0$
of Eq. (\ref{ch3-eq_para}). Moreover, Eq. (\ref{ch3-linear_regr}) is fitted to the values of tidal deformability at a fixed NS mass 1.4$M_\odot$. 
To validate our  parametrized form for $P_{\ms{\rm BEM}}(\rho)$, we have calculated the
values of  $P_{\ms{\rm BEM}}(2\rho_0)$ using Eq. (\ref{ch3-eq_para}) with the help of tidal
deformability for 1.4$M_\odot$ obtained for a large number of mean-field
models, which includes the once considered  in Fig.  \ref{ch3-fig6} along
with those taken from \cite{Tsang:2019vxn, Malik:2022zol, Ferreira:2019bgy}.
The  average deviation of  $P_{\ms{\rm BEM}}(2 \rho_0)$, obtained using
Eq. (\ref{ch3-eq_para}),  from the actual values, is about  $10\%$.
We find marginal improvement when the terms corresponding to quadratic in tidal deformability are included in Eq. (\ref{ch3-eq_para}). 

%\newpage
In Fig.\ref{ch3-fig9}, we display the variations of tidal deformability as a function of mass and pressure for $\beta$-equilibrated matter at $\rho$=1.5, 2.0 and 2.5 $\rho_0$. These results are obtained using the parametrized form for  $P_{\ms{\rm BEM}}(\rho)$ as given by Eq. (\ref{ch3-eq_para}). One can easily estimate the values of $P_{\ms{\rm BEM}}(\rho)$ for $\rho \sim 2\rho_0$ once the values of tidal deformability known in NS mass ranges $1.2-2.0 M_\odot$. 

\section{Conclusion} \label{ch3-summary}
We have used Taylor and $\frac{n}{3}$ expansions of equations of state to construct marginalized posterior
distributions of the nuclear matter parameters, which are consistent with the minimal constraints. Only a few low-order nuclear matter parameters,  such as the energy per nucleon, incompressibility coefficient for the symmetric nuclear matter, and symmetry energy coefficients at the saturation density ($\rho_0$), are constrained in narrow windows along with the low-density pure neutron matter EOS obtained from a precise next-to-next-to-next-to-leading-order (N$^{3}$LO) calculation in chiral effective field theory.  The tidal deformability, radius, and maximum mass are evaluated using large sets of minimally constrained EOSs.

The correlations of neutron star properties over a wide range of mass with various key quantities characterizing the EOS are investigated. We find that the values of tidal  deformability and radius for the neutron star with 1.4$M_\odot$  are strongly correlated with  the pressure for the $\beta$-equilibrated matter at density
$\sim 2\rho_0$. The radius for $2.07M_\odot$ neutron star is strongly correlated with the pressure for $\beta$-equilibrated matter at density $\sim 3\rho_0$.   The maximum mass of a neutron star is correlated with the pressure for the $\beta$-equilibrated matter at density   $\sim 4.5\rho_0$.  These correlation systematics align with those obtained for unified EOSs for the $\beta$-equilibrated matter available for a diverse set of non-relativistic and relativistic mean-field models. We exploit the model independence of correlations to parametrize the pressure for $\beta$-equilibrated matter in the density range $1.5-2.5\rho_0$, in terms of the mass and corresponding tidal deformability of neutron star. Such parametric form may facilitate back-of-the-envelope estimation of the pressure at densities around $2\rho_0$ for a given value of tidal deformability of neutron stars with mass in the range of $1.2-2.0 M_{\odot}$.

%% file: chapter3/chapter3a.tex
%\begin{appendix}
%\begin{figure}[]
%\begin{minipage}{\textwidth}
%\centering .
%\appendix
%\chapter{Supplemental for Chapter \ref{chap3}}
%\end{minipage}
%\end{figure} 
\begin{minipage}{\textwidth}
    \centering 
   \Large\textbf{ Appendix A : Supplemental for Chapter \ref{chap3}}
\end{minipage}

We present our results in tabular form, which are obtained with minimal constraints.  The values of the nuclear matter, properties
of neutron stars, and their correlations with various key quantities
associated with EOS  are listed in Tables \ref{ch3-tab3}–\ref{ch3-tab5}. These results
are depicted in Figs. \ref{ch3-fig1} - \ref{ch3-fig6}.                                            
\setcounter{table}{2} 

\begin{table}[!ht]
    \centering
    \caption{\label{ch3-tab3} The median values and associated 68\%(90\%) uncertainties for the nuclear matter parameters from their marginalized posterior distributions. The results are obtained for  Taylor and $\frac{n}{3}$ expansions with and without pure neutron matter (PNM) constraints.}
    \begin{adjustwidth}{-0.5cm}{}
    \begin{tabular}{ccccc}
    %\toprule
    \hline\hline 
  \multirow{2}{*}{NMPs} & \multicolumn{2}{c}{ without PNM}  & \multicolumn{2}{c}{with PNM} \\ [1.5ex] 
  \cline{2-5}
    {(in MeV)}    & Taylor & $\frac{n}{3}$ & Taylor & $\frac{n}{3}$ \\[1.5ex] 
    \cline{1-5}    
    \hline\hline
    $\varepsilon_0$ & $-16.02^{+0.23(0.41)}_{-0.28(0.56)}$  &$-15.99^{+0.27(0.43)}_{-0.27(0.51)}$  &$-16.00^{+0.27(0.42)}_{-0.30(0.54)}$  & $-16.00^{+0.27(0.44)}_{-0.28(0.56)}$  \\[1.5ex] 
    $K_0$ & $236.42^{+42.78(74.34)}_{-42.58(79.62)}$  &$233.38^{+48.94(76.14)}_{-42.73(83.95)}$ & $237.43^{+44.24(72.25)}_{-45.75(83.22)}$ & $231.96^{+44.80(72.94)}_{-41.33(76.63)}$  \\ [1.5ex] 
    $Q_0$ & $-436.23^{+273.36(419.17)}_{-306.50(603.76)}$ & $-411.84^{+207.53(301.56)}_{-210.88(409.00)}$  & $-419.81^{+262.95(437.69)}_{-272.47(531.58)}$  & $-418.89^{+187.43(300.76)}_{-179.25(377.42)}$   \\ [1.5ex]  
    $Z_0$ & $1441.51^{+792.45(1298.64)}_{-696.39(1381.30)}$ & $1600.07^{+1067.33(1883.00)}_{-1362.28(2615.10)}$  & $1403.84^{+704.56(1133.85)}_{-690.82(1386.25)}$  & $1638.14^{+1241.83(1906.75)}_{-1277.48(2244.23)}$ \\ [1.5ex]  
    $J_0$ & $32.37^{+4.08(6.79)}_{-4.26(8.83)}$ & $32.37^{+4.69(7.22)}_{-4.71(10.23)}$  & $31.88^{+0.87(1.43)}_{-0.92(-1.85)}$  & $31.87^{+0.93(1.49)}_{-0.82(1.68)}$   \\ [1.5ex]  
   $L_0$ & $59.88^{+41.14(65.90)}_{-39.84(78.17)}$ & $55.60^{+37.59(63.89)}_{-43.88(84.62)}$  & $51.25^{+13.32(21.60)}_{-13.91(25.54)}$ & $52.25^{+13.55(22.73)}_{-12.76(23.04)}$  \\ [1.5ex]  
    $K_{\rm sym,0}$ & $-85.86^{+192.67(327.83)}_{-151.57(266.76)}$  & $-40.03^{+161.60(271.89)}_{-135.08(234.67)}$  &
     $-96.65^{+141.41(225.69)}_{-127.49(216.74)}$ & $-67.44^{+127.18(206.09)}_{-114.80(200.38)}$   \\ [1.5ex] 
    $Q_{\rm sym,0}$ & $731.13^{+308.54(543.01)}_{-347.82(669.47)}$  & $705.36^{+311.23(511.39)}_{-352.72(727.86)}$ & $699.56^{+324.38(521.95)}_{-323.52(639.30)}$ & $726.49^{+300.40(510.33)}_{-358.51(631.86)}$  \\ [1.5ex] 
    $Z_{\rm sym,0}$ & $-2.07^{+1190.67(2153.84)}_{-820.92(1473.09)}$ &
    $-1390.39^{+1518.69(2526.53)}_{-1856.18(3623.74)}$  & $55.34^{+1205.62(2255.28)}_{-782.52(1415.84)}$  & $-1622.35^{+1606.61(2788.70)}_{-1911.81(3468.40)}$    \\ [1.5ex] 
   
    \hline\hline 
    \end{tabular}
    \end{adjustwidth}
\end{table}

\begin{table}[!ht]
    \centering
    \caption{\label{ch3-tab4} Similar to Table \ref{ch3-tab3}, but, for the neutron star properties, namely  the tidal deformability ($\Lambda_{1.4}$), radii ($R_{1.4}$ and $R_{2.07}$)  and maximum mass ($M_{\rm max}$) .}
   % \begin{adjustwidth}{-0.5cm}{}
   % \small\addtolength{\tabcolsep}{-1pt}
    \begin{tabular}{ccccc}
    \hline\hline
  \multirow{2}{*}{{NS properties}} & \multicolumn{2}{c}{ without PNM}  & \multicolumn{2}{c}{with PNM} \\ 
  \cline{2-5}
        & Taylor & $\frac{n}{3}$ & Taylor & $\frac{n}{3}$ \\[1.5ex] 
    \cline{1-5}   
   $\Lambda_{1.4}$ & $527.72^{+250.72(477.68)}_{-186.11(292.57)}$  & $455.85^{+223.65(465.72)}_{-163.05(243.23)}$ & $426.20^{+139.93(224.58)}_{-130.32(205.18)}$ & $386.52^{+132.76(213.24)}_{-102.84(199.09)}$  \\[1.5ex]
   $R_{1.4}${(km)} & $14.69^{+1.78(3.43)}_{-1.63(2.74)}$  & $14.15^{+1.87(3.34)}_{-1.69(2.58)}$  & $13.37^{+0.67(1.03)}_{-0.75(1.60)}$  & $13.22^{+0.64(0.99)}_{-0.67(1.59)}$   \\[1.5ex]
   $R_{2.07}${(km)} & $13.24^{+0.82(1.49)}_{-0.82(1.42)}$  & $12.27^{+0.88(1.52)}_{-0.80(1.52)}$  & $12.72^{+0.55(0.85)}_{-0.59(1.07)}$  & $12.02^{+0.54(0.88)}_{-0.58(1.23)}$  \\[1.5ex]
  $ M_{\rm max}${($M_\odot$)} & $2.45^{+0.07(0.11)}_{-0.06(0.13)}$  & $2.19^{+0.10(0.19)}_{-0.09(0.09)}$ & $2.48^{+0.06(0.10)}_{-0.07(0.14)}$  & $2.20^{+0.10(0.16)}_{-0.09(0.11)}$   \\[1.5ex]
\hline\hline 
    \end{tabular}
   % \end{adjustwidth}
\end{table}

\begin{table}[ht!]
    \centering
    \caption{\label{ch3-tab5} The comparison of values for Pearson's correlation coefficient (r) obtained from randomly selected 100 and 1000
    EOSs using both Taylor and $\frac{n}{3}$ expansions. The values of correlation coefficients are also obtained by combining  1000
    EOSs from each of the expansions.
    For comparison,  the values of $r$ obtained for a diverse set of mean-field
models are also presented in 2nd column.    }
    \begin{tabular}{ccccccc}
        \hline\hline
        \multirow{2}{*}{Name of Pairs}& MFMs & \multicolumn{2}{c}{Taylor} & \multicolumn{2}{c}{$\frac{n}{3}$} & combined\\
         & 41 & 100 & 1000 & 100 & 1000  &  2000\\
         \hline
        $\Lambda_{1.4}$-$P_{\ms{\rm BEM}}(2\rho_0)$ & 0.90 &0.98 & 0.98 & 0.99 & 0.98 & 0.98 \\[1.5ex]
        $R_{1.4}$-$P_{\ms{\rm BEM}}(2\rho_0)$ & 0.83 &0.93 & 0.93 & 0.94 & 0.93 & 0.93\\[1.5ex]
        $R_{2.07}$-$P_{\ms{\rm BEM}}(3\rho_0)$ & 0.81 &0.93 & 0.91 & 0.92 & 0.92 & 0.93\\[1.5ex]
        $M_{\rm max}$-$P_{\ms{\rm BEM}}(4.5\rho_0)$ & 0.99 &0.97 & 0.98 & 0.95 & 0.96 & 0.99\\[1.5ex]
        \hline\hline
    \end{tabular}
\end{table}

%\end{appendix}

%% file: chapter4/chapter4.tex
\chapter{Systematic Analysis of impacts of Symmetry energy parameters}\label{chap4}

In this chapter, we have considered a large set of minimally constrained EOSs for the NS matter to examine in detail the correlations between the properties of a neutron star in the mass range $1.2 - 1.6 M_{\odot}$ and the parameters that govern the density dependence of symmetry energy \cite{Patra:2023jvv}. The EOSs at low densities correspond to the nucleonic matter in $\beta$-equilibrium and are described by the nuclear matter parameters evaluated at $\rho_0$ (see Chapter \ref{Taylor}). These EOSs are constrained by empirical values of the low-order nuclear matter parameters determined by the experimental data on the bulk properties of finite nuclei together with the pure neutron matter (PNM) EOS from a precise next-to-next-to-next-to-leading-order (N$^{3}$LO) calculation in chiral effective field theory. The composition of NS matter at high density ($\rho>2\rho_0$) is not very well known due to the possibility of the appearance of various new degrees of freedom such as hyperons, kaons, and quarks \cite{Chatziioannou:2015uea, Chatterjee:2015pua, Stone:2019blq}. Beyond the density ($\rho_{c_s}$), taken to be $1.5 - 2\rho_0$, the EOSs are constructed simply by imposing the causality condition on the speed of sound and are independent of compositions of NS matter (see Chapter \ref{H-eos}). The posterior distributions of nuclear matter parameters that describe the low-density EOSs are obtained within a Bayesian approach with minimal constraints. These constraints introduce correlations among the nuclear matter parameters. The nuclear matter parameters' joint posterior or correlated distribution is employed to study the sensitivity of NS
properties to the parameters that govern the density dependence of the symmetry energy. The calculations are also performed for uncorrelated uniform and Gaussian distributions of the nuclear matter parameters obtained by their marginalized posterior distribution. The influence of the various correlations considered due to a few other factors usually encountered in the literature are investigated.
These factors are as follows, 
\begin{itemize}
\item [(i)] the behavior of the high-density part of the EOS, 
\item [(ii)] the choice of distributions of nuclear matter parameters, their interdependence, and  uncertainties,
\item [(iv)] the value of the density ($\rho_{c_s}$),
\item [(v)] upper bound on the value of tidal deformability.
\end{itemize}

\section{Priors and Posterior distributions of NMPs}\label{ch4-PD}

The posterior distributions of the NMPs are obtained by subjecting the EOSs to a set of minimal constraints, which includes the basic nuclear matter properties at the saturation density and the EOS for the pure neutron matter at low densities from  N$^3$LO calculation in the chiral effective field theory \cite{Hebeler:2013nza, Lattimer:2021emm}. Only a few low-order NMPs are well constrained, such as the binding energy per nucleon in symmetric nuclear matter $e_0 \sim -16.0 $ MeV and symmetry energy coefficient $J_0=32.5 \pm 2.5 $ MeV from the binding energy of finite nuclei over a wide range of nuclear masses~\cite{Chabanat98, Malik:2019whk, Mondal:2015tfa, Mondal:2016roo, Sulaksono:2009rn, Essick:2021ezp}. The nuclear matter incompressibility
coefficient, $K_0=240 \pm 50$ MeV for the symmetric nuclear matter is constrained from the experimental data on the centroid energy of isoscalar giant monopole resonance in a few heavy nuclei~\cite{Garg:2018uam, Agrawal:2005ix}. 
{The value of symmetry energy slope parameter has been deduced by the neutron-skin thickness of $^{48}$Ca nucleus by CREX \cite{CREX:2022kgg} and that in $^{208}$Pb by  PREX-2 collaboration \cite{PREX:2021umo}. The PREX-2 data analysis performed by Reed et al. in Ref. \cite{Reed:2021nqk, Reed:2019ezm} places $L_0=106 \pm 37$ MeV. Other studies combining astronomical observations and PREX-2 data is $L_0=53^{+14}_{-15}$ MeV \cite{Essick:2021kjb}. A smaller value of  $L_0=54 \pm 8$ MeV has also been inferred from PREX-2 data \cite{Reinhard:2021utv}. The CREX data predicts $L_0 = 0 - 51$ MeV \cite{TAGAMI2022}.}
The  remaining nuclear matter parameters, $Q_0$, $K_{\rm sym,0}$ and $Q_{\rm sym,0}$ appearing in Eqs.( \ref{eq:SNM_T}) \& ( \ref{eq:sym_T}) are only weakly constrained ~\cite{Tsang:2020lmb, Ferreira:2021pni, Dutra:2012mb, Dutra:2014qga, Mondal:2017hnh, Patra:2022yqc}. The prior for the binding energy per nucleon is kept fixed to $e_0=-16.0$ MeV throughout. The prior distributions of $J_0$ and $K_0$ are assumed to be uniform with a rather small range, whereas the other higher-order nuclear matter parameters
correspond to uniform distributions with large ranges. We have listed the assumed prior distributions for each nuclear matter parameter in Table \ref{ch4-tab1}. The values of $d$s and $\sigma$s in Eq.(\ref{BA-eq:likelihood}) are taken from Ref. \cite{Lattimer:2021emm, Hebeler:2013nza} for the energy per neutron, and we consider a $6\times$N$^3$LO  uncertainty band for our calculations. In addition to likelihood and priors, we have imposed a few filters on the nuclear matter parameters: (i) pressure for the $\beta$-equilibrated matter should increase monotonically with density (thermodynamic stability),(ii) symmetry energy is positive semi-definite and (iii) maximum mass of neutron star must exceed $2M_\odot$ (observational constraint).

\begin{table}[!ht]
\caption{\label{ch4-tab1}Uniform prior distributions are assumed for all the NMPs except for $e_0$, which is kept fixed to -16.0 MeV. The minimum (min.) and maximum (max.) values of the NMPs are listed in the units of MeV.} 
  \centering
  \begin{tabular}{ccccccc}
  \hline \hline
 &{$K_0$} & {$Q_0$} & {$J_0$} & {$L_0$} &{$K_{\rm sym,0}$} & {$Q_{\rm sym,0}$} \\ [1.3ex]
 \hline
 min. & 190 & -1200 & 30 & 0 & -500 & -250 \\[1.3ex] 
 max. & 290 & 400 & 35 & 100 & 300 & 1350 \\[1.3ex] 
 \hline \hline
  \end{tabular}
\end{table}

\begin{figure}[!ht]
\centering
\includegraphics[width=0.6\textwidth]{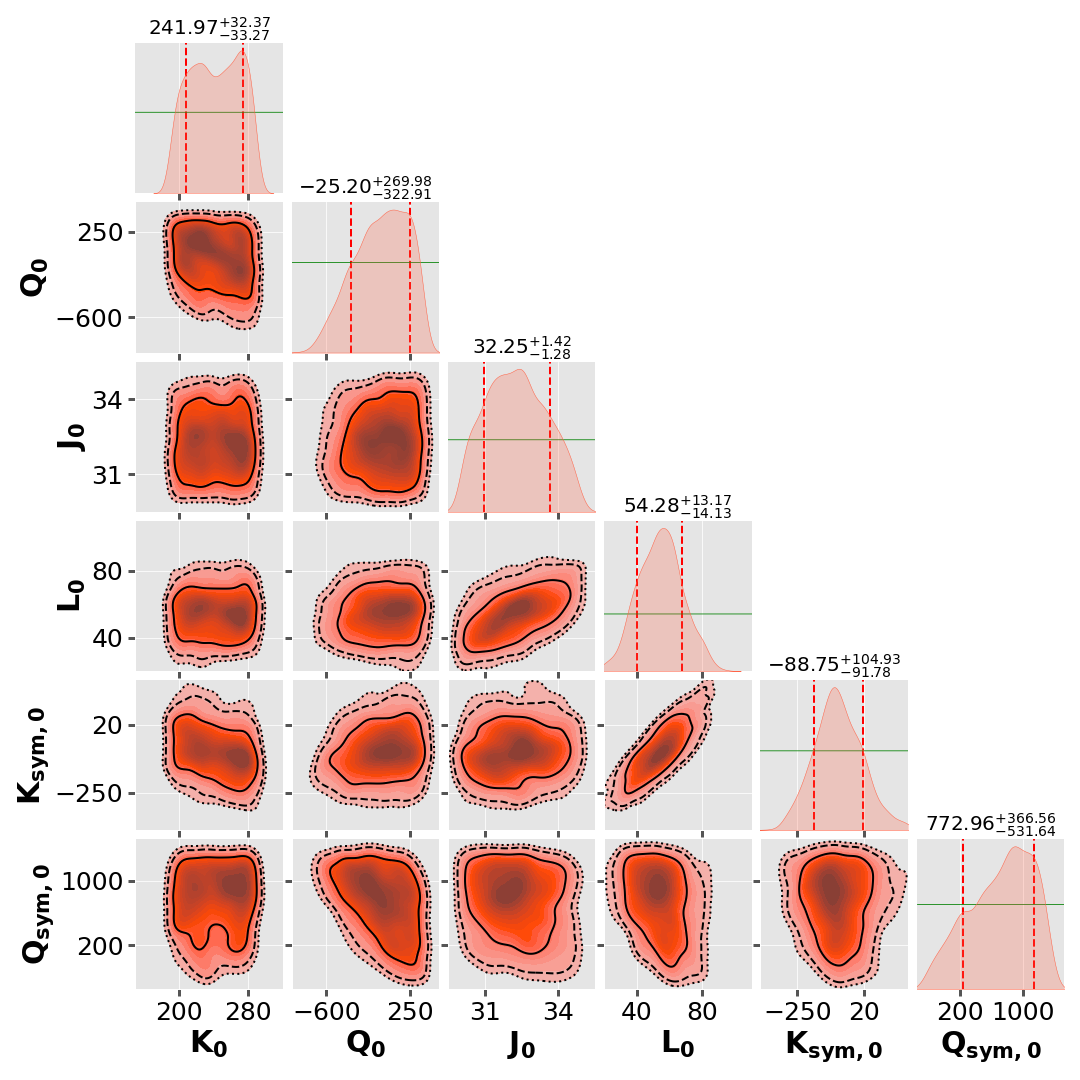}
\caption{Corner plot for the nuclear matter parameters (in MeV). The one-dimensional marginalized posterior distributions (salmon) and the prior distributions (green) lines are displayed along the diagonal plots. The vertical dashed lines indicate 68$\%$ (1$\sigma$) confidence interval. Along the off-diagonal plots, the confidence ellipses for two-dimensional posterior distributions are plotted with confidence intervals of 1$\sigma$, 2$\sigma$, and 3$\sigma$. }\label{ch4-fig1}
\end{figure}
 
The joint posterior distribution of the NMPs for a given model is the product of the likelihood and the prior distribution of NMPs (Eq. (\ref{eq:BT-2})). The posterior distribution of individual parameters can be obtained by marginalizing the joint posterior distribution with the remaining model parameters. If the marginalized posterior distribution of the parameter is narrowed down as compared to the corresponding prior distribution (uniform distribution in this case), then the parameter is said to be constrained by the given data or the likelihood functions. Therefore, the narrower distributions of parameters compared to their prior distributions indicate the importance of the likelihood function. The likelihood function imposes additional constraints on the multi-variate nuclear matter parameters of our model driven by the data. The corner plots for the nuclear matter parameters, which yield the EOSs consistent with the minimal constraints, are shown in Fig.~\ref{ch4-fig1}. The median values of the nuclear matter parameters and the $68\%$ confidence intervals are given in the diagonal plots of the figure. The 68\% confidence intervals for $L_0$, $Q_0$, and $K_{\rm sym,0}$ are significantly smaller than their prior ranges, implying these parameters are well constrained by the low-density EOS for the pure neutron matter. The values of $J_0$,  and $Q_{\rm sym,0}$are also somewhat constrained.  Except for $L_0-K_{\rm sym,0}$ (r = 0.8) and $L_0-J_0$ (r = 0.65), all other pairs of nuclear matter parameters do not show any visible correlations.

\section{Neutron Star properties }\label{ch4-ns}

The EOSs for $\beta$-equilibrated charge neutral matter in the density ranges $0.5\rho_0$ to the density $\rho_{c_s}$  are obtained using Taylor expansion with NMPs corresponding to the posterior distribution as displayed in the Fig.~\ref{ch4-fig1}. The calculations are performed assuming different values for the $\rho_{c_s}$ in the range of $1.5 - 2\rho_0$. Each of the EOSs beyond $\rho_{c_s}$ is smoothly joined by a diverse set of EOSs, which are obtained simply by imposing the causality condition on the speed of sound by following the Eqs.(\ref{eq-vs}-\ref{eq-preE}) in chapter \ref{H-eos}.
The EOS for the density ranges $\rho<0.5\rho_0$ comprises outer and inner crusts. We have. for used the EOS for the outer crust by Baym-Pethick-Sutherland \cite{Baym:1971pw} in the density range $3.9 \times 10^{-11}\rho_0<\rho<0.0016\rho_0$. We have assumed a polytropic form of the EOS for the inner crust  as follows \cite{Carriere:2002bx}, 
 \be
 p(\varepsilon)= \alpha + \beta \varepsilon^{\frac{4}{3}}. \label{eq-ic}
 \ee
Here, the parameters $\alpha$ and $\beta$ are determined so that the EOS for the inner crust matches with the outer crust at one end and the outer core at the other. There is a greater sensitivity to the treatment of crust EOS for neutron stars with mass $\sim 1 M_\odot$~\cite{Fortin:2016hny}. The treatment of crust EOS employed in the present work may introduce the uncertainties of about 50-100 meters in radii of NSs having a mass 1.4$M_\odot$. In Ref. \cite{Piekarewicz2019}, it is shown that the choice of EOS for the inner crust affects both the Love number $k_{2}$ and compactness parameter in such a way that the values of the tidal deformability parameter remain practically unaltered. Once the EOSs for the core and crust are determined, the values of neutron star mass, radius, and tidal deformability corresponding to a given central pressure can be obtained by solving Tolman-Oppenheimer-Volkoff equations~\cite{Oppenheimer:1939ne, Tolman:1939jz}.

\begin{figure}[!ht]
\centering
\includegraphics[width=0.5\textwidth]{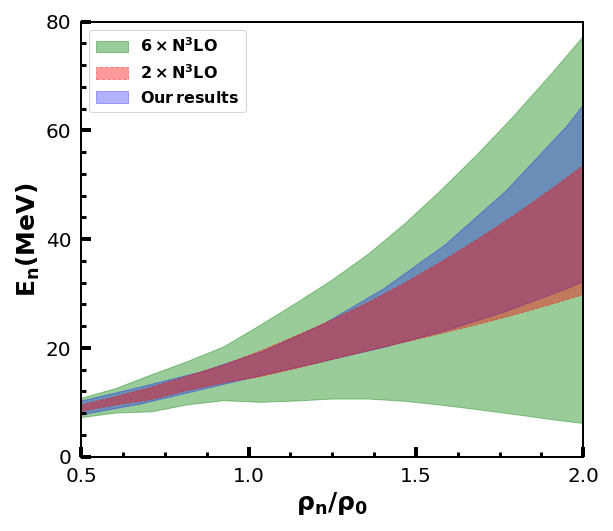}
\caption{ The energy per neutron ($E_n(\rho) = E(\rho,1)$) for pure neutron matter as a function of neutron density. The colored bands correspond to 6 $\times$ N$^3$LO (light green), 2 $\times$ N$^3$LO (light red), and 90\% confidence interval for the EOSs (light blue) obtained in our calculation (see text for details).}\label{ch4-fig2}
\end{figure}

\begin{figure}[!ht]
\centering
\includegraphics[width=0.5\textwidth]{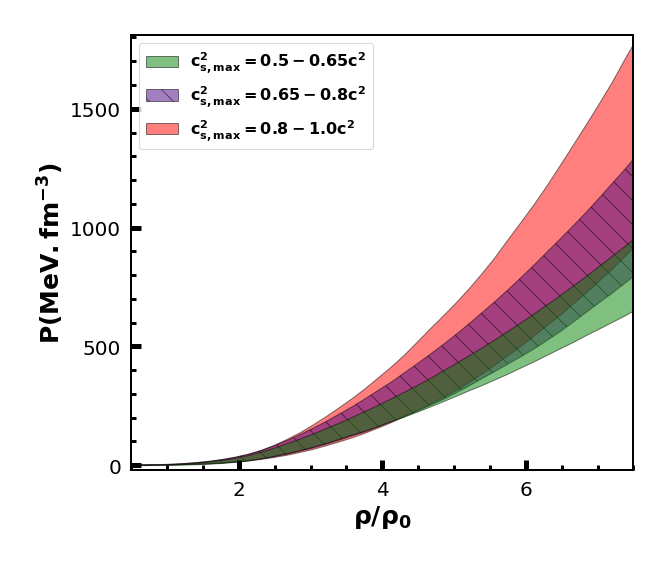}
\caption{ The pressure for $\beta$-equilibrated charge neutral matter as a function of nucleon density. The colored bands correspond to 90\% confidence intervals for the EOSs with different ranges of the square of the speed of sound at the center of NS with maximum mass ( $c_{s,\rm max}^2$)(see text for details).}\label{ch4-fig3}
\end{figure}

{The 90\% confidence interval for pure neutron matter EOSs (light blue) for the density range ( $0.5-2.0\rho_0$) is shown in Fig.~\ref{ch4-fig2}. For comparison, colored bands correspond to 6 $\times$ N$^3$LO (light green), 2 $\times$ N$^3$LO (light red) from chiral effective field theory are displayed. Our EOSs lie almost in the middle of the 6 $\times$ N$^3$LO band and significantly satisfy the 2 $\times$ N$^3$LO band. This indicates that our EoSs are well fitted with the pure neutron matter EOS from a precise next-to-next-to-leading-order calculation in chiral effective field theory.
 The 90$\%$ confidence interval for the pressure of the $\beta$-equilibrated matter is plotted as a function of density in Fig.~\ref{ch4-fig3}. The results are divided into three groups depending upon the square of sound speed ($c_{s,\rm max}^2$) at the centre of NS of its maximum stable mass configuration. The three groups of EOSs correspond to $c_{s,\rm max}^2=0.5-0.65c^2$, $0.65-0.8c^2$, and $0.8-1.0c^2$ are depicted by different colors. All the EOSs in each group are plotted upto the same density 7$\rho_0$. The overall stiffness of the EOS increases with the $c_{s,\rm max}^2$.}  
 
\begin{figure}[htp]
\centering
\includegraphics[width=\textwidth]{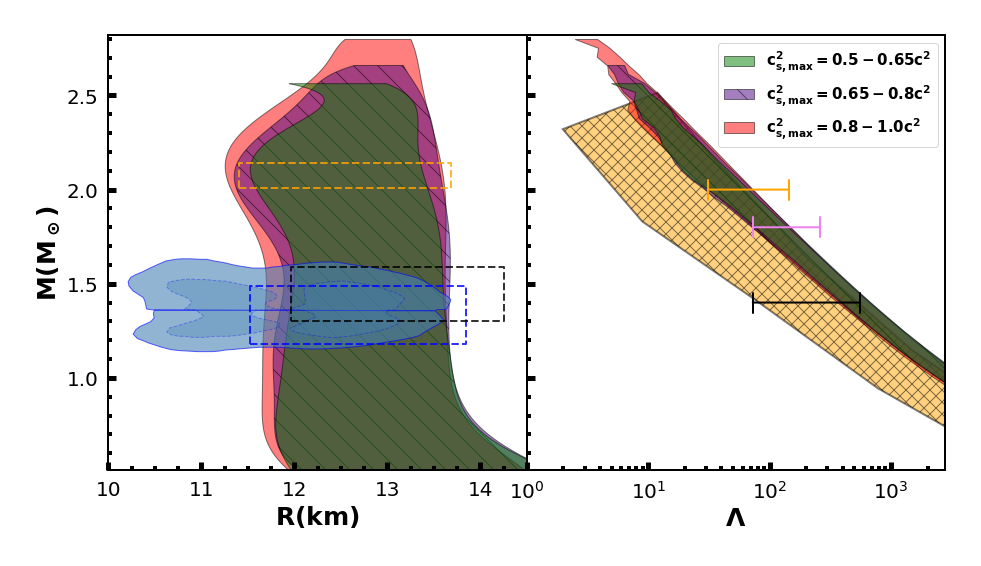}
\caption{ The mass-radius relationship  (left panel)  obtained for the EOSs as shown in Fig.~\ref{ch4-fig3}. 
The 90\% (solid) and 50\% (dashed) confidence intervals of the LIGO-Virgo analysis for the BNS component from the GW170817 event are shown by the outer and inner gray shaded regions ~\cite{Abbott2018, Abbott18a, Abbott2021}. The constraints from the millisecond pulsar PSR J0030+0451 (blue \& black) NICER x-ray data ~\cite{Riley:2019yda, Miller:2019cac} and PSR J0740+6620 (green) ~\cite{Riley:2021pdl} are shown in the rectangular regions enclosed by dotted lines. The right panel displays tidal deformability versus NS mass.   The orange shaded region is the observation for $\Lambda$ with 90\% posterior interval from the LIGO/Virgo Collaboration (GW170817 event) \cite{Abbott2018}. The black line correspond to observational bounds on $\Lambda_{1.4} = 190^{+390}_ {-120}$ ~\cite{Abbott18a}. For the comparison, we have shown violet and gold lines corresponding to $\Lambda_{1.8} = 70-270$, and $\Lambda_{2.0} = 30-150$ obtained from a few well-known theoretical models ~\cite{Piekarewicz2019, Xie2022}}.
\label{ch4-fig4}
\end{figure}

The EOSs displayed in Fig.~\ref{ch4-fig3} are employed to obtain the mass-radius relationship for static neutron stars as presented in Fig.~\ref{ch4-fig4} (see left panel). For the comparison, we also display the constraints obtained from the GW170817 event and NICER x-ray observation. The maximum mass of a neutron star lies in the range of $2.1-2.7M_\odot$, and the radius for a neutron star with mass 1.4$M_\odot$ lies in the range 11.8-14 km. Our mass-radius relationships exclude smaller values of radius for a
given mass, as predicted by the  GW170817 event. This is due to the choice
of priors for the low-order nuclear matter parameters constrained by
the experimental data on bulk properties of the finite nuclei. In the right panel of Fig.~\ref{ch4-fig4}, we plot the variations of tidal deformability as a function of mass. We display the constraints obtained from the GW170817 event for comparison. The value of tidal deformability $\Lambda_{1.4} $ is highlighted~\cite{Abbott18a}.
Further, we depict the constraints on $\Lambda$ for NS of mass  1.8$M_\odot$, and 2.0$M_\odot$  within 90\% CI obtained from ten realistic models that can accurately describe the finite nuclei properties and support the $2M_\odot$ neutron star masses~\cite{Piekarewicz2019, Xie2022}. The values of tidal deformability obtained with our minimally constrained EOSs have a reasonable overlap with the ones inferred from the GW170817 event.

\begin{figure}[!ht]
\includegraphics[height = 20 cm, width=16 cm]{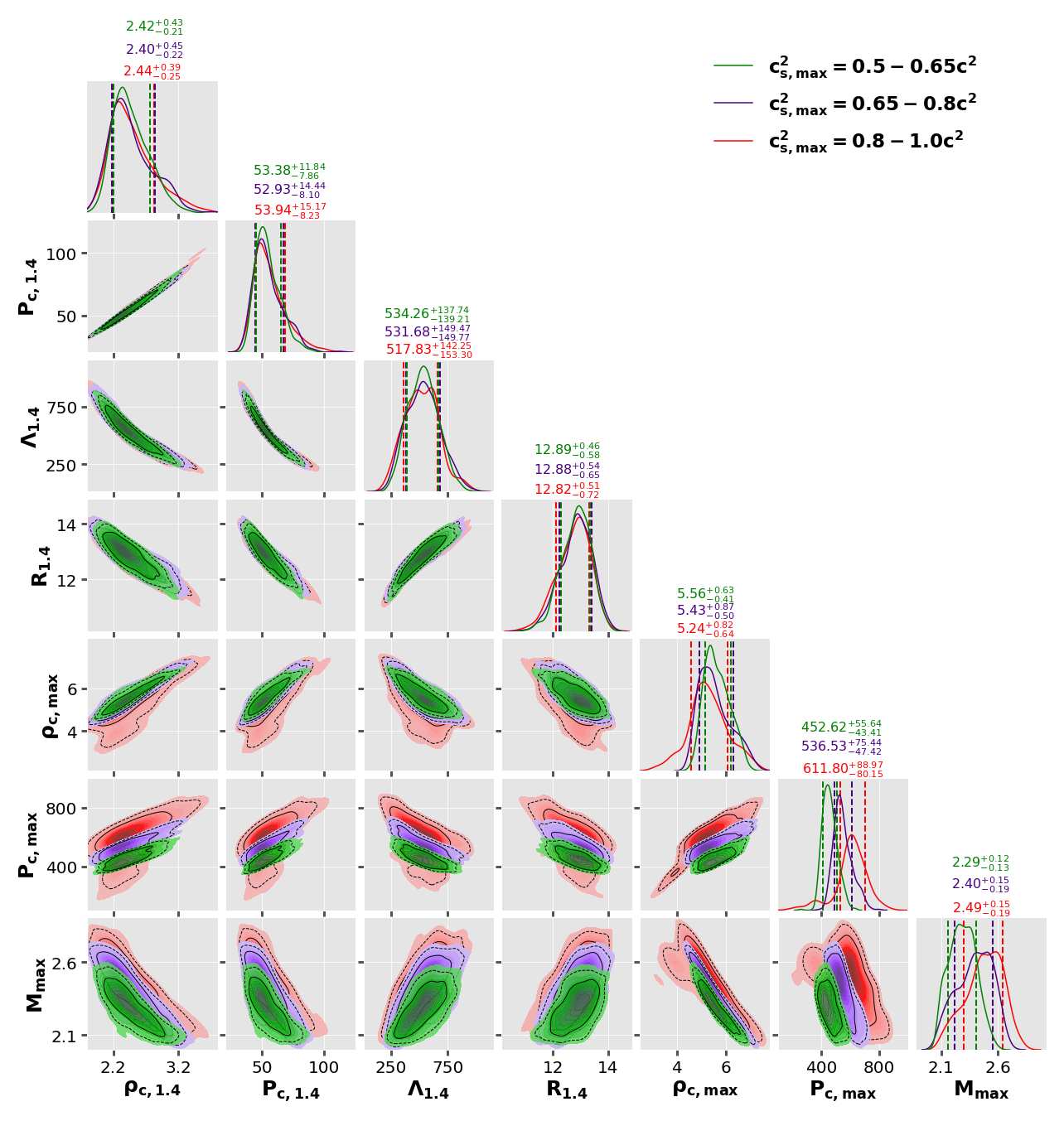}
\caption{Corner plot for the central density ($\rho_c$ in $\rho_0$) and corresponding pressure ($P_c$ in MeV.fm$^{-3}$) for the neutron star with canonical and maximum mass, radius ($R_{1.4}$ in km),  tidal deformability ($\Lambda_{1.4}$) and maximum mass ($M_{\rm max}$ in $M_\odot$) of NS. The confidence ellipses for two-dimensional posterior distributions are plotted with $1\sigma$(solid line) and $2\sigma$ (dashed line) confidence intervals along the off-diagonal plots. The vertical dashed lines indicate 68\% confidence intervals.} \label{ch4-fig5}
\end{figure}

In Fig.~\ref{ch4-fig5}, we display the corner plot describing various quantities associated with neutron stars, such as the central density $\rho_c$ and corresponding pressure $P_c$ for the neutron star with canonical ($1.4M_\odot$) and maximum mass, radius $R_{1.4}$,  tidal deformability $\Lambda_{1.4}$  and maximum mass  $M_{\rm max}$. To understand the impact of the high-density EOS, the distributions of all the quantities are segregated into three different groups of stiffness for the EOSs according to the range of $c_{s,\rm max}^2$ at the higher density part as indicated with different colors. It can be seen from the diagonal plots that the distributions for $\rho_{c,1.4}$, $P_{c,1.4}$, $\Lambda_{1.4}$ and $R_{1.4}$ are more-or-less independent of $c_{s,\rm max}^2$. The median values of $\rho_{c,\rm max}$ decrease with $c_{s,\rm max}^2$, but,  $M_{\rm max}$ and  $P_{c,\rm max}$ increase.
The $\rho_{c,1.4}$, $P_{c,1.4}$, $R_{1.4}$ and  $\Lambda_{1.4}$ are strongly correlated with each other, the absolute values of Pearson's correlation coefficients being r$\sim 0.86-0.98$. 
The $\rho_{c,\rm max}$ and $P_{c,\rm max}$ are moderately correlated with the properties of NS with canonical mass ($\mid r \mid \sim$ 0.70-0.85), except for $\rho_{c,1.4}$. 
The  $M_{\rm max}$ show strong correlations with $\rho_{c,\rm max}$ ($\mid r \mid \sim 0.94$), but,  relatively weakly correlated with $P_{c,\rm max} (\mid r \mid\sim 0.78)$. The maximum mass of a stable neutron star seems to be weakly correlated with the properties of NS with canonical mass.

\section{ Neutron Star properties  and symmetry energy parameters}\label{ch4-cor}

The correlations of NS properties with various symmetry energy parameters and the pressure of $\beta-$equilibrated matter have been extensively investigated earlier using non-parametric~\cite{Essick:2021kjb}, parametric~\cite{Tsang:2020lmb, Biswas:2021yge}, and physics-based
models~\cite{Malik:2020vwo, Reed:2021nqk, Pradhan:2022vdf, Carson:2018xri, Essick:2021kjb} and found to yield the results which are sometimes at variance, as summarized in  Table III of Ref.~\cite{Kunjipurayil:2022zah}. Some of these studies include constraints imposed by bulk nuclear
properties, while others have also imposed those through the constraints on nuclear matter parameters, assuming them to be independent of each other. We now study the correlations of various NS properties with the $L_0$, $K_{\rm sym,0}$ and $P(2\rho_0)$ using our minimally constrained EOSs
and assess how they are affected by the several factors as listed at
the beginning of this section.

\begin{figure}[!ht]
\centering
\includegraphics[width=\textwidth,height=8cm]{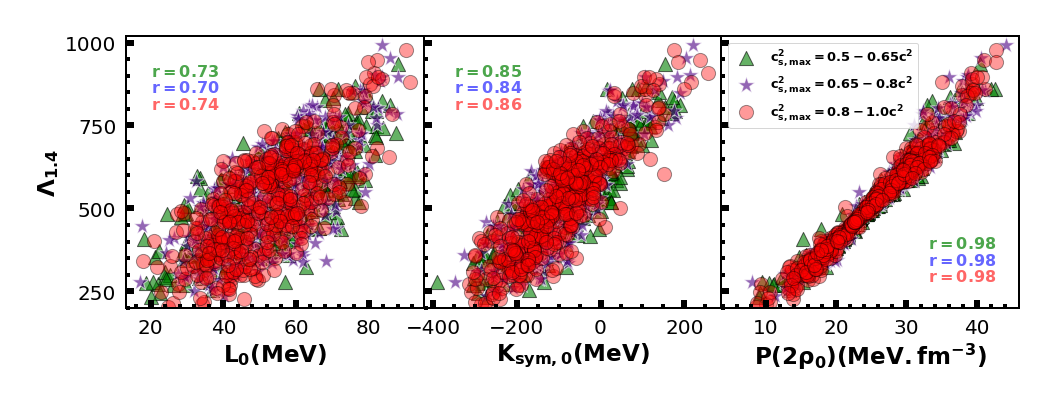}
\caption{ The tidal deformability ($\Lambda_{1.4}$) as a function of slope parameter ($L_0$) , curvature parameter ($K_{\rm sym,0}$) and the pressure of $\beta$-equilibrated matter (P(2$\rho_0$)).}\label{ch4-fig6}
\end{figure}

\begin{figure}[htp]
\centering
\includegraphics[width=\textwidth,height=6cm]{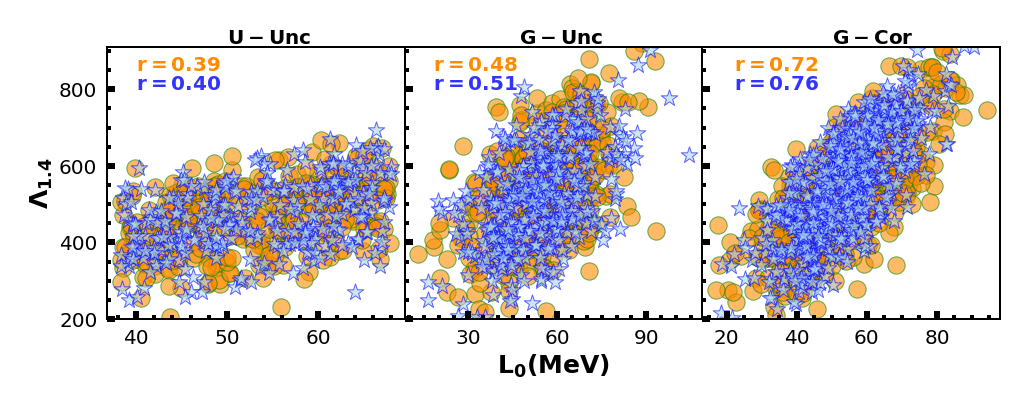}
\caption{ Variations of tidal deformability ($\Lambda_{1.4}$) with slope parameter ($L_0$) for three different distributions of the nuclear matter parameters which are with uniform uncorrelated (U-Unc), Gaussian uncorrelated (G-Unc) and Gaussian correlated posterior distributions (G-Cor) as discussed in the text in details. The results are obtained by considering the EOSs associated with maximum NS mass $M_{\rm max}$ $\geqslant$ 2.1$M_\odot$. The circle (orange) symbols  represent the results obtained by varying all parameters, whereas the (blue) star symbols represent those obtained by fixing $K_0=240$MeV and $J_0=32$ MeV.}\label{ch4-fig7}
\end{figure}

We have seen in the previous subsections that our EOSs yield the various properties of neutron stars within reasonable observational as well as theoretical bounds. As mentioned earlier, the EOSs for $\rho\leqslant\rho_{c_s}$ are obtained with the nuclear matter parameters, which are interdependent due to the minimal constraints. We employ these EOSs to study the variations of radius and tidal deformability with the slope ($L_0$), curvature ($K_{\rm sym,0}$) and $P(2\rho_0)$  for  NS with mass $1.2 - 1.6M_\odot$. The parameters $L_0$ and $K_{\rm sym,0}$ determine the density dependence of symmetry energy. We first consider in detail our results for the tidal deformability corresponding to the NS with canonical mass assuming $\rho_{c_s} = 2\rho_0$. We show the dependence of tidal deformability $\Lambda_{1.4}$ on the $L_0$, $K_{\rm sym,0}$ and $P(2\rho_0)$ in Fig.~\ref{ch4-fig6}. The results for all three groups of EOSs corresponding to different values of $c_{s,\rm max}^2$ overlap with each other, indicating that the values of $\Lambda_{1.4}$ seem to be more-or-less insensitive to the behavior of  EOS at high densities $(\rho > \rho_{c_s}$). The $\Lambda_{1.4}$ tend to increase with $L_0$, $K_{\rm sym,0}$ and $P(2\rho_0)$.  The values of correlation coefficients, as indicated in Fig.~\ref{ch4-fig6}, are practically independent of the choice of $c_{s,\rm max}^2$. Hereafter, we show the results obtained by combining all three groups of the EOSs.

To study the impact of the interdependence of nuclear matter parameters on the results shown in Fig.~\ref{ch4-fig6}, we have generated two different distributions of nuclear matter parameters with the help of their posterior distributions, as shown in Fig.~\ref{ch4-fig1}. These distributions of the nuclear matter parameters are (a) uncorrelated Uniform (U-Unc) and (b) uncorrelated Gaussian (G-Unc). The parameters of U-Unc and G-Unc distributions are obtained from the 95\% confidence interval of the marginalized distributions of Fig.~\ref{ch4-fig1}. Often, U-Unc and G-Unc distributions have been employed to study the correlations of $\Lambda_{1.4}$ with various symmetry energy parameters \cite{Malik:2020vwo, Pradhan:2022vdf, Carson:2018xri, Tsang:2020lmb}.

In Fig.~\ref{ch4-fig7}, we plot the variations of dimensionless tidal deformability $\Lambda_{1.4}$ with $L_0$ for the three different distributions of nuclear matter parameters as indicated (orange circles). The extreme right panels are labelled as G-Cor, corresponding to the ones obtained using correlated or joint posterior distribution of nuclear matter parameters. The results shown are obtained by considering the EOSs associated with the maximum NS mass $M_{\rm max}$ $\geqslant$ 2.1$M_\odot$. For the comparison,  the results obtained by fixing the lower order parameters $K_0=240$MeV and $J_0=32$MeV are also displayed (blue stars). In the figure, Pearson's correlation coefficients are given for all three distributions of NMPs.
The correlations of $\Lambda_{1.4}$ are very sensitive to the choice of the distributions of nuclear matter parameters. The  $\Lambda_{1.4}$ is very weakly correlated with $L_0$ for the case of U-Unc with a correlation coefficient $r\sim 0.39$.. for
The situation somewhat improves for the case of nuclear matter parameters corresponding to uncorrelated  Gaussian distribution as indicated by G-Unc ($r\sim 0.48$). The posterior distribution (G-Cor) of nuclear matter parameters obtained from minimal constraints (See Fig.~\ref{ch4-fig1}) yields relatively stronger correlations of $\Lambda_{1.4}$ with $L_0$ ($r\sim 0.72$).  The correlations also increase marginally when the values of low-order nuclear matter parameters such as $K_0$ and $J_0$ are kept fixed (blue stars). Evidently, the correlations of $\Lambda_{1.4}$ with $L_0$  depend on the
various factors, such as the constraints  imposed on the nuclear matter
parameters that govern the low-density behavior of the EOSs.  

\begin{figure}[!ht]
    \centering
    \includegraphics[width=\textwidth]{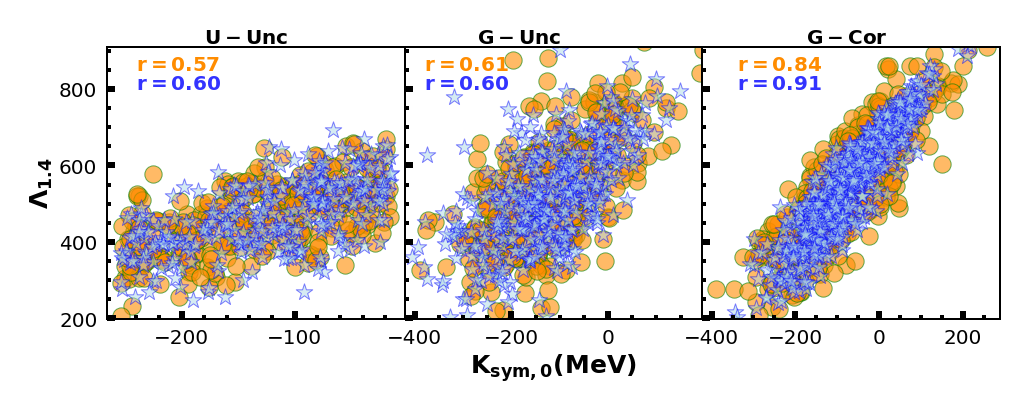}
    \caption{The variation of the tidal deformability ($\Lambda_{1.4}$) with curvature parameter ($K_{\rm sym,0}$) for three different nuclear matter parameter distributions with uniform uncorrelated (U-Unc), Gaussian uncorrelated (G-Unc) and posterior distributions (G-Cor) are discussed in details in the text. The results are shown for those EOSs that are associated with a maximum mass of NS $\geqslant$ 2.1$M_\odot$. The symbols in circles (orange) represent the results obtained by varying all parameters, whereas stars (blue) symbols represent the results obtained when $K_0=240$MeV and $J_0=32$MeV are fixed.}\label{ch4-fig8}
\end{figure}

\begin{figure}[htp]
    \centering
    \includegraphics[width=\textwidth]{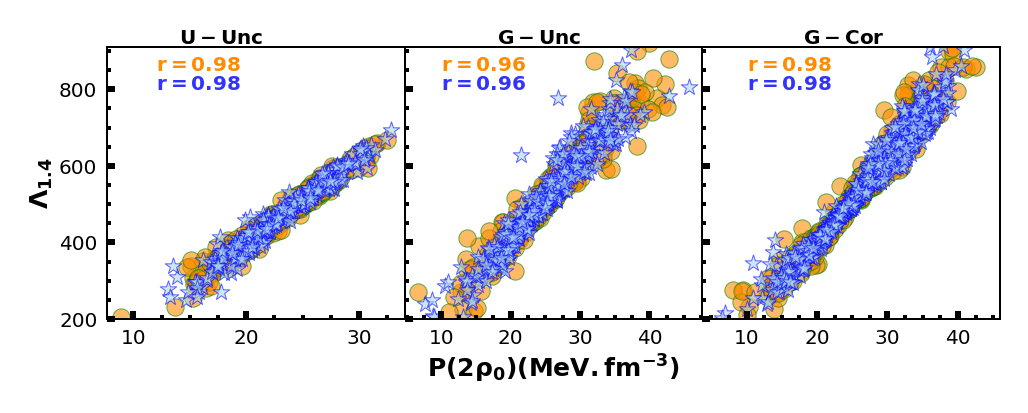}
    \caption{A variation in tidal deformability ($\Lambda_{1.4}$) with the pressure of $\beta$-equilibrated matter at 2$\rho_0$ based on three different nuclear matter parameter distributions, namely uniform uncorrelated (U-Unc), Gaussian uncorrelated (G-Unc) and posterior distributions (G-Cor), are discussed in detail in the text. The results are shown for those EOSs that are associated with a maximum mass of NS $\geqslant$  2.1$M_\odot$. As shown by the circles (orange) symbols, results obtained with all parameters varied, while those obtained with $K_0 = 240$MeV and $J_0 = 32$MeV fixed are represented by the stars (blue) symbols.}\label{ch4-fig9}
\end{figure}

The variations of $\Lambda_{1.4}$ with $K_{\rm sym,0}$ for different distributions of nuclear matter parameters are plotted in Fig.~\ref{ch4-fig8}. The correlations for $\Lambda_{1.4}$ with $K_{\rm sym,0}$ for different cases are stronger in comparison to those obtained with $L_0$. It appears that the correlations of $\Lambda_{1.4}$ with various symmetry energy parameters are quite sensitive to the distributions of nuclear matter parameters employed.  Our results for the correlations for the U-Unc case are qualitatively similar to those obtained in Refs. \cite{Pradhan:2022vdf, Carson:2018xri} with a similar strategy for nuclear matter parameters but with different models. Similar qualitative trends are observed for the G-Unc \cite{Tsang:2020lmb}. The correlations significantly improve even with the inclusion of minimal constraints, as indicated by G-Cor.  Our results for the case of G-Cor are in harmony with those obtained very recently using about 400 non-relativistic and relativistic mean field models \cite{Carlson:2022nfb}, which demonstrates the impact of low-density EOSs on the properties of NS of canonical mass. It was found that tighter constraints on the bulk properties of finite nuclei, such as binding energy, charge radii, and isoscalar giant monopole resonance energy, yield stronger correlations of $\Lambda_{1.4}$ with $L_0$ and $K_{\rm sym,0}$. The correlations of $\Lambda_{1.4}$ with various symmetry energy parameters are stronger only when the nuclear matter parameters evaluated at saturation densities that govern the low-density behavior of EOSs are appropriately constrained. 
\begin{table}[htp]
\caption{\label{ch4-tab2} 
The Pearson's correlation coefficients for  $\Lambda_{1.4}$ with the slope parameter ($L_0$), curvature parameter ($K_{\rm sym,0}$) of the symmetry energy and the pressure for $\beta$-equilibrated matter at a density $2\rho_0$ ($P(2\rho_0)$) obtained using joint posterior distribution of the nuclear matter parameters. The results are obtained assuming different upper bounds on $\Lambda_{1.4}$ and those are associated with a maximum mass of NS $\geqslant$  2.1$M_\odot$.}
\centering
\begin{tabular}{cccc}
\hline\hline
{Upper bound} & $\Lambda_{1.4}-L_0$ & $\Lambda_{1.4}-K_{\rm sym,0}$&$\Lambda_{1.4}-P(2\rho_0)$ \\ 
on $\Lambda_{1.4}$& &  & \\
\hline
400 &0.29 & 0.42&0.88 \\[1.2ex]
600 & 0.54 & 0.71&0.97 \\[1.2ex]
800 & 0.67 &0.81 &0.98 \\[1.2ex]
1000 & 0.72 & 0.84&0.98 \\[1.2ex]
\hline\hline
\end{tabular}
\end{table}

The variations of $\Lambda_{1.4}$ with the pressure of $\beta$-equilibrated charge neutral matter at twice the saturation densities $P(2\rho_0)$ are plotted in Fig.~\ref{ch4-fig9}. The correlations of $\Lambda_{1.4}$ with $P(2\rho_0)$ are quite robust, independent of distributions of nuclear matter parameters. It may be pointed out that the pressure is related to the density derivative of the energy per particle and would depend on several
nuclear matter parameters, except for $\rho=\rho_0$. Its value  at
$\rho_0$ is  mainly governed by the $L_0$.
The strong $\Lambda_{1.4}- P(2\rho_0)$  correlations may not allow one to reconstruct the EOS of symmetric nuclear matter and the density-dependent symmetry energy very accurately, though they are highly desirable~\cite{Imam:2021dbe}. The accurate determination of lower-order nuclear matter parameters from the bulk properties of finite nuclei in conjunction with tighter constraints on $P(2\rho_0)$ may shed some light on the value of high-order nuclear matter parameters.

The results presented in Figs \ref{ch4-fig7} - \ref{ch4-fig9} may also be sensitive
to the range of values for the $\Lambda_{1.4}$ as well as to the bounds
on the nuclear matter parameters. In Table \ref{ch4-tab2},   we have listed
the values of Pearson's correlation coefficient for $\Lambda_{1.4} - L_0$,
$\Lambda_{1.4} - K_{\rm sym,0}$ and $\Lambda_{1.4} - P(2\rho_0)$ obtained
for the joint posterior distribution of nuclear matter parameters. The
correlation coefficients increase with the increase in the upper bound
on $\Lambda_{1.4}$. In particular, these effects are quite strong for
the case of $\Lambda_{1.4} - L_0$ and  $\Lambda_{1.4} - K_{\rm sym,0}$.

\begin{table}[!ht]
\caption{\label{ch4-tab3} 
The Pearson's correlation coefficients for  $\Lambda_{M}$ and $R_{M}$,  corresponding to different neutron star mass $M$  with the slope parameter ($L_0$), curvature parameter ($K_{\rm sym,0}$) of the symmetry energy and  the pressure for $\beta$-equilibrated matter at a density $2\rho_0$ ($P(2\rho_0)$). The correlation coefficients are obtained  at different $\rho_{c_s}$ and for the lower bound on  maximum neutron star mass $M_{\rm max}$ $\geqslant$  2.1$M_\odot$.} 
  \centering
\begin{tabular}{ccccc|ccc}
\hline\hline
\multirow{2}{*}{$\frac{\rho_{c_s}}{\rho_0}$} & \multirow{2}{*}{$\frac{M}{M_\odot}$} &  & {$\Lambda_M$}& & &{$R_M$}&  \\
 \cline{3-8}
&  & $L_0$ & $K_{\rm sym,0}$ & $P(2\rho_0)$ & $L_0$ & $K_{\rm sym,0}$ & $P(2\rho_0)$ \\
  \cline{3-8}
\hline
\multirow{3}{*}{1.5}& 1.2 &0.76 & 0.85&0.91&0.83 & 0.67&0.76 \\[1.2ex]
&1.4 & 0.70 & 0.81&0.93 &0.79 &0.70 &0.81 \\[1.2ex]
&1.6 & 0.60 &0.76 &0.91 & 0.72& 0.70&0.86 \\[1.2ex]
\multirow{3}{*}{1.75}& 1.2 & 0.77 & 0.87&0.94 &0.84 &0.70 &0.81 \\[1.2ex]
&1.4 &0.71  &0.83 &0.96 &0.80 &0.74 &0.88 \\[1.2ex]
&1.6 & 0.61 &0.77 &0.97 &0.73 &0.72 & 0.91\\[1.2ex]
\multirow{3}{*}{2.0}& 1.2 & 0.80 &0.88 &0.97 & 0.85&0.73 &0.86 \\[1.2ex]
&1.4 & 0.72 &0.84 &0.98 &0.81 &0.75 & 0.91\\[1.2ex]
&1.6 & 0.65 & 0.80&0.98 & 0.75&0.75 &0.94 \\[1.2ex]
\hline\hline
\end{tabular}
\end{table}

So far, we have considered only the case of tidal deformability for a neutron star with the canonical mass obtained for the  EOSs assuming $\rho_{c_s} = 2.0\rho_0$, beyond which the high-density part of the EOS is switched on. We now consider the tidal deformability and radius with neutron star masses M$= 1.2, 1.4$, and $1.6M_\odot$ and study their correlations with symmetry energy slope $L_0$, curvature parameters $K_{\rm sym,0}$ and pressure  $P(2.0\rho_0)$. In Table~\ref{ch4-tab3}, we present the values of Pearson's correlation coefficients for $\Lambda_{M}$ and $R_{M}$ with $L_0$, $K_{\rm sym,0}$ and $P(2\rho_0)$. These values of correlation coefficients are calculated  for $\rho_{c_s}=1.5, 1.75$ and $2.0\rho_0$ and $M_{\rm max} \geqslant 2.1M_\odot$. The results are obtained for the joint posterior distribution of nuclear matter parameters (Fig.~\ref{ch4-fig1}), which determine the EOS at $\rho \leqslant \rho_{c_s}$. The correlations increase a little with the $\rho_{c_s}=1.5 - 2.0\rho_0$. It is interesting to note that the $\Lambda - L_0$ correlations are moderate, while the $\Lambda - K_{\rm sym,0}$ correlations become relatively stronger which further increase for the $\Lambda -
P(2\rho_0)$ case. The correlations involving the NS radius display little different trends; the $R - L_0$ correlations are stronger than those for $R- K_{\rm sym,0}$. The correlations of $R-P(2\rho_0)$ are weaker than those for $\Lambda - P(2\rho_0)$. The values of the correlation coefficient presented in Table~\ref{ch4-tab3} may not be significantly affected by the inclusion of any exotic degrees of freedom beyond 2$\rho_0$, since the EOSs beyond the $\rho_{c_s}$ have diverse behavior as they
are obtained simply by imposing the causality condition.  We have also repeated the calculations to obtain the joint posterior distribution of nuclear matter parameters by reducing the uncertainty on the pure neutron matter EOS by a factor of two.  The values of the median and the   68\%  confidence interval for the nuclear 
matter parameters with the reduced uncertainties are $K_0=239_{-32}^{+34}$, $Q_0=-5_{-343}^{+258}$, $J_0=32.19_{-0.9}^{+0.9}$, $L_0=52_{-8}^{+8}$, $K_{\rm sym,0}=-102_{-60}^{+63}$ and $Q_{\rm sym,0}=722_{-510}^{+379}$.
The values of symmetry energy coefficient $J_0$, slope parameter $L_0$, and curvature parameter $K_{\rm sym,0}$ are now more constrained as compared to those in Fig.~\ref{ch4-fig1}.  Similar to Table \ref{ch4-tab3}, we have listed the values of Pearson's correlation coefficients for $\Lambda_{M}$ and $R_{M}$ with $L_0$, $K_{\rm sym,0}$ and $P(2\rho_0)$ in Table \ref{ch4-tab4}. The values of correlation coefficients show similar trends as listed in Table \ref{ch4-tab3}, but the values of correlation coefficients for $\Lambda_{M}$ and $R_{M}$ with $L_0$, $K_{\rm sym,0}$  are reduced. But, the  correlations of the $\Lambda_M$ and $R_M$ with $P(2\rho_0)$
seem to be more-or-less independent of the ranges of the nuclear matter parameters,  which once again implies that the NS properties may be more sensitive to the combination of several nuclear matter parameters
rather than the individual ones.
We have also examined the influence of the lower bound on the maximum mass of NS for 2.1 $M_\odot$ to 2.4 $M_\odot$ on our correlation systematics. The correlations do not change significantly; for example, the correlation coefficient value for $\Lambda_{1.4}$ with $L_0$ and $K_{\rm sym,0}$ changes from 0.72 to 0.78 and from 0.84 to 0.86, respectively.

\begin{table}[!ht]
\caption{\label{ch4-tab4} 
Same as Table \ref{ch4-tab3} but, the results are obtained 
using the uncertainty on the PNM EOS reduced to 3$\times$ N$^3$L0.}
  \centering
\begin{tabular}{ccccc|ccc}
\hline\hline 
\multirow{2}{*}{$\frac{\rho_{c_s}}{\rho_0}$} & \multirow{2}{*}{$\frac{M}{M_\odot}$} &  & {$\Lambda_M$}& & &{$R_M$}&  \\
 \cline{3-8}
&  & $L_0$ & $K_{\rm sym,0}$ & $P(2\rho_0)$ & $L_0$ & $K_{\rm sym,0}$ & $P(2\rho_0)$ \\
  \cline{3-8}
\hline
\multirow{3}{*}{1.5}& 1.2 &0.47 & 0.73&0.92&0.68 & 0.65&0.74  \\[1.2ex]
&1.4 & 0.36 & 0.64&0.93 &0.57 &0.66 &0.83 \\[1.2ex]
&1.6 & 0.26 &0.58 &0.88 & 0.45& 0.65&0.84 \\[1.2ex]
\multirow{3}{*}{1.75}& 1.2 & 0.50 & 0.76&0.95 &0.70 &0.67 &0.81 \\[1.2ex]
&1.4 &0.39  &0.72 &0.96 &0.59 &0.68 &0.88 \\[1.2ex]
&1.6 & 0.31 &0.67 &0.96 &0.49 &0.66 & 0.91\\[1.2ex]
\multirow{3}{*}{2.0}& 1.2 & 0.51 &0.77 &0.97 & 0.70&0.70 &0.84 \\[1.2ex]
&1.4 & 0.40 &0.73 &0.98 &0.59 &0.70 & 0.90\\[1.2ex]
&1.6 & 0.31 & 0.69 & 0.98 & 0.49 & 0.68 & 0.94 \\[1.2ex]
\hline\hline
\end{tabular}
\end{table}

We have assessed the influence of crust EOS on our correlation systematics. We have repeated our calculations for the correlations of $\Lambda_{1.4}$ and $R_{1.4}$ with $L_0$, $K_{\rm sym,0}$ and $P(2\rho_0)$ for $\rho_{c_s}=2\rho_0$ 
 by using different crust EOSs such as NL3$\omega\rho$-L55~\cite{Grill:2014aea, Pais:2016xiu} and TM1e~\cite{Grill:2014aea, Boukari:2020iut} available in the \href{https://compose.obspm.fr}{CompOSE} \cite{CompOSECoreTeam:2022ddl}. The results for the correlations involving $\Lambda_{1.4}$ change almost by 1\%. The correlation of $R_{1.4}$ with $L_0$ also remains practically unaffected. However, the correlations of $R_{1.4}$ with $K_{\rm sym,0}$ and $P(2\rho_0)$ improve by 5-10 \%.

\section{Conclusion} \label{ch4-summary}

We have constructed a large set of minimally constrained EOSs for the NS matter and performed a detailed investigation of the correlations of NS properties with several nuclear matter parameters that determine the density dependence of symmetry energy.
The joint posterior distribution of nuclear matter parameters that determine the EOSs at low densities ($\rho \leq \rho_{c_s}$) is obtained by employing our minimal constraints within a Bayesian approach. 
These  EOSs are consistent with (i) the pure neutron matter EOS from a precise next-to-next-to-next-to-leading-order ( N$^{3}$LO) calculation in chiral effective field theory and (ii) empirical ranges of low-order nuclear matter parameters determined by the experimental data on the bulk properties of finite nuclei. 
The EOSs beyond $\rho_{c_s}=1.5-2\rho_0$ are constrained only by imposing the causality condition on the speed of sound. The large set of EOSs so obtained is employed to study the sensitivity of NS properties to the symmetry energy slope parameter $L_0$ and curvature parameter $K_{\rm sym,0}$ as well as to the pressure of $\beta-$equilibrated matter at 2$\rho_0$. The calculations are also performed with uncorrelated Uniform and Gaussian distributions of NMPs that ignore the interdependence among them as present in their joint posterior distribution.

The tidal deformability and radius of NS, as a function of mass, are evaluated using our minimally constrained EOSs.
The NS properties at canonical mass and the maximum NS mass are found to be consistent with the observational constraints. The correlations of tidal deformability and radius for different NS masses ($1.2-1.6 M_\odot$) with the slope and curvature parameter of symmetry energy and the pressure for $\beta-$equilibrated charge neutral matter at $2\rho_0$ are studied.
We have examined the sensitivity of these correlations 
to several factors, such as, 
(i) the behavior of the high-density part of the EOS,(ii) the choice of distributions of nuclear matter parameters, their interdependence, and  uncertainties,(iii) the lower bound on the maximum mass of the stable neutron stars, (iv)  the value of $\rho_{c_s}$ beyond which the low-density EOSs are smoothly joined with a diverse set of EOSs constrained only by the causality condition on the speed of sound,
(v) the upper bound on the value of tidal deformability.

The tidal deformability is reasonably correlated with the symmetry energy slope parameter $L_0$ for the EOSs obtained from the joint posterior distribution of nuclear matter parameters. The correlation of tidal deformability with $K_{\rm sym,0}$ is slightly stronger. These correlations become even stronger when the priors of lower-order nuclear matter parameters corresponding to incompressibility and symmetry energy coefficients are kept fixed. The correlations become noticeably weaker in the absence of interdependence among nuclear matter parameters. For instance, the Pearson's correlations coefficients for $\Lambda_{1.4}-L_0$ ( $\Lambda_{1.4}-K_{\rm sym,0}$) are $r\sim 0.4(0.6)$ for independent distribution of nuclear matter parameters which become $r\sim0.8(0.9)$ for the joint posterior distribution. This implies that the correlations of NS properties with individual symmetry energy parameters are masked in the absence of appropriate constraints on the EOSs at low densities.  It also partly explains why the outcome of the similar correlations studied in the earlier publications~\cite{Malik:2020vwo, Pradhan:2022vdf, Carson:2018xri, Tsang:2020lmb, Kunjipurayil:2022zah}  are at variance. These correlations improve a little bit with the increase in $\rho_{c_s}$. The diverse behavior of EOSs at high-density ($\rho \geqslant \rho_{c_s}$), as modeled by wide variations of the speed of sound within the causal limit, do not affect the sensitivity of the NS properties to the symmetry energy parameters evaluated at saturation density. 
The results for the correlation of tidal deformability with the pressure at 2$\rho_0$ are robust and practically independent of all the factors considered.
The correlations of NS radius with the symmetry energy parameters are also sensitive to various factors considered. The vulnerability of correlations of NS properties with individual parameters of symmetry energy and, on the contrary, the robustness of their correlations with $P(2\rho_0)$ needs to be further investigated to pin down the combination of the optimum number of nuclear matter parameters required to describe the NS properties for the masses $\sim 1.4M_\odot$. With the observations of heavier NS, the correlation between NS properties and symmetry energy parameters presented in this paper may be improved.
 

%% file: chapter5/chapter5.tex
\chapter{Multivariate Analysis of Neutron Star Properties}
\label{chap5}

In this Chapter, we perform a Principal Component Analysis (PCA) ( see Chapter \ref{PCA}) to investigate the connection between multiple nuclear matter parameters and the tidal deformability as well as the radius of neutron stars across a wide range of masses \cite{Patra:2023jbz}. The analysis is carried out using two different sets of  EOSs. The EOSs at low densities (using Taylor expansion \ref{Taylor}) have been derived using the uncorrelated uniform and joint posterior distributions of nuclear matter parameters. The joint posterior distributions of nuclear matter parameters are obtained by imposing the constraints on the low-order nuclear matter parameters determined by the experimental data on the bulk properties of finite nuclei together with the pure neutron matter (PNM) EOS from a precise next-to-next-to-next-to-leading-order (N$^{3}$LO) calculation in chiral effective field theory \cite{Hebeler:2013nza, Lattimer:2021emm} within the Bayesian Inference (see the chapter \ref{BA}). The EOSs at high density ($\rho>2\rho_0$) are constructed by imposing the causality condition on the speed of sound and are independent of compositions of NS matter (see the chapter \ref{H-eos}). Our analysis reveals that more than one principal component is necessary to appropriately describe the NS properties, such as tidal deformability and radius. The role of iso-scalar nuclear matter parameters becomes increasingly important with neutron star mass.

\section{Distributions of Nuclear Matter Parameters}
\begin{table}[htp]
\centering
\caption{\label{ch5-tab1} The lower (Min.) and upper (Max.) bounds of the uniform distributions (D1) along with the means ($\mu$) and uncertainties ($2\sigma$) of marginalized distributions (D2) for each nuclear matter parameter except for $e_0$, which is kept fixed to -16.0 MeV are listed.} 
\begin{tabular}{ccccc}
\hline \hline
\multirow{2}{*}{NMPs} & \multicolumn{2}{c}{D1}  &\multicolumn{2}{c}{D2} \\ [1.5ex] 
  \cline{2-5}
{(in MeV)}    & Min. & Max. & $\mu$ & 2$\sigma$ \\[1.5ex] 
\cline{1-5}    
 $K_0$ & 200 & 290 & 242 & 45 \\[1.5ex]
 $Q_0$ & -500 & 450 & -25 & 466 \\[1.5ex]
 $J_0$ & 30 & 35 & 32.2 & 2 \\[1.5ex]
 $L_0$ & 30 & 80 & 54.2 & 24 \\[1.5ex]
 $K_{\rm sym,0}$ &-300 & 100 & -89 & 180 \\[1.5ex]
 $Q_{\rm sym,0}$ & 0 & 1000 & 772 & 700 \\[1.5ex] 
 \hline \hline
\end{tabular}
\end{table}

We have generated two distinct sets of EOSs that correspond to different distributions of nuclear matter parameters.  One of these sets is based on uncorrelated uniform distributions of nuclear matter parameters, and the other one is obtained from their joint posterior distribution. The joint posterior distribution of the nuclear matter parameters is taken from Ref.~\cite{Patra:2023jvv}, which was obtained by imposing minimal constraints that include some selected basic nuclear matter properties at the saturation density and the EOS for the pure neutron matter at low densities from  N$^3$LO calculation in the chiral effective field theory ~\cite{Hebeler:2013nza, Lattimer:2021emm}. The conditions of thermodynamic stability, causality speed of sound on the EOS, and the resulting maximum mass of neutron star larger than 2$M_\odot$ were also imposed. Some of the nuclear matter parameters are correlated due to minimal constraints from the chiral effective field theory applied within the Bayesian statistical method ~\cite{Gelman2013, Buchner2014, Ashton2019}. The marginalized distributions of each nuclear matter parameter are derived from the joint posterior distribution in order to put the bounds on their uniform distributions. The bounds on uniform distributions of nuclear matter parameters roughly the 90\% confidence intervals of the corresponding marginalized distributions.  In Table \ref{ch5-tab1}, we have listed the lower and upper bounds of the uniform distributions, as well as the means and variances of the marginalized distributions for each nuclear matter parameter. The iso-scalar nuclear matter parameters, the binding energy per nucleon $e_0$, and the saturation density $\rho_0$ for symmetric nuclear matter remain fixed at -16.0 MeV and 0.16 fm$^{-3}$, respectively. References ~\cite{Patra:2023jvv, Kunjipurayil:2022zah} have shown that the dependence of neutron star properties on individual nuclear matter parameters is sensitive to the choice of the distributions of nuclear matter parameters. We would like to explore how multivariate analysis is sensitive to different distributions of these nuclear matter parameters.

\begin{figure}[!ht]
\centering
\includegraphics[width=\textwidth]{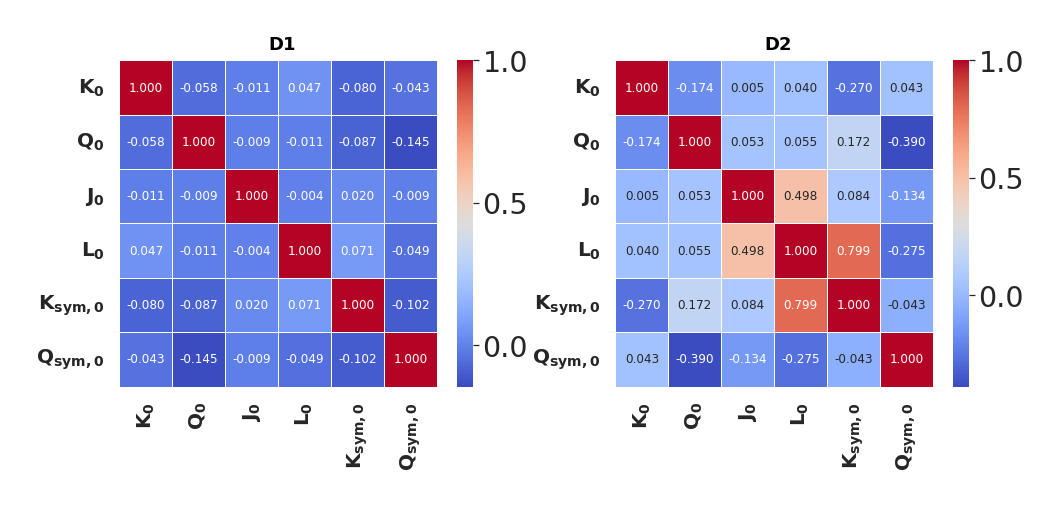}
\caption{ The correlation among various nuclear matter parameters is shown. The results for the left and right panels are for uncorrelated (D1) and correlated (D2) distributions of nuclear matter parameters, respectively. Color codes indicate the values of Pearson’s correlation coefficients among the various NMPs.}\label{ch5-fig1}
\end{figure} 

We have generated  10,000 samples of nuclear matter parameters for each of the uniform uncorrelated and joint posterior distributions, hereafter referred to as  D1 and D2, respectively. For each set of nuclear matter parameters, the equations of the state beyond $2\rho_0$ are obtained for a random set of the speed of sound parameters $n_b, h_p, w_p,$ and $n_p$ as discussed in the previous Chapter \ref{H-eos}.
Out of these, around 2,500 samples from both distributions have been selected after applying the filters mentioned earlier.
In Fig. \ref{ch5-fig1}, we have illustrated the correlations among nuclear matter parameters for both D1 and D2. The distribution D1 displays poor correlations among the nuclear matter parameters, whereas D2 reveals stronger correlations among some of the parameters due to the minimal constraints, such as the notable correlation coefficient r($L_0$, $K_{\rm sym,0}$) = 0.80. The significant correlations between $L_0$ and $K_{\rm sym,0}$ have been previously documented as well ~\cite{Patra:2022yqc, Patra:2023jvv}.

\section{Fit of NS Properties to Nuclear Matter Parameters}

\begin{figure}[htp]
\centering
\includegraphics[width=0.5\textwidth]{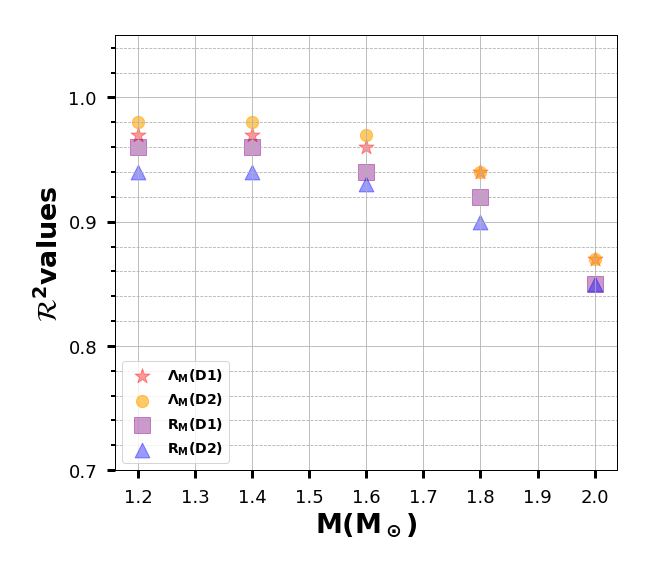}
\caption{
The $\mathcal{R}^2$ values of the neutron star properties, such as tidal deformability and radius in the mass range of $1.2-2 M_\odot$, are shown. The different color symbols correspond to different NS properties and the data.}\label{ch5-fig2}
\end{figure}

The EOSs for $\beta$-equilibrated charge neutral matter in the density ranges $0.5\rho_0$ to $2\rho_0$ are constructed using Taylor expansion with the D1 and D2 distributions of nuclear matter parameters.  Each of these low-density EOS is smoothly joined by a diverse set of EOSs
that satisfy causality conditions as given by Eqs. (\ref{eq-vs}-\ref{eq-preE}). The EoS for outer and inner crusts for density ranges $\rho<0.5\rho_0$ is used as follows. We employed Baym-Pethick-Sutherland's EOS in the density range $3.9 \times 10^{-11}\rho_0<\rho<0.0016\rho_0$ for the outer crust. The inner crust EOS is polytropic ~\cite{Carriere:2002bx}, 
\bea 
p(\varepsilon)= \alpha + \beta \varepsilon^{\frac{4}{3}}. \label{eq-ic} 
\eea
The inner crust EOS is matched with the outer crust at one end and the outer core at the end by appropriately adjusting the values of the coefficients $\alpha$ and $\beta$. The inner crust's EOS affects the Love number $k_{2}$ and compactness parameter, but not the tidal deformability parameter ~\cite{Piekarewicz2019}. After determining the core and crust EOSs,  the neutron star mass, radius, and tidal deformability for a given central pressure can be computed using TOV equations.  In order to demonstrate our approach,  we use a linear fit function  of nuclear matter parameters as listed in Table~\ref{ch5-tab1} to calculate the tidal deformability ($\Lambda_M$) and radius ($R_M$) for a given NS mass as,
\bea
{\Lambda_M} &=& \sum_{i} W_i P_i + b=\sum_{i}\Lambda_i + b \label{eq-lam}\\
{R_M} &=& \sum_{i} W^\prime_i P^\prime_i + b^\prime=\sum_{i}R_i + b^\prime\label{eq-rad}
\eea
where $W_i$ and $W^\prime_i$ are the weight factors of given nuclear matter parameters. The $b$ and $b^\prime$ are the biases. The $P \in$ \{ $K_0$, $Q_0$, $J_0$, $L_0$, $K_{\rm sym,0}$, $Q_{\rm sym,0}$\} and M stands for the neutron star masses.  The $\Lambda_i$ and $R_i$ in the right part of above Eqs.(\ref{eq-lam}) and (\ref{eq-rad}) correspond to the $W_i  P_i$ and $W^\prime_i  P^\prime_i$, respectively. We assess the quality of fit for our regression model using the $\mathcal{R}^2$ value as a measure ~\cite{Dennis2002}. The $\mathcal{R}^2$ value ranges from 0 to 1, where 0 signifies no variability in the dependent variable, and 1 represents complete variance. The  $\mathcal{R}^2$ close to unity indicates that the  NS properties from the solutions of TOV equations are almost equal to the ones obtained by their linear fit to the nuclear matter parameters. In Fig. \ref{ch5-fig2}, we present the $\mathcal{R}^2$  values obtained from fitting the tidal deformability and radius of neutron stars within the mass range of $M=1.2 - 2.0 M_\odot$. As the neutron star mass increases, the associated $\mathcal{R}^2$ value tends to decrease. Specifically, the $\mathcal{R}^2$ values for the fitted properties of neutron stars with a mass of 2$M_\odot$ fall below 0.9. Consequently, in what follows,  we focus on the multivariate analysis of neutron stars within the mass range of $1.2 - 1.8 M_\odot$, guided by the $\mathcal{R}^2$ value considerations. 

\begin{figure}[htp]
\centering
\includegraphics[width=0.9\textwidth,height=10cm]{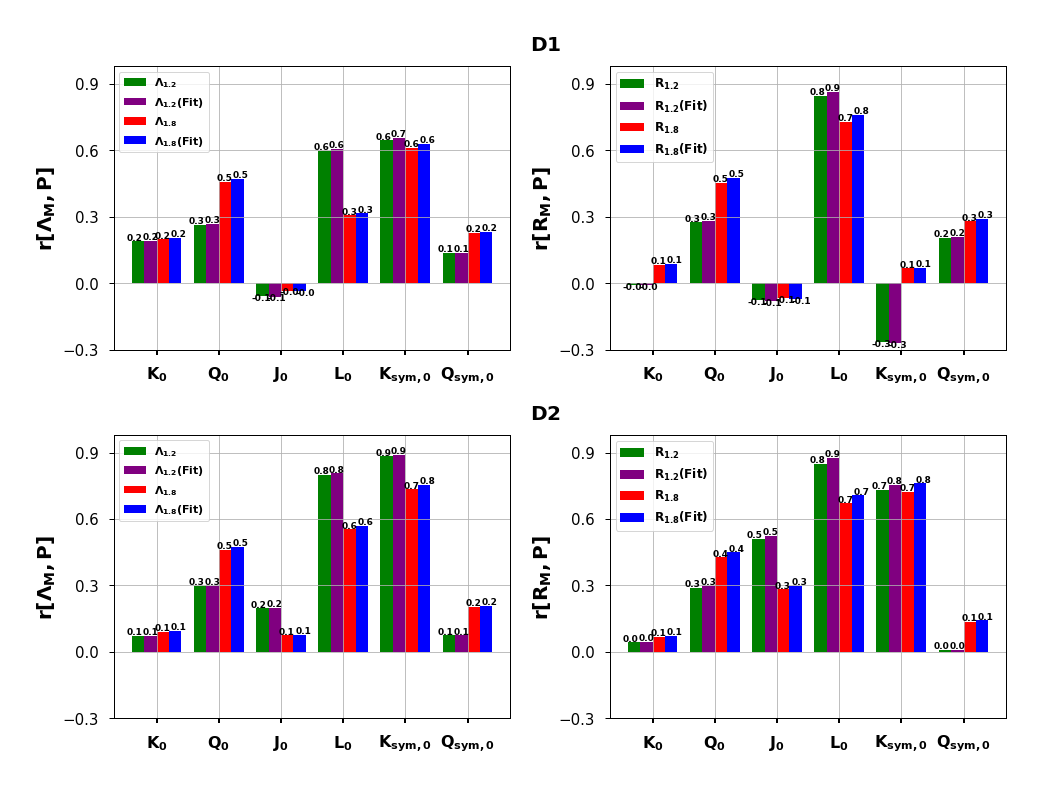}
\caption{ The bar plot shows correlation coefficient values of NS properties with nuclear matter parameters only for mass 1.2 and 1.8 $M_\odot$. For the comparison, the values of correlation coefficients from the fitted model are shown with different color bars. The upper and lower panels correspond to uncorrelated (D1) and correlated (D2) NMPs distributions.}\label{ch5-fig3}
\end{figure}

In Fig.\ref{ch5-fig3}, we display the correlation coefficient values obtained for exact values of tidal deformability and radius with individual nuclear matter parameters for the masses 1.2 and 1.8  $M_\odot$  by green and red bars, respectively. We also juxtapose similar results obtained from the fitted values of tidal deformability and radius through Eqs. (\ref{eq-lam}) and (\ref{eq-rad}) for the masses 1.2 and 1.8  $M_\odot$  by purple and blue  bars, respectively. The results are presented for two different distributions of NMPs as indicated by labels  D1 (Upper) and D2 (Lower).  This visual representation demonstrates that all the correlation coefficient values remain consistent even after the fitting process. The application of the fit does not alter these correlation trends. The correlations of the tidal deformability with NMPs are qualitatively similar for the D1 and D2 cases. In general, the D2 distributions of NMPs result in stronger correlations. The correlations of the radius with the NMPs appear to be sensitive to the choice of their distributions. In particular, the radius is very weakly correlated with $K_{\rm sym,0}$ for the D1 case and shows reasonably strong correlations for the D2 case. It is also interesting to note that the correlations of tidal deformability and radius with  $Q_0$ increase somewhat with an increase in NS mass. 

\section{Principal Component Analysis of NS Properties}
As discussed in Chapter \ref{PCA}, the methodology of PCA analysis comprises the following steps: (1) Construction of Covariance matrix 
(2) Diagonalization of the covariance matrix.
(3) The principal components' proportionate variance captured by the eigenvectors and eigenvalues. First, we calculate the values of weights $W_i$s and $W^\prime_i$s appear in Eqs. (\ref{eq-lam}) and (\ref{eq-rad}). The PCA is performed using $\Lambda_i$s and $R_i$s as features corresponding to the target variables $\Lambda$ and R, respectively. Then, $6\times 6$ covariance matrices for the tidal deformability and radius of NS are constructed for a given mass. The eigenvalues and the corresponding eigenvectors obtained by diagonalizing the covariance matrices are arranged in descending order. The most important principal component, PC1 corresponds to the eigenvector with the highest eigenvalue.  The succeding eigenvalues and the eigenvectors are labelled as  PC2, PC3, etc.  The nuclear matter parameter having the largest contribution to the eigenvector for PC1 is the most dominant one in determining the  NS property in consideration. Likewise, the eigenvectors associated with remaining PCs together with eigenvalues, can be used to identify other important nuclear matter parameters for the NS properties.

\begin{table}[!ht]
\caption{\label{ch5-tab2} The normalized eigenvalues associated with the principal components are listed for the NS properties with masses range $1.2-1.8M_\odot$. D1 and D2 correspond to the uncorrelated and correlated distributions of the nuclear matter parameters.
}
\centering

\begin{tabular}{cccccccc|cccccc}
\hline \hline
\multirow{2}{*}{NS} & \multirow{2}{*}{$\frac{M}{M_\odot}$} &  \multicolumn{6}{c}{D1} & \multicolumn{6}{c}{D2} \\
 \cline{3-14}
&  & PC1 & PC2 & PC3 & PC4 & PC5 & PC6 & PC1 & PC2 & PC3 & PC4 & PC5 & PC6 \\[1.5ex]
\cline{3-14}
\hline
\multirow{5}{*}{{$\Lambda_M$}}& 1.2 & 1.00 & 0.62 & 0.32 & 0.17 & 0.13 & 0.01 & 1.00 &0.35 & 0.23 & 0.13 & 0.04 &0.00  \\[1.5ex]
&1.4 & 1.00 & 0.40 & 0.35 & 0.20 & 0.14 & 0.00 & 1.00 & 0.52 & 0.27 & 0.16 & 0.04 & 0.00 \\[1.5ex]
&1.6 & 1.00 & 0.53 & 0.25 & 0.22 & 0.15 & 0.00 & 1.00 & 0.67 & 0.33 & 0.20 & 0.03 & 0.00  \\[1.5ex]
&1.8 & 1.00 & 0.70 & 0.31 & 0.16 & 0.13 & 0.00 & 1.00 & 0.71 & 0.36 & 0.21 & 0.01 & 0.00  \\[1.5ex]
\multirow{5}{*}{{R$_M$}}& 1.2 & 1.00 & 0.12 & 0.10 & 0.07 & 0.01 &0.00 & 1.00 &0.18 &0.09 & 0.02 & 0.00 &0.00\\[1.5ex]
&1.4 & 1.00 & 0.19 & 0.12 & 0.03 & 0.00 & 0.00 & 1.00 & 0.29 & 0.13 & 0.01 & 0.00 & 0.00\\[1.5ex]
&1.6 & 1.00 & 0.31 & 0.18 & 0.01 &0.00 &0.00 & 1.00 & 0.46 & 0.20 &0.01 & 0.00 &0.00\\[1.5ex]
&1.8 & 1.00 & 0.54 & 0.29 & 0.03 & 0.02 &0.00 &1.00& 0.70& 0.32& 0.04& 0.00&0.00\\[1.5ex]
\hline \hline
\end{tabular}
\end{table}

\begin{figure}[!ht]
\centering
\includegraphics[width=0.9\textwidth,height=12cm]{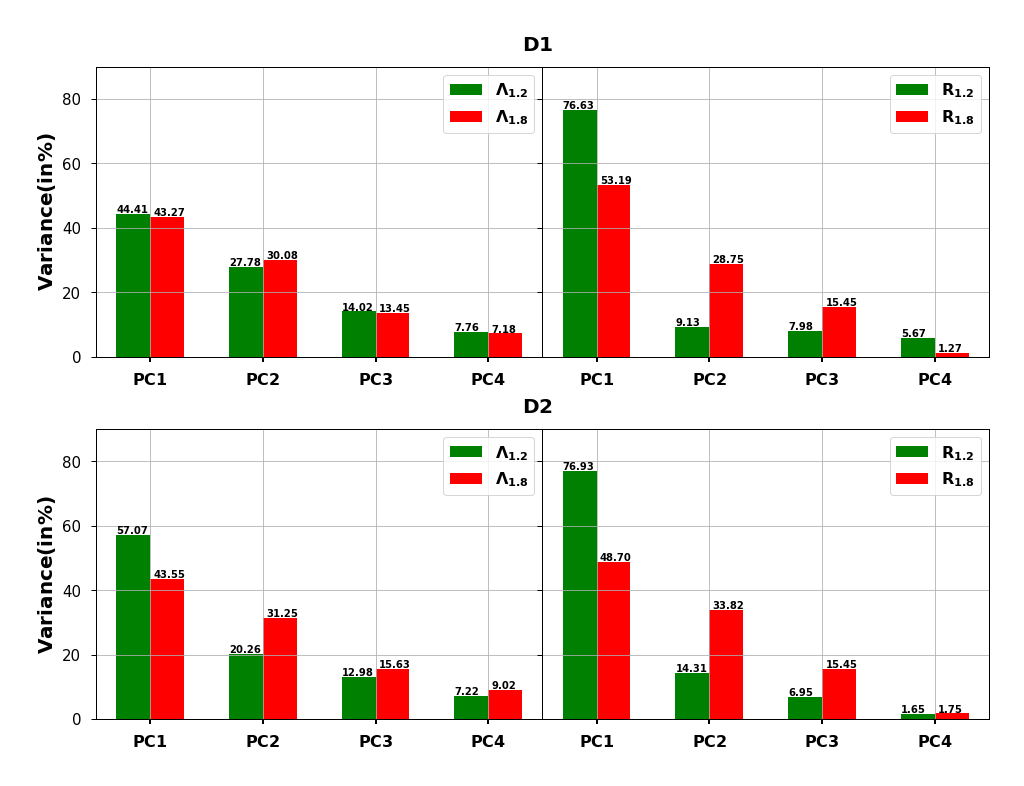}
\caption{ The variance (in \%) of different principal components corresponding to the NS properties is displayed. The results in the upper and lower panels represent the uncorrelated (D1) and correlated (D2) nuclear matter parameter distributions, respectively. }\label{ch5-fig4}
\end{figure}
In Table \ref{ch5-tab2}, we provide a list of eigenvalues that were obtained by diagonalizing the covariance matrix for a specific property and mass of a neutron star. The highest eigenvalues are normalized to unity, representing PC1. The principal component with a normalized eigenvalue less than 0.1 does not contribute significantly.
It is evident that a larger number of principal components contribute to tidal deformability for most of the cases than those for the corresponding radius. For instance, consider a neutron star with a mass of $1.2M_\odot$; there are four significant principal components with normalized eigenvalues higher than 0.1 for tidal deformability, while only three for the radius.

In Fig. \ref{ch5-fig4}, we display the percentage of variation in neutron star properties explained by the first four principal components. The upper and lower panels illustrate the outcomes for uncorrelated (D1) and joint posterior (D2) distributions of nuclear matter parameters. The bars in green correspond to properties at a neutron star mass of $1.2M_\odot$, while the red bars represent $1.8M_\odot$. This figure depicts how much each PC contributes to capturing the variance. From PC1 to PC4, the portion of variance decreases. The total number of PCs needed to account for over 90\% of the variance is smaller for the radius than for the tidal deformability. As an example, consider $\Lambda_{1.2}$ in cases D1 and D2: to achieve over 94\% and 97\% variance, respectively, at least four PCs are necessary. On the other hand, for $R_{1.2}$, only three PCs are sufficient to achieve more than 94\% and 98\% variance in the respective cases.
\begin{figure}[htp]
\centering
\includegraphics[width=0.9\textwidth,height=10cm]{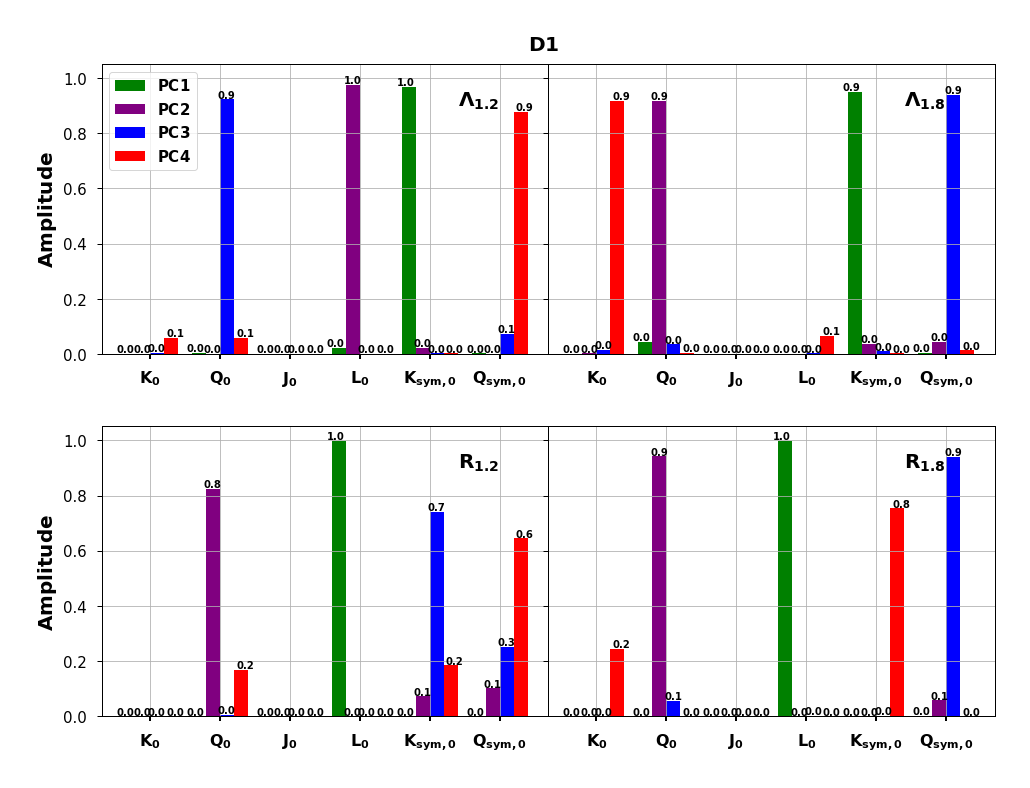}
\caption{ The square of the amplitude values of various nuclear matter parameters for each principal component. The results are presented for the uncorrelated nuclear matter parameter distributions (D1). The different PCs are indicated by different color bars.}\label{ch5-fig5}
\end{figure}

\begin{figure}[htp]
\centering
\includegraphics[width=0.9\textwidth,height=10cm]{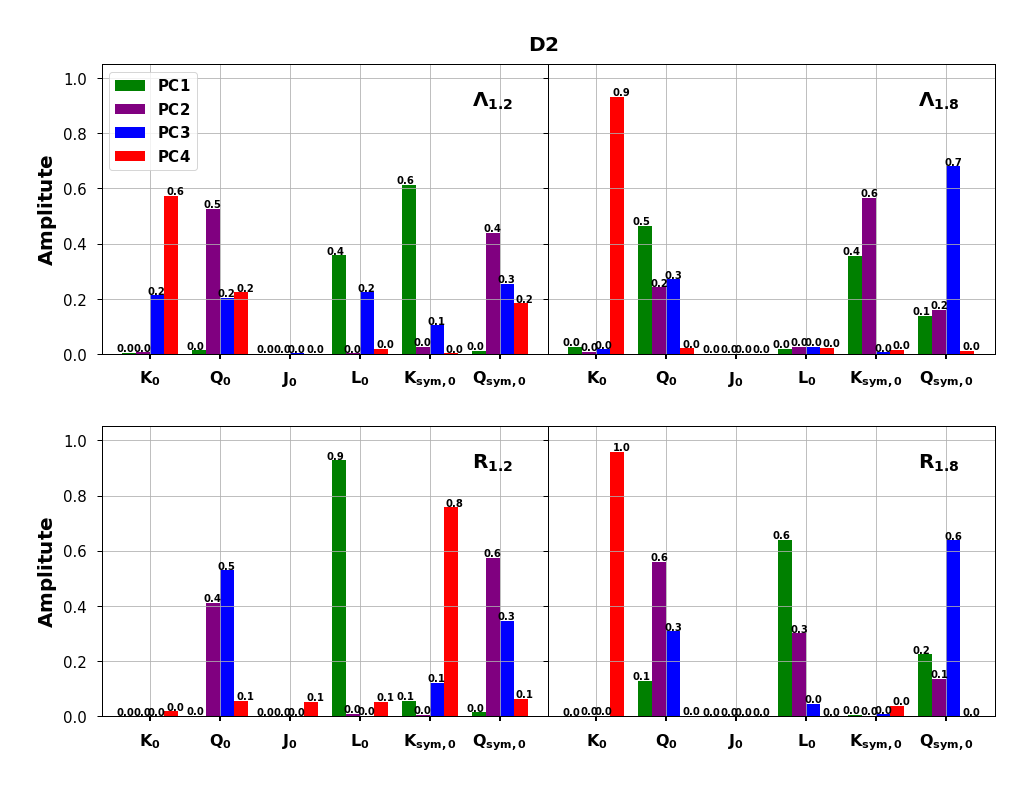}
\caption{ Same as Fig.\ref{ch5-fig5} but for  correlated nuclear matter parameters distributions (D2).}\label{ch5-fig6}
\end{figure}

The squared amplitude components of eigenvectors for a specific principal component provide insights into the contributions from different nuclear matter parameters. This is illustrated in Fig. \ref{ch5-fig5} and \ref{ch5-fig6}, corresponding to D1 and D2 distributions, respectively. The upper panels pertain to tidal deformability, while the lower panels relate to radii. Various nuclear matter parameters contribute differently to the principal components, and this sensitivity depends on the chosen distribution of nuclear matter parameters. In the case of D1 distribution (Fig. \ref{ch5-fig5}), PC1 is primarily composed of a single parameter, $K_{\rm sym,0}$, for both $\Lambda_{1.2}$ and $\Lambda_{1.8}$. The PC2 is dominated by $L_0$ for $\Lambda_{1.2}$ and by $Q_0$ for $\Lambda_{1.8}$. 
The contribution to PC3 for $\Lambda_{1.2}$ is mainly  from $Q_0$, while for $\Lambda_{1.8}$, it is primarily $Q_{\rm sym,0}$. In PC4, $Q_{\rm sym,0}$ has the most significant contribution for $\Lambda_{1.2}$, while for $\Lambda_{1.8}$, it is  from $K_{0}$. Combining the outcomes of  figures \ref{ch5-fig4} - \ref{ch5-fig6}, we can infer that the behavior of $\Lambda_{1.2}$ can be primarily explained by the linear combination of $K_{\rm sym,0}$, $L_0$, and $Q_0$ in the D1 distribution. However, in the case of the D2 distribution, the influence of $K_0$ becomes evident as an additional factor. Shifting the focus to $\Lambda_{1.8}$, the impact of $L_0$ diminishes, while the contributions from iso-scalar parameters like $K_0$ and $Q_0$, along with the iso-vector parameter $Q_{\rm sym,0}$, increase. This suggests that $\Lambda_{1.8}$ is
influenced by a simultaneous interplay of various nuclear matter parameters.  For radii ($R_{1.2}$ and $R_{1.8}$), PC1 is dominated by $L_0$, while PC2 by $Q_0$. In PC3 for $R_{1.2}$, contributions from $K_{\rm sym,0}$ and $Q_{\rm sym,0}$ are notable, whereas for $R_{1.8}$, it's mainly $Q_{\rm sym,0}$.
In Fig. \ref{ch5-fig6}, similar outcomes are depicted, but for D2 distributions, which are notably different from those in Fig. \ref{ch5-fig5}. Multiple nuclear matter parameters contribute to most PCs due to correlations in the D2 distribution. Strong correlations, such as between $L_0$ and $K_{\rm sym,0}$, lead to their combined contributions. For instance, in PC1, both $\Lambda_{1.2}$ and R$_{1.2}$ are influenced by $L_0$ and $K_{\rm sym,0}$, while $\Lambda_{1.8}$ is influenced by $Q_0$, $K_{\rm sym,0}$, $Q_{\rm sym,0}$  and $R_{1.8}$ by $Q_0$, $L_{0}$, and $Q_{\rm sym,0}$. The PC2's contributions include $Q_0$ and $Q_{\rm sym,0}$ for $\Lambda_{1.2}$ and $R_{1.2}$, $Q_0$, $K_{\rm sym,0}$, and $Q_{\rm sym,0}$ for $\Lambda_{1.8}$, and $Q_0$, $L_0$, and $Q_{\rm sym,0}$ for $R_{1.8}$. The PC3 encompasses all NMPs except $J_0$ for $\Lambda_{1.2}$, $Q_0$ and $Q_{\rm sym,0}$ for $\Lambda_{1.8}$, $Q_0$, $K_{\rm sym,0}$, and $Q_{\rm sym,0}$ for $R_{1.2}$, and $Q_0$, $Q_{\rm sym,0}$ for $R_{1.8}$. Similarly, PC4 involves $K_0$, $Q_{0}$, and $Q_{\rm sym,0}$ for $\Lambda_{1.2}$, while $K_0$ dominates PC4 for $\Lambda_{1.8}$. The contributions of PC4 for radii are negligible.

\begin{figure}[!ht]
\centering   
\includegraphics[width=0.94\textwidth,height=12cm]{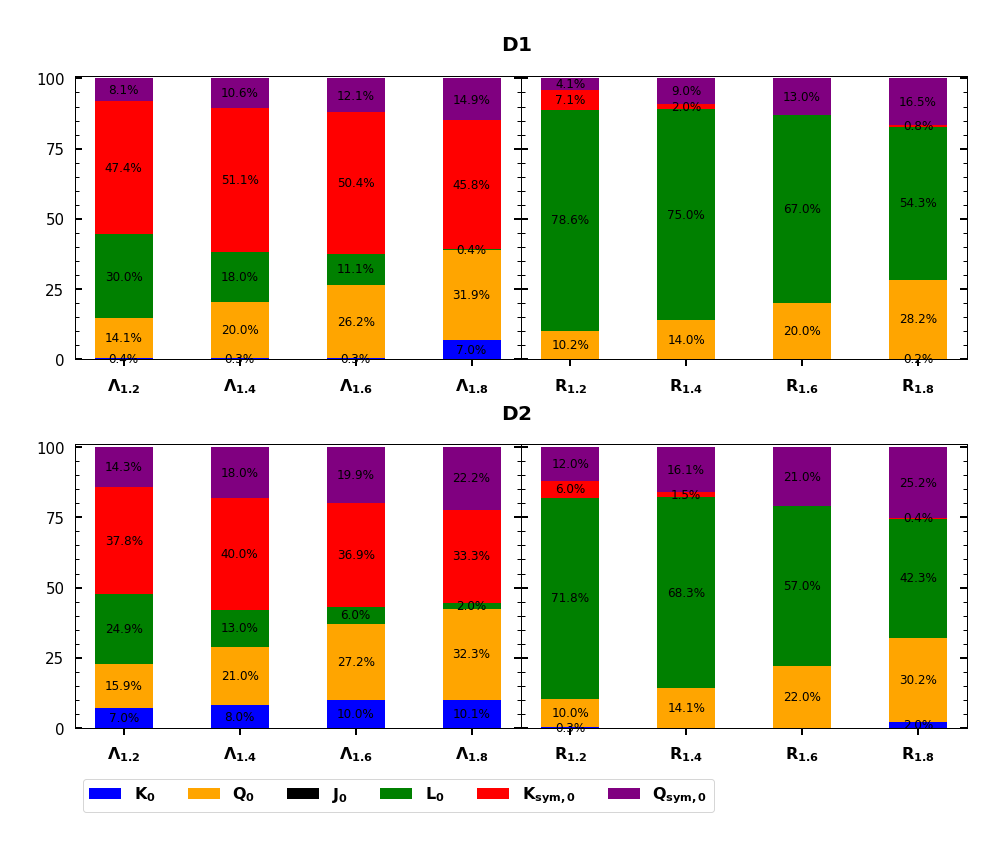}
\caption{ The values of the percentage contributions of nuclear matter parameters to the NS properties with masses range $1.2-1.8M_\odot$. The results in the upper panels and lower panels correspond to the D1 and D2 distributions of nuclear matter parameters, respectively.  The colors of the bar are related to the nuclear matter parameters.}\label{ch5-fig7}
\end{figure}

We computed the total contributions for each of the nuclear matter parameters by summing their contributions to each of the PCs.  The weights for this calculation are derived from the product of the square of amplitude for each nuclear matter parameter and the corresponding eigenvalues of the PCs listed in Table \ref{ch5-tab2}. These weights are then normalized so that the total sum of contributions from all nuclear matter parameters equals unity. The percentages representing the contributions of individual nuclear matter parameters to neutron star properties within the mass range of $1.2-1.8M_\odot$ are shown in Fig. \ref{ch5-fig7}.  The upper and lower panels show the results for D1 and D2 distributions, respectively. Different colors are used to depict the percentage contributions of each nuclear matter parameter. It is important to note that the choice of nuclear matter parameter distributions influences the contributions of different nuclear matter parameters to neutron star properties. For both D1 and D2 distributions, the contributions of specific nuclear matter parameters to tidal deformability and radius can differ by up to 20\%. However, these differences diminish when categorized broadly into iso-scalar and iso-vector parameters. Iso-scalar parameters,   $K_0$ and $Q_0$, contribute together, while the remaining parameters contribute to the iso-vector category. The total contributions from iso-scalar parameters increase, while they decrease for the iso-vector parameters with the increase in NS mass.
For instance, the  contribution from iso-scalar parameters in the case
of D1 (D2) distributions increases  from 15\% to 40\% (24\% to  44\%)
for tidal deformability with the NS mass increasing from 1.2$M_\odot$ to
1.8$M_\odot$.  Concerning radius, iso-scalar parameter contributions
increase from approximately 10\% to 30\% as the neutron star mass
increases from $1.2M_\odot$ to $1.8M_\odot$.

Finally, it may be emphasized that the present work highlights the necessity of multivariate analysis of neutron star properties through PCA as one of the tools. A more comprehensive investigation addressing improved treatment of crust EOSs and high-density EOSs is warranted for a more realistic assessment \cite{Carreau:2019zdy}.

\section{Conclusion} \label{ch5-summary}
We have addressed an unresolved issue of connecting the nuclear matter parameters to the key neutron star properties, such as tidal deformability and radius. 
The outcomes of the majority of the investigations exploring the correlations between properties of neutron stars and individual nuclear matter parameters, which describe the equations of state, are at variance.  We have exploited the efficacy of Principal Component Analysis, a sophisticated analytical tool, in order to establish a comprehensive connection between multiple nuclear matter parameters and the key properties of neutron stars, with the ultimate aim of shedding some light on the existing issue. 
The EOSs essential for describing neutron star matter within the core region up to a density of $2\rho_0$ have been derived. This was accomplished by utilizing both uncorrelated uniform and joint posterior distributions of nuclear matter parameters. To ensure continuity and consistency, each of these distinct EOSs is joined smoothly at $2\rho_0$ by a diverse set of the EOSs obtained by parameterizing the speed of sound such that it remains causal and approaches the conformal limit gradually.

We have found that the variability in the considered neutron star properties requires the incorporation of more than one principal component. These observations emphasize that the neutron star properties depend on multiple nuclear matter parameters. In particular, tidal deformability demands the inclusion of three or more principal components to account for over 90\% of its variations, while for the radius, two principal components suffice to explain similar variations.   
The dominance of nuclear matter parameters contributing to the principal components depends on the specific NS properties and their mass.
For instance, in the case of the tidal deformability of a neutron star with a mass of 1.2$M_\odot$, the symmetry energy curvature parameter $K_{\rm sym,0}$ emerges as the primary contributor to the first principal component.  The second and third principal components are significantly influenced by the symmetry energy slope parameter $L_0$ and the skewness parameter $Q_0$ for symmetric nuclear matter, respectively. The significance of iso-scalar nuclear matter parameters, specifically the incompressibility coefficient K$_0$ and the skewness parameter $Q_0$ of symmetric nuclear matter, becomes more pronounced with an increase in the mass of the neutron star. When the NS mass reaches 1.8$M\odot$, the incompressibility coefficient $K_0$ surpasses the importance of the symmetry energy slope parameter $L_0$. When considering the radius of a neutron star at lower masses, the symmetry energy slope parameter $L_0$ stands out as the primary driver behind the observed variations. Overall, when analyzing the collective impact of iso-scalar parameters ($K_0$ and $Q_0$), these contributions exhibit an approximately 25\% increase with the increase in neutron star mass from 1.2$M_\odot$ to 1.8$M_\odot$. 

%% file: chapter6/conclusion.tex
\chapter{Summary and Conclusion}
\label{chap6}

This thesis investigated the connection of nuclear physics inputs to the neutron star properties. 
For this analysis, we employed several statistical tools. 
The Bayesian analysis is used to construct the equation of the state of neutron stars by using recently available astrophysical observational constraints. The Principal Component Analysis is employed to connect the various neutron star properties directly to nuclear physics inputs.

Bayesian analysis is a common statistical approach for estimating the joint posterior distribution of model parameters. It is based on Bayes's theorem, which combines the prior, likelihood, and evidence to compute a posterior probability distribution, representing updated beliefs based on observed data. The nuclear part of EOS is expressed in terms of nuclear matter parameters. We use the nuclear part of EOS to construct marginalized posterior distributions of the nuclear
matter parameters that are consistent with the minimal constraints. A few low-order nuclear matter parameters, such
as the binding energy per nucleon, the incompressibility coefficient for the symmetric nuclear matter, and symmetry energy coefficients
at the saturation density ($\rho_0$), are constrained in narrow windows along with the low-density pure neutron matter EOS
obtained from a precise next-to-next-to-next-to-leading-order
(N$^3$LO) calculation in chiral effective field theory.
Principal Component Analysis (PCA) is a statistical method used for dimensionality reduction of a dataset.
The process begins with standardizing the data, ensuring variables are on a comparable scale. The diagonalized covariance matrix is then computed, and its eigenvalues and eigenvectors are calculated. Sorting the eigenvalues in descending order allows for the selection of the top eigenvectors, known as principal components. The largest eigenvalue associated with the first PC captures the largest variance, and so on. PCs are the composition of linear combinations of original variables to capture the shared variation patterns. By analyzing the correlation between key principal components and the target variable, we can identify the key parameters essential for the properties of neutron stars.

In Chapter-\ref{chap3}, we used Taylor and $\frac{n}{3}$ expansions to obtain the marginalized posterior
distributions of the nuclear matter parameters by applying a Bayesian approach to both expansions. The nuclear matter parameters or the corresponding EOSs are consistent with a set of minimal constraints that includes basic properties of saturated nuclear matter and low-density ($\rho=0.08-0.16$fm$^{-3}$) EOS for the pure neutron matter from  (N$^{3}$LO) calculation in chiral effective field theory \cite{Hebeler:2013nza}. The NS properties, such as tidal deformability, radius, and maximum mass, are evaluated using large sets of minimally constrained EOSs. The correlations of the $\beta-$equilibrium EOSs as a function of density with various neutron star properties over a wide mass range are investigated.
We found that the pressure for the $\beta$-equilibrated matter at density $\sim 2\rho_0$ is strongly correlated with NS properties such as tidal deformability and radius at 1.4$M_\odot$. The pressure for $\beta$-equilibrated matter at density $\sim 3\rho_0$ substantially correlates with the radius of $2.07M_\odot$ neutron star.  The maximum mass of a neutron star is linked to the pressure of $\beta$-equilibrated matter at density $\sim 4.5\rho_0$.  These correlation systematics match those of unified EOSs for $\beta$-equilibrated matter in several non-relativistic and relativistic mean-field models. We found model independence in the correlations of pressure for $\beta$-equilibrated matter with NS properties. The variations of  NS properties with the pressure of $\beta-$equilibrated matter at $2\rho_0$ remain quite robust, which may be due to the pressure depending on the combination of multiple nuclear matter parameters that describe the symmetric nuclear matter as well as the density dependence of the symmetry energy.

There have been several attempts to study the correlations of radius and tidal deformability of a neutron star with individual nuclear matter parameters which determine the density-dependence of symmetry energy~\cite{Alam:2016cli, Carson:2018xri, Malik2018, Tsang:2019vxn, Guven:2020dok, Malik:2020vwo, Tsang:2020lmb, Malik_book, Reed:2021nqk, Pradhan:2022vdf, Pradhan:2022txg, Ghosh:2022lam, Ghosh:2021bvw, Beznogov:2022rri}. Depending upon the distribution of nuclear matter parameters, the correlations of NS properties with the nuclear matter parameters are also summarized in Table III of Ref.~\cite{Kunjipurayil:2022zah}. There may be other factors that affect these correlations such as (i) the behavior of the high-density part of the EOS, (ii) the choice of distributions of nuclear matter parameters, their interdependence, and uncertainties, (iii) the lower bound on the maximum mass of the stable neutron stars, (iv) the value of the density ($\rho_{c_s}$), (v) upper bound on the value of tidal deformability, are studied in Chapter-\ref{chap4}. The low-density ($\rho \leq \rho_{c_s}$) EOSs are constructed by employing our minimal constraints within a Bayesian approach. The EOSs beyond $\rho_{c_s}=1.5-2\rho_0$ are constrained only by imposing the causality condition on the speed of sound. The correlation of tidal deformability is found to be stronger with $K_{\rm sym,0}$ than with slope parameter $L_0$. These correlations become even stronger when the priors of lower-order nuclear matter parameters corresponding to incompressibility and symmetry energy coefficients are kept fixed. The correlations become noticeably weaker in the absence of interdependence among nuclear matter parameters.  The high-density part of the EOSs does not influence to the correlation of NS properties with symmetry energy parameters. With an increase in the upper bound of $\Lambda_{1.4}$, these correlations improved much.
These correlations improve a little bit with the increase in the density $\rho_{c_s}$. 
The results for the correlation of tidal deformability with the pressure at 2$\rho_0$ are robust and practically independent of all the factors considered. This is because the pressure is not determined by single parameters. It is composed of multivariate parameters.

A multivariate analysis is required to identify the key nuclear matter parameters that have the greatest impact on the neutron star's properties, such as tidal deformability and radius at canonical mass 1.4 $M_\odot$. In Chapter-\ref{chap5}, we perform a Principal Component Analysis (PCA) to investigate the connection between multiple nuclear matter parameters and the tidal deformability as well as the radius of neutron stars within the mass range of $1.2-1.8M_\odot$. The analysis is carried out using two different sets of  EOSs. The EOSs at low densities (using Taylor expansion \ref{Taylor}) have been derived using the uncorrelated uniform and joint posterior distributions of nuclear matter parameters. The EOSs at high density ($\rho>2\rho_0$) are constructed by imposing the causality condition on the speed of sound and are independent of compositions of NS matter (see the chapter \ref{H-eos}). 
We have found that the variability in the considered neutron star properties requires the incorporation of more than one principal component. These observations emphasize that the neutron star properties depend on multiple nuclear matter parameters. 
In particular, tidal deformability demands the inclusion of three or more principal components to account for over 90\% of its variations, while for the radius, two principal components suffice to explain similar variations. 
In the case of the tidal deformability of a neutron star, the symmetry energy curvature parameter $K_{\rm sym,0}$ emerges as the primary contributor to the first principal component. When considering the radius of a neutron star, the symmetry energy slope parameter $L_0$ stands out as the primary driver behind the observed variations. Overall, when analyzing the collective impact of iso-scalar parameters ($K_0$ and $Q_0$), these contributions exhibit an approximately 25\% increase with the increase in neutron star mass from 1.2$M_\odot$ to 1.8$M_\odot$. 

%% file: chapter6/future-work.tex
\chapter{Future Prospective}
\label{chap7}
In recent years, our understanding of the equation of state for dense matter has made significant progress. However, there is still space for expansion of the research presented in this thesis, particularly with regard to neutron stars. We have already started work on potential expansions, taking into consideration recent and upcoming developments in the available facilities around the world. In addition, multiple theoretical, observational, experimental, and empirical perspectives are being considered. Detailed below are some of our ongoing initiatives:-

\begin{enumerate}
    \item \textbf{Constraints the EOS in a model-independent way:-}\\
    Using different nuclear models, researchers have been looking into the possibility of discovering links between the parameters of nuclear matter and the properties of neutron stars. However, these association investigations have shown that there is a strong model reliance. This dependence on the model is caused by the fact that different models might result in distinct equations of state, even though the nuclear matter properties of the models are comparable. We plan to use a Bayesian analysis to study further the correlations between nuclear matter parameters and various astrophysical observables in order to overcome this issue and get a more model-independent point of view. This will allow us to better understand how various astrophysical observables relate to one another. We can evaluate these associations using a method that is less dependent on particular nuclear models as a result of this approach.
    \item \textbf{Machine Learning approach to nuclear matter studies:-}\\
    The correlation analysis is mostly used to find factors that are linearly dependent on each other. It is important to note, though, that the connection between observables in astrophysics and properties of nuclear matter might be complicated and not follow a straight line. Supervised machine learning methods can be used to find these non-linear connections. In neutron star physics, for example, a Deep Neural Network (DNN) has been used as a useful way to connect a limited set of mass-radius observed data to the equation of state plane. This methodology facilitates the investigation of intricate and non-linear correlations that might not be apparent alone through conventional correlation analysis.
    \item  \textbf{Neutron stars with high magnetic field:-}\\
    The magnetic fields of neutron stars are extraordinarily powerful, usually between $10^{12}$ and $10^{15}$ Gauss. They are created over millions of years from interstellar gases. Anomalous X-ray pulse stars are frequently linked to magnetars, which are neutron stars with considerably stronger fields, typically between $10^{12}$ and $10^{15}$  Gauss. Significant to the physics of compact star mergers, these magnetic fields also affect the structure and composition of neutron stars. In addition to the fact that magneto-rotational instability can cause magnetic fields to grow extraordinarily enormous, which can have an impact on the dense matter equations of state, gravitational waves released during mergers can also shed light on these equations.
    \item \textbf{{Exploring Neutron Star Oscillations and Hybrid Star Phase Transitions:-}} \\
    There is significant potential for future research on phase transitions in hybrid stars and oscillations in neutron stars. It is still an exciting and difficult study path to look at the equation of state governing these oscillation modes and to look for phase transitions within hybrid stars. Comprehending the basic characteristics of neutron stars, like their internal makeup and behavior, will contribute to our understanding of extreme astrophysical environments and clarify the complex physics of matter in extreme circumstances. In the end, this investigation might provide significant new information on the high-density matter domain and further our knowledge of the cosmos.

\end{enumerate}